\documentclass[11pt,a4paper]{article}
\usepackage{jcappub}
\usepackage{slashed}
\usepackage{braket}
\usepackage{hyperref}
\usepackage{float}
\usepackage{amssymb}
\usepackage{mathrsfs}
\usepackage{afterpage}

\newcommand{\Aa}[2]{A_{\alpha,#1}^{#2}}

\newcommand{\ain}{a_\mathrm{in}}
\newcommand{\arctanh}{{\mathrm{arctanh}}}
\newcommand{\Ba}[2]{B_{\alpha #1}^{#2}}
\newcommand{\bNi}[2]{b^{(#1)}_{#2}}
\newcommand{\Ca}[1]{C_{\alpha,#1}}
\newcommand{\ca}[1]{c_{\alpha,#1}}
\newcommand{\Da}[2]{D_{\alpha, #1}^{#2}}
\newcommand{\delD}{\delta^\mathrm{(D)}}
\newcommand{\delK}{\delta^\mathrm{(K)}}
\newcommand{\Dk}{{\mathcal D}_k}
\newcommand{\Ea}[2]{E_{\alpha #1}^{#2}}
\newcommand{\Even}{{\mathcal E}}
\newcommand{\EQ}[1]{Eq.~(\ref{#1})}
\newcommand{\EQS}[2]{Eqs.~(\ref{#1}-\ref{#2})}
\newcommand{\fcb}{f_\mathrm{cb}}
\newcommand{\FFTM}{{\tt FlowsForTheMasses}}

\newcommand{\fnu}{f_\nu}
\newcommand{\Hc}{{\mathcal H}}
\newcommand{\Hco}{{\mathcal H}_0}
\newcommand{\Ia}[2]{I_{\alpha, #1}^{#2}}
\newcommand{\intq}{\int_{\vec q}}
\newcommand{\Ja}[2]{J_{\alpha,#1}^{#2}}
\newcommand{\kfs}{k_\mathrm{FS}}
\newcommand{\Lc}{{\mathcal L}}
\newcommand{\Lx}{{\mathscr L}}
\newcommand{\Mc}{{\mathcal M}}
\newcommand{\MLeg}{{\mathcal R}}
\newcommand{\mua}{\mu_\alpha}
\newcommand{\muap}{\mu_{\alpha p}}
\newcommand{\muaq}{\mu_{\alpha q}}
\newcommand{\mupq}{\mu_{pq}}
\newcommand{\Nk}{N_k}
\newcommand{\Nkp}{N_{k\mathrm{p}}}

\newcommand{\Nmunl}{N_{\mu,\mathrm{NL}}}
\newcommand{\nua}{\nu_\alpha}
\newcommand{\nufpt}{\nu_\mathrm{FFT}}

\newcommand{\Obo}{\Omega_{\mathrm{b},0}}
\newcommand{\Ocb}{\Omega_\mathrm{cb}}
\newcommand{\Ocbo}{\Omega_{\mathrm{cb},0}}

\newcommand{\Odd}{{\mathcal O}}

\newcommand{\Omo}{\Omega_{\mathrm{m},0}}

\newcommand{\Ono}{\Omega_{\nu,0}}
\newcommand{\Pa}[2]{P_{\alpha #1}^{#2}}
\newcommand{\PLeg}{{\mathcal P}}
\newcommand{\Pot}{P^{(1,3)}}
\newcommand{\psia}[2]{\psi_{\alpha #1}^{#2}}
\newcommand{\psiba}[2]{\breve \psi_{\alpha #1}^{#2}}
\newcommand{\Ptt}{P^{(2,2)}}
\newcommand{\QLeg}{{\mathcal Q}}
\newcommand{\Sa}[2]{S_{\alpha,#1}^{#2}}
\newcommand{\sigmamnj}[3]{\sigma^{(#1,#2)}_{#3}}
\newcommand{\sign}{{\mathrm{sign}}}

\newcommand{\Xia}[2]{\Xi_{\alpha #1}^{#2}}

\newcommand{\Xita}[2]{{\tilde \Xi}_{\alpha #1}^{#2}}
\newcommand{\va}{v_\alpha}
\newcommand{\vtc}{c_\vartheta}
\newcommand{\vtct}{\tilde c_\vartheta}
\newcommand{\vts}{s_\vartheta}
\newcommand{\vtst}{\tilde s_\vartheta}
\newcommand{\Wa}[2]{W_{\alpha, #1}^{#2}}
\newcommand{\Za}[2]{Z_{\alpha, #1}^{#2}}

\title{Flows For The Masses: \\A multi-fluid non-linear perturbation theory for massive neutrinos}

\author[a]{Joe Zhiyu Chen,}
\author[b,a]{Amol Upadhye,}
\author[a]{Yvonne Y.~Y.~Wong}

\affiliation[a]{Sydney Consortium for Particle Physics and Cosmology,
  School of Physics, The University of New South Wales, 
  Sydney NSW 2052, Australia} 
\affiliation[b]{Astrophysics Research Institute,
  Liverpool John Moores University,
  146 Brownlow Hill, Liverpool L3 5RF, United Kingdom}

\emailAdd{a.r.upadhye@ljmu.ac.uk}

\abstract{
  Velocity dispersion of the massive neutrinos presents a daunting challenge for non-linear cosmological perturbation theory.  We consider the neutrino population as a collection of non-linear fluids, each with uniform initial momentum, through an extension of the Time Renormalization Group perturbation theory.  Employing recently-developed Fast Fourier Transform techniques, we accelerate our non-linear perturbation theory by more than two orders of magnitude, making it quick enough for practical use.  After verifying that the neutrino mode-coupling integrals and power spectra converge, we show that our perturbation theory agrees with N-body neutrino simulations to within $10\%$ for neutrino fractions $\Ono h^2 \leq 0.005$ up to wave numbers of $k = 1~h/$Mpc, an accuracy consistent with $\leq 2.5\%$ errors in the neutrino mass determination.  Non-linear growth represents a $>10\%$ correction to the neutrino power spectrum even for density fractions as low as $\Ono h^2 = 0.001$, demonstrating the limits of linear theory for accurate neutrino power spectrum predictions.  Our code \FFTM{} is avaliable online at {\tt github.com/upadhye/FlowsForTheMasses}~.
}

\begin{document}
\maketitle

\section{Introduction}
\label{sec:introduction}

Cosmic surveys are rapidly closing in on the sum of neutrino masses $\sum m_\nu$, one of the final unmeasured parameters of the Standard Model of particle physics.  Joint analyses using restrictive assumptions about the dark energy find $95\%$~CL upper bounds on $\sum m_\nu$ ranging from $0.09$~eV to $0.14$~eV~\cite{Planck:2018vyg,Palanque-Delabrouille:2019iyz,eBOSS:2020yzd,DES:2022ccp}, about twice the lower bound from neutrino oscillation experiments~\cite{deSalas:2017kay,Esteban:2018azc}.  Yet the case for considering larger neutrino masses remains compelling. Allowing a time-varying dark energy equation of state, or analyzing different combinations of cosmological data, can weaken the upper bound on the sum of neutrino masses to $\approx 0.5$~eV or larger~\cite{Upadhye:2017hdl,KiDS:2020ghu,Sgier:2021bzf}.  Neutrinos and other hot dark matter may help resolve long-running tensions in the cosmic expansion rate~\cite{Ghosh:2019tab,Lyu:2020lps,Escudero:2021rfi,Gu:2021lni,Cai:2022dkh}, measures of the clustering amplitude~\cite{Leauthaud:2016jdb,Lange:2020mnl,Amon:2022ycy}, and laboratory oscillation experiments~\cite{Abazajian:2012ys,Arguelles:2021meu,MiniBooNE:2022emn}, though others cast doubt upon such resolutions~\cite{Brinckmann:2020bcn,RoyChoudhury:2020dmd}.  Independently of these tensions, constraints are weak on the distribution functions of muon, tau, and hypothetical sterile neutrinos,  while substantial deviations from the Fermi-Dirac distribution can make larger-mass neutrinos consistent with observations~\cite{Oldengott:2019lke,Alvey:2021sji}.

Neutrinos, especially those with larger masses, present several unique cosmic signatures which may allow for their definitive detection.  They introduce a scale-dependence into the dark matter halo bias~\cite{LoVerde:2014pxa,Chiang:2017vuk,Chiang:2018laa,Banerjee:2019omr}; a dipole distortion in galaxy cross-correlations due to their relative velocities~\cite{Zhu:2013tma,Inman:2016prk,Zhu:2019kzb,Zhou:2021sgl}; a long-range correlation in galactic rotation directions~\cite{Yu:2018llx};  ``wakes'' caused by neutrinos coherently free-streaming past collapsed halos~\cite{Zhu:2014qma,Inman:2015pfa}; non-Gaussianity due to their clustering in voids~\cite{Liu:2020mzl}; a suppression in mass accretion by dark matter halos~\cite{Wong:2021ats}; and an environment-dependence of the halo mass function due to differential capture of neutrinos by halos~\cite{Yu:2016yfe}.  Quantifying these effects requires accurate computations of the formation of large-scale structure in the non-linear regime.

Chief among the obstacles to such a computation is the large thermal velocity dispersion of the neutrino population, which requires that we follow its full six-dimensional phase space, rather than the three dimensions characterizing cold species such as the baryons.  The most accurate approach is an N-body computer simulation realizing the cold dark matter (CDM) and baryons as a set of point particles, with neutrinos treated through either perturbation theory responding to non-linear CDM+baryon clustering~\cite{Brandbyge:2008js,AliHaimoud:2012vj,Archidiacono:2015ota,Dakin:2017idt,Mccarthy:2017yqf,Bird:2018all,Chen:2020bdf,Chen:2020kxi,Inman:2020oda} or as point particles themselves~\cite{Brandbyge:2008rv,Viel:2010bn,Villaescusa-Navarro:2013pva,Castorina:2015bma,Banerjee:2016zaa,Banerjee:2018bxy,Brandbyge:2018tvk,Bayer:2020tko,Elbers:2020lbn,Elbers:2022xid,Parimbelli:2022pmr}.  Since perturbation theory is more accurate for weakly-clustering neutrinos with high thermal velocities, while an N-body treatment more accurately captures non-linear neutrino clustering, hybrid simulations combine the two approaches, selectively realizing a low-velocity portion of the neutrino population as particles~\cite{Brandbyge:2009ce}.  In the absence of a complete non-linear neutrino perturbation theory, N-body and hybrid techniques have provided the only accurate computations of the non-linear neutrino clustering power.

We present the first fully non-linear perturbative calculation of the massive neutrino power spectrum.  Our theoretical advance represents the convergence of three distinct currents in cosmological perturbation theory: non-linear perturbation theory, multi-fluid treatments of hot dark matter, and Fast Fourier Transform (FFT) acceleration techniques.  Firstly, in response to observations of the matter and galaxy power spectra pushing to ever-smaller length scales, non-linear perturbation theory was developed to extend the reach of cosmological perturbation theory~\cite{Crocce:2005xy,Crocce:2005xz,McDonald:2006hf,Valageas:2006bi,Crocce:2007dt,Matsubara:2007wj,Taruya:2007xy,Matsubara:2008wx,Pietroni:2008jx,Lesgourgues:2009am,Garny:2020ilv}.  The Time-Renormalization Group (Time-RG) perturbation theory of Refs.~\cite{Pietroni:2008jx,Lesgourgues:2009am} and the bispectrum of Ref.~\cite{Fuhrer:2014zka} are of particular interest to us here.

Secondly, Dupuy and Bernardeau formulated a multi-fluid linear perturbation theory for massive neutrinos~\cite{Dupuy:2013jaa,Dupuy:2014vea,Dupuy:2015ega}.  The hot neutrino gas, with a thermal velocity dispersion described by a Fermi-Dirac distribution of momenta, is subdivided into bins, each characterized by a uniform initial velocity.  Importantly, this theory is Lagrangian in momentum space, meaning that neutrinos cannot jump from one velocity bin to another, simplifying the theoretical description.  Reference~\cite{Chen:2020bdf} developed a multi-fluid linear response calculation in which a non-relativistic version of this binned perturbation theory was used to track the response of massive neutrinos to the clustering of the cold dark matter (CDM) and baryons.

Thirdly, FFT techniques were applied by Refs.~\cite{McEwen:2016fjn,Fang:2016wcf,Schmittfull:2016jsw,Upadhye:2017hdl} to reducing the computational expense of non-linear perturbation theory by orders of magnitude.  By contrast to linear theory, non-linear perturbation theory couples different wave numbers.  It quantifies these couplings through multi-dimensional mode-coupling integrals which represent the bulk of its computational cost.  Time-RG perturbation theory is especially expensive due to requiring the computation of mode-coupling integrals at each time step. Its advantage is its ready applicability to multiple species which cluster differently and to expansion histories very different from that of the Einstein-de Sitter cosmology.   Reference~\cite{Upadhye:2017hdl} used FFTs to accelerate Time-RG and the galaxy bias of Ref.~\cite{McDonald:2009dh}, then applied them to a data analysis.

In this article, we construct an FFT-accelerated non-linear multi-fluid perturbation theory for massive neutrinos, which we call \FFTM{}.  We begin by extending Time-RG perturbation theory to a fluid with a spatially-uniform velocity $\vec v$.  Rather than depending solely upon the wave number $k$, the linear multi-fluid theory has a preferred direction $\hat v$, with perturbations dependent upon both $k$ and $\hat k \cdot \hat v$, the cosine of the angle between the velocity and the Fourier vector.  The bispectrum $B(\vec k, \vec q, \vec p)$, or Fourier-transformed three-point correlation function, is an object of central importance in Time-RG.  In the case of a non-zero velocity, the bipsectrum also gains a dependence upon the angles between its three wave vectors and the velocity.  We generalize the bispectrum integrals of Time-RG, which quantify non-linear corrections to the power spectrum, to apply to a fluid with a uniform background velocity.  Then we derive the equations of motion of these generalized bispectrum integrals, whose non-linear contribution, in turn, is sourced by generalized mode-coupling integrals.

Next, we proceed to a systematic study of these mode-coupling integrals.  In ordinary Time-RG, assuming a power spectrum tracked using $N_k \sim 100$ wave number bins, the computation of a two-dimensional mode-coupling integral at each of the $N_k$ points involves a computational cost ${\mathcal O}(N_k^3)$.  In our extension to multi-fluid perturbation theory, the number of these mode-coupling integrals also grows as a high power of $N_\mu$, the number of angular modes which we use to track the dependence of perturbations upon the angle between $\vec k$ and $\vec v$.  Thus FFT acceleration, which reduces the computational cost of a single mode-coupling integral from ${\mathcal O}(N_k^3)$ to ${\mathcal O}(N_k \log N_k)$, is essential to our method.  After considerable algebra, we derive mode-coupling integrals that are percent-level accurate over our region of interest, and we use FFT acceleration to reduce their computational cost by a factor of four hundred.

Finally, we put our perturbation theory to the test against the hybrid neutrino N-body simulation of Ref.~\cite{Chen:2022dsv}, which itself is based upon a multi-fluid treatment of the neutrinos.  This hybrid simulation can track individual neutrino fluids, either using the multi-fluid linear response of Ref.~\cite{Chen:2020bdf}, or by realizing them as massive particles, each beginning with a velocity $\vec v$ in addition to any velocity kicks it receives from the local gravitational potential.  We confirm the accuracy of our perturbation theory, particularly at early times, $z \gtrsim 1$, and higher velocities, $v/c \gtrsim 0.2\%$.  Our results accord with our intuition that non-linear perturbation theory will lose accuracy as the non-linear corrections come to dominate the power.  Nevertheless, we find that even our fully perturbative computation agrees with simulations to $<10\%$ at $z=0$ for neutrino fractions up to $\Ono h^2 = 0.005$ and $k\leq 1~h/$Mpc.

High-velocity neutrinos are particularly difficult for N-body particle simulations.  Firstly, they require small time steps, since they move quickly past collapsed structures which could deflect them.  Secondly, momentum shells at higher $|\vec p|$ occupy more phase space volume, meaning that more particles are necessary in order to keep shot noise under control.  Thus our work points the way to a hybrid multi-fluid non-linear-response simulation in which each neutrino fluid would be evolved linearly until its dimensionless power spectrum exceeded a predetermined threshold.  Then it would be evolved non-linearly until the non-linear corrections themselves became a significant portion of the total power, at which point it could be converted into simulation particles.  Our non-linear multi-fluid perturbation theory of massive neutrinos, when integrated into a hybrid simulation, promises several advantages:
\begin{enumerate}
\item Non-linear corrections make perturbative neutrino calculations more
  accurate.
\item Error in the perturbative calculation may be estimated using the ratio of
  non-linear corrections to the linear power.
\item Multi-fluid perturbation theory offers fine-grained information about the
  velocity distribution of neutrino particles.
\item Conversion to particles may be delayed relative to linear theory,
  reducing their velocities and alleviating the resulting shot noise.
\item Several neutrino fluids may be handled entirely using non-linear
  perturbation theory, avoiding a particle treatment altogether.
\end{enumerate}

This paper is organized as follows.  Section~\ref{sec:background} describes the three strands woven together in the rest of the paper: multi-fluid linear theory, Time-RG perturbation theory, and FFT acceleration of mode-coupling integrals.  These three are joined in Sec.~\ref{sec:time-rg_for_massive_nu}, which derives the equations of motion in our \FFTM{} perturbation theory and defines the necessary mode-coupling integrals.  In Secs.~\ref{sec:mode-coupling_P22} and \ref{sec:mode-coupling_P13}, we compute these FFT-accelerated integrals and test them against a direct numerical integration.  Section~\ref{sec:nu_NL_enhancement} assembles these pieces into a complete non-linear perturbation theory, tests its convergence, and then compares it to the results of a hybrid N-body simulation, while Sec.~\ref{sec:conclusion} concludes.

\section{Background}
\label{sec:background}

\subsection{Multi-fluid perturbation theory}
\label{subsec:bkg_mflr}

Cosmological neutrinos are characterized by a Fermi-Dirac distribution of velocities in the early universe after they have decoupled from electrons, positrons, and photons during big bang nucleosynthesis.  However, cosmological perturbation theory is founded upon the continuity and Euler equations governing the dynamics of fluids, which have well-defined velocity fields.  Multi-fluid perturbation theory resolves this conflict by discretizing the Fermi-Dirac distribution, with each bin corresponding to a uniform zeroth-order comoving velocity field.

Let $\breve\psi_{\alpha a}(\eta, \vec x, \vec v_\alpha)$ be a Gaussian random perturbation variable for a neutrino fluid $\alpha$ with velocity $\vec v_\alpha = v_\alpha \hat v_\alpha$ relative to the total matter density.  The index $a$ is $0$ for the density contrast $\delta_\alpha$ and $1$ for $\theta_\alpha := -\vec\nabla\cdot\vec V_\alpha / \Hc$, the dimensionless divergence of the peculiar velocity field $\vec V_\alpha$.  The time variable $\eta = \ln(a/\ain)$ for some initial scale factor $\ain$.  

Reference~\cite{Ma:1995ey} expands its Fourier transform $\breve\psi_{\alpha,a}(\eta, \vec k, \vec v_\alpha)$ in Legendre moments as
\begin{equation}
  \breve\psi_{\alpha a}(\eta, \vec k, \vec v_\alpha)
  =
  \sum_\ell (-i)^\ell \mathcal{P}_\ell(\hat k \cdot \hat v_\alpha)
  \breve\psi_{\alpha a \ell}(\eta, \vec k, v_\alpha)
  \label{e:Legendre_expansion}
\end{equation}
where the $\breve\psi_{\alpha,a,\ell}$ are random variables; note a difference in the Legendre expansion convention.  These can further be expanded as
\begin{equation}
  \breve\psi_{\alpha a \ell}(\eta, \vec k, v_\alpha)
  =
  \sum_m \breve C_{\alpha a \ell m}(k) Y_{\ell m}(\hat k)
  \psi_{\alpha a \ell}(\eta, k, v_\alpha)
  \label{e:Ylm_expansion}
\end{equation}
where the $\breve C_{\alpha a \ell m}$ are random variables specified at the initial time, and the $\psi_{\alpha a \ell}(\eta, k, v_\alpha)$ are transfer functions whose normalizations we define later.  This is a spherical harmonic expansion of $\breve\psi_{\alpha a}$ in which the coefficient $ \mathcal{P}_\ell(\hat k \cdot \hat v_\alpha) \psi_{\alpha a \ell} \breve C_{\alpha a \ell m}$ of $Y_{\ell m}$ is explicitly dependent upon the angle between the Fourier vector and the velocity direction.

In the subhorizon limit of linear perturbation theory, the Fourier space equations of motion obeyed by the $\breve\psi_{\alpha a}(\eta, \vec k, \vec v_\alpha)$ are Eqs.~(3.2,~3.4) of Ref.~\cite{Chen:2020bdf},
\begin{eqnarray}
  \breve\psi_{\alpha 0}'
  &=&
  -\frac{i \vec k \cdot \vec v_\alpha}{\Hc} \breve\psi_{\alpha 0}
  + \breve\psi_{\alpha 1}
  \label{e:eom_delta_mflr_0}
  \\
  \breve\psi_{\alpha 1}'
  &=&
  -\left(1 + \frac{\Hc'}{\Hc}\right) \breve\psi_{\alpha 1}
  - \frac{i \vec k \cdot \vec v_\alpha}{\Hc} \breve\psi_{\alpha 1} 
  - \frac{k^2}{\Hc^2} \breve\Phi
  \label{e:eom_theta_mflr_0}
\end{eqnarray}
where primes denote derivatives with respect to $\eta$.  The gravitational potential $\breve \Phi$ is a random variable proportional to the density-weighted sums of all density contrasts, hence it is independent of $\hat v_\alpha$ and only depends upon the monopole perturbations $\breve\psi_{\alpha a 0}$.

By substituting the Legendre polynomial expansion of Eqs.~(\ref{e:Legendre_expansion},~\ref{e:Ylm_expansion}) and noting that $(2\ell + 1) \mu \mathcal{P}_\ell(\mu) = \ell \mathcal{P}_{\ell-1}(\mu) + (\ell+1) \mathcal{P}_{\ell+1}(\mu)$, we find the evolution equations of the $\psia{a\ell}{}(\eta,k,\va)$:
\begin{eqnarray}
  (\psia{0\ell}{})'
  &=&
  \psia{1\ell}{}
  + \frac{k\va}{\Hc}\left(
  \frac{\ell}{2\ell-1}\psia{0,\ell-1}{}
  - \frac{\ell+1}{2\ell+3}\psia{0,\ell+1}{}
  \right)
  \label{e:eom_delta_mflr_1}
  \\
  (\psia{1\ell}{})'
  &=&
  -\delK_{0\ell} \frac{k^2}{\Hc^2}\Phi
  - \left(1+\frac{\Hc'}{\Hc}\right) \psia{1,\ell}{}
   + \frac{k\va}{\Hc}\left(
  \frac{\ell}{2\ell-1}\psia{1,\ell-1}{}
  - \frac{\ell+1}{2\ell+3}\psia{1,\ell+1}{}
  \right)
  \label{e:eom_theta_mflr_1}
  \\
  k^2\Phi
  &=&
  -\frac{3}{2}\Hc^2
  \left( \Ocb(\eta) \delta_\mathrm{cb}
  + \sum_{\alpha=0}^{N_\tau-1} \Omega_\alpha(\eta) \psia{00}{} \right)
  \label{e:eom_Phi}
\end{eqnarray}
where $\delK$ is the Kronecker delta. Time-dependent density fractions in \EQ{e:eom_Phi} are $\Ocb(\eta) = \Ocbo \Hco^2 \Hc^{-2} a(\eta)^{-1}$ and $\Omega_\alpha(\eta) = \Omega_{\alpha,0} \Hco^2 \Hc^{-2} a(\eta)^{-1} (1-\va^2)^{-1/2}$~\cite{Chen:2020bdf}. Coupling between different Legendre moments occurs only through free-streaming terms proportional to $k v_\alpha / \Hc$.

Note that the direction of the velocity $\vec \va$ has disappeared from the equations of motion, \EQS{e:eom_delta_mflr_1}{e:eom_Phi}, in favor of the Legendre index $\ell$ for the angle between $\vec k$ and $\vec \va$.  This reflects the fact that perturbations with different velocities, $\vec \va \neq \vec v_{\alpha'}$ but identical speeds, $\va = v_{\alpha '}$, evolve in the same way.  Henceforth we employ the term {\em flow} to denote the set of all fluids with the same speed $v$. The index $\alpha$ in the above equations of motion is then reinterpreted as a flow index, describing all fluids making up the flow characterized by the speed $\va$.  The multi-flow perturbation theory developed here is a special case of multi-fluid perturbation theories, which could in general treat the different fluids making up a flow differently.  Following Ref.~\cite{Chen:2020bdf}, we discretize the Fermi-Dirac distribution into bins of equal number density.

For compactness, we recast \EQS{e:eom_delta_mflr_1}{e:eom_theta_mflr_1} using a $2\times2$ linear evolution matrix $\Xia{ab\ell}{}(\eta, \vec k)$.  Furthermore, we use a shorthand in which the time-dependence of perturbations is suppressed, and their dependence on a wave number is expressed as a superscript.  Thus $\breve \psi_{\alpha a}^{\vec k}(\vec \va)$ represents $\breve \psi_{\alpha a}(\eta,\vec k, \vec v_\alpha)$ and $\psia{a \ell}{k}$ represents $\psi_{\alpha a \ell}(\eta,k,v_\alpha)$; in the latter case, $\va$ is already specified by $\alpha$. In this notation, the linear equation of motion for the transfer function is
\begin{eqnarray}
  {\psia{a \ell}{k}}'
  &=&
  -\Xia{ac\ell}{k}\psia{c\ell}{k}
  \label{e:eom_lin_psi_Xi4}
\\
  \Xia{ab\ell}{k}
  &=&
  \left[
    \begin{array}{cc}
      0 & -1 \\
      \frac{k^2 \Phi^k}{\Hc^2 \psia{0 0}{k}}\delK_{\ell 0}
      & 1 + \frac{\Hc'}{\Hc}
    \end{array}
    \right]
  -\frac{k v_\alpha}{\Hc}
  \!\left(\!\frac{\ell}{2\ell-1}\frac{\psia{a,\ell-1}{k}}{\psia{a\ell}{k}}
  \!-\! \frac{\ell+1}{2\ell+3}\frac{\psia{a,\ell+1}{k}}{\psia{a\ell}{k}}
  \!\right) \pmb{1}\qquad
  \label{e:Xia4}
\end{eqnarray}
where summation over the repeated index $c$ in \EQ{e:eom_lin_psi_Xi4} is implicit, and $\pmb{1}$ is the $2\times2$ identity matrix.  Rows and columns refer, respectively, to the $a$ and $b$ indices of $\Xia{ab\ell}{k}$.

We define the power spectrum of flow $\alpha$ as
\begin{equation}
  (2\pi)^3 \delD(\vec k - \vec k') P_{\alpha a b}(\eta, \vec k, \vec v_\alpha)
  =
  (2\pi)^3 \delD(\vec k - \vec k') \Pa{ab}{\vec k}
  :=
  \left<
  \breve\psi_{\alpha a}^{\vec k *}
  \breve\psi_{\alpha b}^{\vec k'}
  \right>
  =
  \left<
  \breve\psi_{\alpha a}^{-\vec k}
  \breve\psi_{\alpha b}^{\vec k'}
  \right>.
  \label{e:power_spectrum}
\end{equation}
The third equality follows from the fact that $\breve\psi_{\alpha a}(\eta, \vec x, \vec v_\alpha)$ is real, implying $(\psiba{a}{\vec k})^* = \psiba{a}{-\vec k}$.  We have yet to choose normalizations for the random variables $\breve C^k_{\alpha a \ell m}$ and the transfer functions $\psi^k_{\alpha a \ell}$.  Perturbations with different $\ell$ should be uncorrelated, while in linear theory, $\delta$ and $\theta$ are perfectly correlated, so $\left< \breve C^{k*}_{\alpha a \ell m} \breve C^k_{\alpha b \ell' m'}\right>$ should be proportional to $\delK_{\ell,\ell'} \delK_{m,m'}$.  We choose
\begin{equation}
\left< \breve C^{k*}_{\alpha a \ell m} \breve C^k_{\alpha b \ell' m'}\right>
=
\frac{4\pi}{2\ell + 1} \delK_{\ell,\ell'} \delK_{m,m'} V
\end{equation}
pushing all non-random dependence on $k$ into the transfer functions.  With this definition, we may substitute $\breve\psi_{\alpha a}$ into Eq.~(\ref{e:power_spectrum}) to find the power and its linear evolution:
\begin{eqnarray}
  \Pa{ab}{\vec k}
  &=&
  \sum_\ell \mathcal{P}_\ell(\mua)^2
  \psi^k_{\alpha a \ell}\psi^k_{\alpha b \ell}
  =:
  \sum_\ell \mathcal{P}_\ell(\mua)^2 P^k_{\alpha a b \ell}
  \label{e:power_Legendre2_expansion}
  \\
  {\Pa{ab\ell}{k}}'
  &=&
  -\Xia{ac\ell}{k} \Pa{cb\ell}{k} - \Xia{bc\ell}{k} \Pa{ac\ell}{k}.
  \label{e:eom_lin_Pabl}
\end{eqnarray}
where we have defined $\mua := \hat k \cdot \hat v_\alpha$ and  $P_{\alpha a b \ell} := \psi_{\alpha a \ell} \psi_{\alpha b \ell}$.  Thus the transfer functions in our convention also contain the linear growth factors and the power spectrum normalization.  Since the $\psia{a\ell}{k}$ may be negative while the $\Pa{aa\ell}{k}$ are positive by definition, we cannot express $\Xia{ab\ell}{k}$ as functions of the $\Pa{ab\ell}{k}$ alone.  Thus we regard the $\psia{a\ell}{k}$ as more fundamental to multi-fluid perturbation theory than the power spectra $\Pa{ab\ell}{k}$.

\subsection{Time-Renormalization Group perturbation theory}
\label{subsec:bkg_time-rg}

Above we considered linear evolution, but the continuity and Euler equations of fluid dynamics are quadratic in the density and velocity.  In the case of interest here, with an irrotational peculiar velocity field $\vec\nabla \times \vec V_\alpha = 0$ allowing the representation of the peculiar velocity by a scalar perturbation $\theta_\alpha$, these equations take the form
\begin{eqnarray}
  (\psiba{a}{\vec k})'
  &=&
  -\breve \Xi_{\alpha ab}^{\vec k} \psiba{b}{\vec k}
  + \int \frac{d^3q}{(2\pi)^3}\frac{d^3p}{(2\pi)^3}
  (2\pi)^3 \delD(\vec k - \vec q - \vec p)
  \gamma_{abc}^{\vec k \vec q \vec p} \psiba{b}{\vec q} \psiba{c}{\vec p}
  \nonumber\\
  &=&
  -\breve \Xi_{\alpha ab}^{\vec k} \psiba{b}{\vec k}
  + \intq \gamma_{abc}^{\vec k \vec q \vec p} \psiba{b}{\vec q} \psiba{c}{\vec p}
  \label{e:eom_nonlin_psi}
  \\
  \gamma_{001}^{\vec k \vec q \vec p}
  &=&
  \frac{(\vec q + \vec p)\cdot \vec p}{2p^2},
  \quad
  \gamma_{010}^{\vec k \vec q \vec p}
  =
  \frac{(\vec q + \vec p)\cdot \vec q}{2q^2}
  =
  \gamma_{001}^{\vec k \vec p \vec q},
  \quad
  \gamma_{111}^{\vec k \vec q \vec p}
  =
  \frac{(\vec q + \vec p)^2 \vec q \cdot \vec p}{2 q^2 p^2}
  \label{e:gamma}
\end{eqnarray}
and all other $\gamma_{abc}^{\vec k \vec q \vec p}$ zero.  Here we have employed the shorthand $\intq X(\vec k, \vec q, \vec p)$, given any function $X$, for $\int \frac{d^3q}{(2\pi)^3}\frac{d^3p}{(2\pi)^3} (2\pi)^3 \delD(\vec k - \vec q - \vec p) X(\vec k, \vec q, \vec p)$.

In this subsection we restrict our consideration to cold fluids for which only the monopole power spectrum $\Pa{ab0}{k}$ is nonzero.  Evolution of this power spectrum is governed by
\begin{equation}
{\Pa{ab}{\vec k}}'
=
- \Xia{ac0}{k} \Pa{cb}{\vec k} - \Xia{bc0}{k}\Pa{ac}{\vec k}
+ \Ia{acd,bcd}{\vec k} + \Ia{bcd,acd}{\vec k}.
\label{e:eom_P_cb}
\end{equation}
Here, the $I$ are integrals over the bispectrum $\Ba{abc}{\vec k \vec q \vec p}$:
\begin{eqnarray}
  (2\pi^3)^3 \delD(\vec k - \vec q - \vec p)\Ba{abc}{\vec k \vec q \vec p}
  &=&
  \left< \psiba{a}{\vec k *} \psiba{b}{\vec q} \psiba{c}{\vec p}\right>
  \label{e:def_B}
  \\
  \Ia{acd,bef}{\vec k}
  &:=&
  \intq \gamma_{acd}^{\vec k \vec q \vec p} \Ba{bef}{\vec k \vec q \vec p}.
  \label{e:def_I}
\end{eqnarray}
Further application of the non-linear evolution \EQ{e:eom_nonlin_psi} shows that the bispectrum depends upon the four-point function, which in turn depends upon the five-point function, and so forth.  The continuity and Euler equations give rise to an infinite tower of evolution equations.

Time-RG perturbation theory makes two approximations:
\begin{enumerate}
\item {\em truncation} of the infinite hierarchy of evolution equations by
  neglecting the trispectrum, the connected part of the four-point function; and
\item {\em closure} of the linear evolution of the $\Ia{acd,bef}{\vec k}$, in
  the sense that $(\Ia{acd,bef}{\vec k})'$ depends upon the bispectrum only
  through the set of $\Ia{acd,bef}{\vec k}$.
\end{enumerate}
Specifically, truncation is the approximation that
\begin{equation}
  \left<\psiba{a}{\vec k_1} \psiba{b}{\vec k_2}
  \psiba{c}{\vec k_3} \psiba{d}{\vec k_4}\right>
  \approx
  \left<\psiba{a}{\vec k_1} \psiba{b}{\vec k_2}\right>
  \left<\psiba{c}{\vec k_3} \psiba{d}{\vec k_4}\right>
  +
   \left<\psiba{a}{\vec k_1} \psiba{c}{\vec k_3}\right>
  \left<\psiba{b}{\vec k_2} \psiba{d}{\vec k_4}\right>
  +
   \left<\psiba{a}{\vec k_1} \psiba{d}{\vec k_4}\right>
   \left<\psiba{b}{\vec k_2} \psiba{c}{\vec k_3}\right>.
   \label{e:trg_truncation}
\end{equation}
Using \EQ{e:power_spectrum} we may reduce the right hand side to a series of power spectrum pairs.  Thus the non-linear evolution of the bispectrum is governed by
\begin{eqnarray}
  (\Ba{abc}{\vec k \vec q \vec p})'
  &=&
  -\Xia{ad0}{k} \Ba{dbc}{\vec k \vec q \vec p}
  - \Xia{bd0}{q} \Ba{adc}{\vec k \vec q \vec p}
  - \Xia{cd0}{p} \Ba{abd}{\vec k \vec q \vec p}
  \ldots
  \nonumber\\
  &~&
  + 2\gamma_{ade}^{\vec k \vec q \vec p} \Pa{db}{\vec q} \Pa{ec}{\vec p}
  + 2\gamma_{bde}^{\vec q, \vec p, -\vec k} \Pa{dc}{\vec p} \Pa{ea}{\vec k}
  + 2\gamma_{cde}^{\vec p, -\vec k, \vec q} \Pa{da}{\vec k} \Pa{eb}{\vec q}
  \label{e:eom_B_cb}
\end{eqnarray}
in the Time-RG truncation approximation.

Evolution of the bispectrum integral $\Ia{acd,bef}{\vec k}$ is determined by the substitution of \EQ{e:eom_B_cb} into the derivative of \EQ{e:def_I}, that is, $(\Ia{acd,bef}{\vec k})' = \intq \gamma_{acd}^{\vec k \vec q \vec p} (\Ba{bef}{\vec k \vec q \vec p})'$.  The first term on the right hand side of \EQ{e:eom_B_cb} presents no problems, since the linear evolution matrix can be pulled outside of the integral, resulting in $-\Xia{bg0}{k} \Ia{acd,gef}{\vec k}$.  However, in the second and third terms, $\Xia{}{}$ depends upon one of the integration variables and cannot be factored out.  The Time-RG closure approximation, $\Xia{bd0}{q} \approx \Xia{bd0}{k}$ and $\Xia{cd0}{p} \approx \Xia{cd0}{k}$ inside these integrals, allows us to treat the second and third terms in $(\Ia{acd,bef}{\vec k})'$ just like the first, resulting in the linear terms $-\Xia{bg0}{k} \Ia{acd,gef}{\vec k} - \Xia{eg0}{k} \Ia{acd,bgf}{\vec k} - \Xia{fg0}{k} \Ia{acd,beg}{\vec k}$. 

Scale-dependence of $\Xia{ab0}{k}$ is limited to its $\Xia{100}{k}$ component.  Furthermore, the most common application of Time-RG is to the combined CDM+baryon fluid, making up a fraction $\fcb \approx 1$ of the total matter, in the presence of a massive neutrino population making up a fraction $\fnu \ll 1$ which is treated linearly.  In this case, the scale dependence of $\Xia{100}{k}$ is suppressed by the ratio $\fnu/\fcb \ll 1$.  Thus $\Xia{ab0}{q}$ is approximately $\Xia{ab0}{k}$ across the full range of $q$.  Reference~\cite{Upadhye:2013ndm} finds the error resulting from the closure approximation to be negligible.

Finally, Time-RG evolution of $\Ia{acd,bef}{\vec k}$ includes a non-linear correction arising from the final three terms on the right side of \EQ{e:eom_B_cb}.  We define the mode-coupling integral
\begin{equation}
  \Aa{acd,bef}{\vec k}
  :=
  \intq \gamma_{acd}^{\vec k \vec q \vec p}
  \left[ \gamma_{bgh}^{\vec k \vec q \vec p} \Pa{ge}{\vec q} \Pa{hf}{\vec p}
    + \gamma_{egh}^{\vec q,-\vec p,\vec k} \Pa{gf}{\vec p} \Pa{hb}{\vec k}
    + \gamma_{fgh}^{\vec p,\vec k,-\vec q} \Pa{gb}{\vec k} \Pa{he}{\vec q} \right].
  \label{e:def_A}
\end{equation}
Summation over the repeated $g$ and $h$ indices is implicit.  With this definition, along with the closure approximation, the non-linear evolution of $\Ia{acd,bef}{\vec k}$ is quantified by
\begin{eqnarray}
  (\Ia{acd,bef}{\vec k})'
  &=&
  -\Xia{bg0}{k} \Ia{acd,gef}{\vec k}
  - \Xia{eg0}{k} \Ia{acd,bgf}{\vec k}
  - \Xia{fg0}{k} \Ia{acd,beg}{\vec k}
  + 2\Aa{acd,bef}{\vec k}.
  \label{e:eom_I_cb}
\end{eqnarray}
Time-RG evolution of cold matter perturbations is thus governed by \EQ{e:eom_P_cb} and \EQ{e:eom_I_cb}.  Initial conditions set $\Pa{ab}{\vec k}$ to its linear approximation and $\Ia{acd,bef}{\vec k}$ to zero or $2\Aa{acd,bef}{\vec k}$~\cite{Audren:2011ne}.  By contrast to linear perturbation theory, in which the power at each wave number $\vec k$ evolves independently, the mode-coupling integrals couple power spectra at different wave numbers.

\subsection{Fast Fourier Transform acceleration}
\label{subsec:bkg_FFT_accel}

Numerical calculation of the mode-coupling integrals of \EQ{e:def_A} is the most computationally expensive step in Time-RG perturbation theory.  Over the past several years, FFT-based techniques such as FAST-PT have been developed by Refs.~\cite{McEwen:2016fjn,Fang:2016wcf,Schmittfull:2016jsw} to accelerate the computation of perturbation theory integrals.  Reference~\cite{Upadhye:2017hdl} applied these techniques to Time-RG in redshift space, whose line of sight direction is conceptually similar to the velocity direction $\hat \va$ for a moving fluid.  Here we briefly summarize the FAST-PT method for Time-RG.

Note that out of the three terms on the right hand side of \EQ{e:def_A}, the second can be transformed into the third by simultaneously switching the indices $c \leftrightarrow d$ and $e \leftrightarrow f$ as well as the wave numbers $\vec q \leftrightarrow \vec p$.  This follows from the symmetry $\gamma_{acb}^{\vec k \vec p \vec q} = \gamma_{abc}^{\vec k \vec q \vec p}$ of \EQ{e:gamma}.  Thus the integrand of $\Aa{acd,bef}{\vec k}$ contains two types of terms, those in which both power spectra depend upon the integration variables and those in whicn only one power spectrum does.  We refer to these respectively as $\Ptt$-type and $\Pot$-type terms by analogy with Standard Perturbation Theory, which expands $\delta(\vec k) = \sum_n \delta^{[n]}(\vec k)$ in perturbative orders and finds two leading-order corrections to the power, $\Ptt(k) \propto \left< \delta^{[2]}(\vec k)^* \delta^{[2]}(\vec k) \right>$ and $\Pot(k) \propto \left< \delta^{[1]}(\vec k)^* \delta^{[3]}(\vec k) + \delta^{[3]}(\vec k)^* \delta^{[1]}(\vec k)\right>$, respectively integrating over two and one power spectra.  

The FAST-PT technique of Ref.~\cite{McEwen:2016fjn} computes $\Ptt$-type terms, in which both power spectra in the integrand depend upon integration variables, by expanding them in terms of
\begin{equation}
  J_{\alpha\beta\ell}(k)
  =
  \intq q^\alpha p^\beta \PLeg_\ell(\mupq) P(q) P(p)
  \label{e:def_J}
\end{equation}
with $\mupq := \hat p \cdot \hat q$ and both power spectra inside the integrand being the linear $\delta$ power spectra.  Reference~\cite{McEwen:2016fjn} demonstrates that $J_{\alpha\beta\ell}(k)$ can be computed rapidly through a discrete convolution followed by an inverse discrete Fourier transform.  

Meanwhile, $\Pot$-type terms such as the third term on the right side of \EQ{e:def_A} can be reduced to integrals over a single variable $x = q/k$ by pulling all factors independent of $\vec q$ and $\vec p$ outside of their integrands.  The result is an integral of the form $k^3 P_{gb}(k) \int dx \, x^2 P_{he}(kx) G(x)$ with $G$ determined by the indices $a$, $c$, $d$, $b$, $e$, $f$, $g$, and $h$.  Defining $w = -\log(x)$, ${\tilde P}(w) = P_{he}(e^w)$, and ${\tilde G}(w) = G(e^{-w})e^{-3w}$, we recognize this as the convolution of $\tilde G$ and $\tilde P$, which may be computed directly or using the FFT algorithm, as discussed in  Ref.~\cite{McEwen:2016fjn}.

While a brute-force numerical computation such as that described in Ref.~\cite{Pietroni:2008jx} requires ${\mathcal O}(N_k^3)$ operations, for $N_k$ wave numbers, the FAST-PT computation needs only ${\mathcal O}(N_k \log(N_k))$, representing a significant acceleration of the most computationally-expensive integrals in Time-RG perturbation theory.  FAST-PT computation of all $\Ptt$-type and $\Pot$-type components of the Time-RG mode-coupling integral of \EQ{e:def_A} was worked out in detail in Ref.~\cite{Upadhye:2017hdl} for the purpose of constraining the neutrino masses.  In order to study redshift-space distortions, that reference extended Time-RG to the case of a preferred direction, which is directly relevant to the massive neutrino streams considered here.

\section{Non-linear perturbation theory for massive neutrinos}
\label{sec:time-rg_for_massive_nu}

\subsection{Extension of Time-RG for hot dark matter}
\label{subsec:trg_hdm}

Time-RG perturbation theory for cold matter is based upon the non-linear evolution equation for a single perturbation, \EQ{e:eom_nonlin_psi}, applied to the power spectrum and bispectrum.  Time-RG as formulated in Ref.~\cite{Pietroni:2008jx} evolved the three CDM+baryon power spectra, $P_{00}$, $P_{01}$, and $P_{11}$, along with the bispectrum integrals.  Though the linear $P_{01}$ is simply the geometric mean of $P_{00}$ and $P_{11}$, non-linear evolution allows for the density-velocity correlation coefficient $P_{01} / \sqrt{P_{00} P_{11}}$ to fall below unity.  Thus the two perturbations $\delta$ and $\theta$ are no longer sufficient for capturing all information contained in the power spectra.

In this work we evolve the perturbations $\delta$ and $\theta$ themselves, rather than the power spectra, in order to define $\Xia{ab\ell}{k}$ as in \EQ{e:Xia4}.  As a result we must define a third perturbation $\chi := 1 - P_{01} / \sqrt{P_{00} P_{11}}$ quantifying this density-velocity correlation. Generalizing to the angle-dependent $\psia{a\ell}{k}$ of \EQ{e:Ylm_expansion}, we define
\begin{equation}
  \psia{2\ell}{k} := \chi_{\alpha \ell}^k
  :=
  1 - \Pa{01,\ell}{k} / \sqrt{ \Pa{00,\ell}{k} \Pa{11,\ell}{k} }
  \quad\Longleftrightarrow\quad
  \Pa{01,\ell}{k}
  =
  \left(1-\chi_{\alpha \ell}^k\right) \psia{0\ell}{k} \psia{1\ell}{k}.
  \label{e:def_chi}
\end{equation}
We thus replace the linear power spectrum of \EQ{e:power_Legendre2_expansion} by
\begin{eqnarray}
  \Pa{ab}{\vec k}
  &=&
  \sum_\ell \mathcal{P}_\ell(\mua)^2
  \left[ 1 - \left(1-\delK_{ab}\right)\chi_{\alpha \ell}^k  \right]
  \psi^k_{\alpha a \ell}\psi^k_{\alpha b \ell}
  =:
  \sum_\ell \mathcal{P}_\ell(\mua)^2 P^k_{\alpha a b \ell}
  \label{e:power_Legendre2_expansion_NL}
  \\
  \Rightarrow
  \Pa{ab\ell}{k}
  &=&
  \left[ 1 - \left(1-\delK_{ab}\right)\chi_{\alpha \ell}^k  \right]
  \psi^k_{\alpha a \ell}\psi^k_{\alpha b \ell}
  \label{e:def_Pabl_NL}
\end{eqnarray}
where $a,b \in \{0,1\}$.  The quantity in square brackets is unity if $a$ equals $b$ and $1-\chi_{\alpha \ell}^k$ otherwise.  Derivatives of the perturbations follow directly from \EQ{e:def_Pabl_NL},
\begin{eqnarray}
  (\psia{a\ell}{k})'
  &=&
  \frac{(\Pa{aa\ell}{k})'}{2 \psia{a\ell}{k}}
  \quad
  \textrm{if}~a\in\{0,1\}
  \label{e:eom_delta_theta_NL_0}
  \\
  (\psia{2\ell}{k})'
  &=&
  (\chi_{\alpha \ell}^{k})'
  =
  -\frac{(\Pa{01\ell}{k})'}{\delta_{\alpha\ell}^{k}\theta_{\alpha\ell}^{k}}
  + (1-\chi_{\alpha \ell}^k) \frac{(\delta_{\alpha\ell}^k)'}{\delta_{\alpha\ell}^k}
  + (1-\chi_{\alpha \ell}^k) \frac{(\theta_{\alpha\ell}^k)'}{\theta_{\alpha\ell}^k},
  \label{e:eom_chi_NL_0}
\end{eqnarray}
given Time-RG calculations of the power spectrum derivatives.

Large flow velocities $\vec \va$ directly affect the non-linear evolution of massive neutrinos and other hot dark matter.  Dupuy and Bernardeau argue in Refs.~\cite{Dupuy:2013jaa,Dupuy:2014vea,Dupuy:2015ega} that the vertex function $\gamma_{acd}^{\vec k \vec q \vec p}$ in \EQ{e:eom_nonlin_psi} should be replaced by
\begin{eqnarray}
  {\tilde \gamma}_{acd}^{\vec k \vec q \vec p}
  &:=&
  \sqrt{1-\va^2} \gamma_{acd}^{\vec k \vec q \vec p}
  - \va^2 \sqrt{1-\va^2} \Delta_{acd}^{\vec k \vec q \vec p}
  \\
  \Delta_{001}^{\vec k \vec q \vec p}
  &:=&
  \Delta_{010}^{\vec k \vec p \vec q}
  :=
  \frac{[(\vec q + \vec p) \cdot \hat \va] (\vec p \cdot \hat \va)}{2 p^2},
  \quad
  \Delta_{111}^{\vec k \vec q \vec p}
  :=
  \frac{\left|\vec q + \vec p\right|^2
    (\vec q \cdot \hat \va)(\vec p \cdot \hat\va)
  }{ 2q^2p^2}
\end{eqnarray}
with all other $\Delta_{acd}^{\vec k \vec q \vec p}$ zero.  Each splits into $\eta$-dependent but wave-number-independent factors, $\sqrt{1-\va^2}$ and $\va^2$, and $\eta$-independent but wave-number-dependent factors, $\gamma_{acd}^{\vec k \vec q \vec p}$ and $\Delta_{acd}^{\vec k \vec q \vec p}$.

Time-dependent factors can be pulled outside of integrals over wave numbers.  Thus \EQS{e:eom_P_cb}{e:def_I} readily generalize to 
\begin{eqnarray}
  {\Pa{ab}{\vec k}}'
  &=&
  -\sum_{j=0} \PLeg_j(\mua)^2
  \left[ \Xia{acj}{k} \Pa{cbj}{k} + \Xia{bcj}{k}\Pa{acj}{k} \right]
  + \sqrt{1-\va^2} \left(\Ia{acd,bcd}{\vec k} + \Ia{bcd,acd}{\vec k}\right)
  \ldots
  \nonumber\\
  &~&
  - \va^2 \sqrt{1-\va^2} \left(\Da{acd,bcd}{\vec k}
  + \Da{bcd,acd}{\vec k}\right)
  \label{e:eom_P_rel}
  \\
  \Da{acd,bef}{\vec k}
  &:=&
  \intq \Delta_{acd}^{\vec k \vec q \vec p} \Ba{bef}{\vec k \vec q \vec p}
\end{eqnarray}
where $\Da{acd,bef}{}$ is a second bispectrum integral.  Since we now consider a non-zero velocity $\vec \va$, the power spectrum of \EQS{e:power_Legendre2_expansion_NL}{e:def_Pabl_NL} and the bispectrum of \EQ{e:def_B}, hence also the bispectrum integrals $\Ia{acd,bef}{\vec k}$ and  $\Da{acd,bef}{\vec k}$, depend upon $\mua = \hat k \cdot \hat \va$.

Mode-coupling integrals defined in \EQ{e:def_A} are also modified through the replacement of $\gamma_{acd}^{\vec k \vec q \vec p}$ by ${\tilde \gamma}_{acd}^{\vec k \vec q \vec p}$.  Each term in the integrand of $\Aa{acd,bef}{}$ has two vertex functions, one with indices $a$, $c$, and $d$, and the second whose indices are contracted with those of the power spectra, while ${\tilde \gamma}_{acd}^{\vec k \vec q \vec p}$ itself has two terms. Thus replacing $\gamma_{acd}^{\vec k \vec q \vec p} \rightarrow {\tilde \gamma}_{acd}^{\vec k \vec q \vec p}$ and pulling wave-number-independent factors outside of the integral leads to four sets of mode-coupling integrals, a significant increase in computational cost.  Further complications arise from the fact that the $\Delta_{acd}^{\vec k \vec q \vec p}$, unlike the $\gamma_{acd}^{\vec k \vec q \vec p}$, depend upon the direction of $\vec \va$.

However, our aim in this article is not a fully relativistic non-linear perturbation theory.  Relativistic corrections to linear theory are important only around horizon scales $k \sim \Hc$, while non-linear corrections are important only at far larger $k$.  Thus we proceed as in Ref.~\cite{Chen:2020bdf} by neglecting ${\mathcal O}(\va^2)$ non-linear terms.  Since $\Delta_{acd}^{\vec k \vec q \vec p}$ itself enters into ${\tilde \gamma}_{acd}^{\vec k \vec q \vec p}$ with a factor $\va^2$, we may entirely neglect $\Da{acd,bef}{\vec k}$ and all mode-coupling integrals containing $\Delta_{acd}^{\vec k \vec q \vec p}$ from our evolution equations.  The resulting power spectrum evolution equation, and the perturbation equations implied by \EQS{e:eom_delta_theta_NL_0}{e:eom_chi_NL_0}, with $\Ia{acd,bef}{\vec k}$ defined as in \EQ{e:def_I}, are
\begin{eqnarray}
  (\Pa{ab}{\vec k})'
  &=&
  -\sum_{j=0} \PLeg_j(\mua)^2
  \left[ \Xia{acj}{k} \Pa{cbj}{k} + \Xia{bcj}{k}\Pa{acj}{k} \right]
  + \left(\Ia{acd,bcd}{\vec k} + \Ia{bcd,acd}{\vec k}\right)
  \label{e:eom_P}
  \\
  (\delta_{\alpha\ell}^k)'
  &=&
  \tfrac{k\va}{\Hc}\left(\tfrac{\ell}{2\ell-1}\delta_{\alpha,\ell-1}^k
  - \tfrac{\ell+1}{2\ell+3}\delta_{\alpha,\ell+1}^k\right)
  + \theta_{\alpha\ell}^k
  + \tfrac{2}{\delta_{\alpha\ell}^k} \Ia{001,001,\ell}{k}
  \label{e:eom_delta}
  \\
  (\theta_{\alpha\ell}^k)'
  &=&
  -\delK_{\ell0}\tfrac{k^2\Phi^k}{\Hc^2}
  - \left(1+\tfrac{\Hc'}{\Hc}\right)\theta_{\alpha\ell}^k
  + \tfrac{k\va}{\Hc}\left(\tfrac{\ell}{2\ell-1}\theta_{\alpha,\ell-1}^k
  - \tfrac{\ell+1}{2\ell+3}\theta_{\alpha,\ell+1}^k\right)
  + \tfrac{1}{\theta_{\alpha\ell}^k} \Ia{111,111,\ell}{k}\qquad
  \label{e:eom_theta}
  \\
  (\chi_{\alpha\ell}^k)'
  &=&
  \tfrac{2(1-\chi_{\alpha\ell}^k)}{(\delta_{\alpha\ell}^k)^2}\Ia{001,001,\ell}{k}
  + \tfrac{1-\chi_{\alpha\ell}^k}{(\theta_{\alpha\ell}^k)^2} \Ia{111,111,\ell}{k}
  - \tfrac{2}{\delta_{\alpha\ell}^k \theta_{\alpha\ell}^k} \Ia{001,101,\ell}{k}
  - \tfrac{1}{\delta_{\alpha\ell}^k \theta_{\alpha\ell}^k} \Ia{111,011,\ell}{k}.
  \label{e:eom_chi}
\end{eqnarray}
These are the principal results of this subsection.  As expected, the derivative of $\chi_{\alpha\ell}^k$ is nonzero only when the bispectrum integrals are as well.

\subsection{Evolution of the $\Ia{acd,bef,j}{k}$}
\label{subsec:trg_lin_evol_I}

Next, we extend the evolution equations of the bispectrum and bispectrum integral, respectively \EQ{e:eom_B_cb} and \EQ{e:eom_I_cb}, to the case of a neutrino flow with non-zero $\va$. Since the bispectrum is negligible in linear perturbation theory, applicable at early times, the dependence of the bispectrum upon the direction of $\vec \va$ is due entirely to the power spectrum pairs on the right side of \EQ{e:eom_B_cb}.  Thus the bispectrum depends upon $k$, $q$, $p$, $\mua:=\hat k \cdot \hat \va$, $\muaq := \hat q \cdot \hat \va$, and $\muap := \hat p \cdot \hat \va$, with $\vec k = \vec q + \vec p$ implying the constraint that $k \mua = q \muaq + p \muap$.

References~\cite{Scoccimarro:1999ed,Yankelevich:2018uaz} study the angle-dependence of the bispectrum in the presence of a preferred direction $\hat s$.  They find that $B(\vec k_1, \vec k_2, \vec k_3)$, with $\vec k_1 + \vec k_2 + \vec k_3=0$, can be parameterized in terms of $k_1$, $k_2$, $k_3$, and two angles, $\varpi_i$ and $\varphi_{ij}$, for any given $i,j \in \{1,~2,~3\}$ such that $i \neq j$. The first is the angle between $\vec k_i$ and $\hat s$, that is, $\varpi_i := \cos^{-1}(\hat k_i \cdot \hat s)$.  The second, $\varphi_{ij}$, is the azimuthal angle of $\vec k_j$ in the plane perpendicular to $\vec k_i$, that is, the angle between $\vec k_j  - (\vec k_j \cdot \hat k_i) \hat k_i$ and $\hat s - (\hat s \cdot \hat k_i)\hat k_i$.  In terms of these, the bispectrum may be expressed as
\begin{equation}
  B^{\vec k_1 \vec k_2 \vec k_3}
  =
  \sum_{\ell=0}^\infty \sum_{m=-\ell}^{\ell}
  Y_{\ell m}(\varpi_i, \varphi_{ij}) B^{k_1 k_2 k_3}_{\ell m, i j}.
\end{equation}
Here, $B^{k_1 k_2 k_3}_{\ell m, i j}$ is a function of the wave numbers $k_1$, $k_2$, and $k_3$ and the integer indices $\ell$ and $m$ that is specific to our choices of $i$ and $j$.

Defining $\muaq := \hat q \cdot \hat \va$ and $\muap := \hat p \cdot \hat \va$, let $\varpi_k:=\cos^{-1}(\mua)$, $\varpi_q:=\cos^{-1}(\muaq)$, and $\varpi_p:=\cos^{-1}(\muap)$.  In each case, the integer $j$ used to define $\varphi_{ij}$ may be chosen arbitrarily.  Then we have three equivalent representations of the bispectrum,
\begin{equation}
  \sum_{\ell=0}^\infty \sum_{m=-\ell}^{\ell} \!\!
  Y_{\ell m}(\!\varpi_k, \varphi_k\!) B^{kqp}_{\ell m, (k) j}
  =
  \sum_{\ell=0}^\infty \sum_{m=-\ell}^{\ell} \!\!
  Y_{\ell m}(\!\varpi_q, \varphi_q\!) B^{kqp}_{\ell m, (q) j}
  =
  \sum_{\ell=0}^\infty \sum_{m=-\ell}^{\ell} \!\!
  Y_{\ell m}(\!\varpi_p, \varphi_p\!) B^{kqp}_{\ell m, (p) j}.
\end{equation}
The  first, second, and third forms, respectively, are useful for extending the first, second, and third terms on the right hand side of \EQ{e:eom_B_cb}.  With the angle-dependent power spectrum of \EQ{e:power_Legendre2_expansion} obeying the evolution \EQ{e:eom_P}, 
\begin{eqnarray}
  (\Ba{abc}{\vec k \vec q \vec p})'
  &=&
  -\sum_{\ell=0}^\infty \sum_{m=-\ell}^{\ell}
  \Xia{ad\ell}{k} Y_{\ell m}(\varpi_k, \varphi_k) B^{kqp}_{dbc,\ell m, (k) j}
  -\sum_{\ell=0}^\infty \sum_{m=-\ell}^{\ell}
  \Xia{bd\ell}{q} Y_{\ell m}(\varpi_q, \varphi_q) B^{kqp}_{adc,\ell m, (q) j}
  \ldots\nonumber\\
  &~&
  -\sum_{\ell=0}^\infty \sum_{m=-\ell}^{\ell}
  \Xia{cd\ell}{p} Y_{\ell m}(\varpi_p, \varphi_p) B^{kqp}_{abd,\ell m, (p) j}
  \ldots\nonumber\\
  &~&
  + 2\gamma_{ade}^{\vec k \vec q \vec p} \Pa{db}{\vec q} \Pa{ec}{\vec p}
  + 2\gamma_{bde}^{\vec q, \vec p, -\vec k} \Pa{dc}{\vec p} \Pa{ea}{\vec k}
  + 2\gamma_{cde}^{\vec p, -\vec k, \vec q} \Pa{da}{\vec k} \Pa{eb}{\vec q}.
  \label{e:eom_B}
\end{eqnarray}
Since $\mua$, $\muaq$, and $\muap$ are related by $k\mua = q\muaq + p\muap$, \EQ{e:eom_B} is consistent with the result of Ref.~\cite{Scoccimarro:1999ed} that $\Ba{abc}{\vec k \vec q \vec p}$ may be expressed in terms of $k$, $q$, $p$, and two angles.  Equivalently, since $p$ may be expressed as a function of $k$, $q$, and $\hat k \cdot \hat q$ using $p^2 = k^2 + q^2 - 2 k q \hat k \cdot \hat q$, the bispectrum may be expressed as a function of $k$, $\mua$, and $\vec q$, which allows us to extend the definition of $\Ia{acd,bef}{\vec k}$ of \EQ{e:def_I} to the case of massive neutrino flows.

Similarly, since $\Ia{acd,bef}{\vec k}$ is initially negligible, the dependence of $\Ia{acd,bef}{\vec k}$ upon the direction of $\vec \va$ through $\mua$ is determined by the mode-coupling integrals $\Aa{acd,bef}{\vec k}$.  We will show in Sec.~\ref{sec:mode-coupling_P22} that $\Aa{acd,bef}{\vec k}$ can itself be expanded in squares of Legendre polynomials, $\Aa{acd,bef}{\vec k} = \sum_{j=0}^\infty \PLeg_j(\mua)^2 \Aa{acd,bef,j}{k}$.  Thus $\Ia{acd,bef}{\vec k}$ must have a similar expansion,
\begin{equation}
  \Ia{acd,bef}{\vec k}
  =
  \sum_{j=0}^\infty \PLeg_j(\mua)^2 \Ia{acd,bef,j}{k},
\end{equation}
with the evolution of $\Ia{acd,bef,j}{k}$ depending upon $\Aa{acd,bef,j}{k}$.  

The Time-RG closure approximation discussed in Sec.~\ref{subsec:bkg_time-rg} is complicated by the fact that neutrinos are not the dominant clustering species.  Thus neglecting the difference between $\Xia{ab\ell}{k}$ and $\Xia{ab\ell}{q}$ for $q \neq k$ is no longer an ${\mathcal O}(\fnu/\fcb)$ approximation.  On the other hand, neutrinos are significantly more linear than the CDM and baryons.  The wave number at which non-linear corrections dominate the neutrino power spectrum is typically much larger than the neutrino free-streaming wave number $\sim \Hc / \va$.  Thus we seek a closure approximation appropriate to the free-streaming limit $k \va / \Hc \gg 1$.

Free streaming washes out the clustering sourced by the gravitational potential $\Phi$.  Thus our closure approximation in the free-streaming limit is to neglect both the free-streaming terms and $\Phi$ in $\Xia{abj}{q}$.  We adopt the following equation of motion for the bispectrum integral:
\begin{eqnarray}
  (\Ia{acd,bef,j}{k})'
  &=&
  -\Xia{bgj}{k} \Ia{acd,gef,j}{k}
  - \Xita{egj}{k} \Ia{acd,bgf,j}{k}
  - \Xita{fgj}{k} \Ia{acd,beg,j}{k}
  + 2\Aa{acd,bef,j}{k}
  \label{e:eom_I}
  \\
  \Xia{abj}{k}
  &=&
  \left[
    \begin{array}{cc}
      0 & -1 \\
      \frac{k^2 \Phi^k}{\Hc^2 \psia{0 0}{k}}\delK_{j 0}~~
      & 1 \!+\! \frac{\Hc'}{\Hc}
    \end{array}
    \right]
  -
  \frac{k v_\alpha}{\Hc}
  \!\left(\!\frac{j}{2j-1}\frac{\psia{a,j-1}{k}}{\psia{aj}{k}}
  \!-\! \frac{j+1}{2j+3}\frac{\psia{a,j+1}{k}}{\psia{aj}{k}}
  \!\right)
  \left[ \begin{array}{cc} 1 & 0 \\ 0 & 1 \end{array} \right] \qquad
  \label{e:def_Xia}
  \\
  \Xita{abj}{k}
  &=&
  \left[
    \begin{array}{cc}
      0~~~ & -1 \\
      0~~~ & 1 + \frac{\Hc'}{\Hc}
    \end{array}
    \right]
  \label{e:def_Xita}
\end{eqnarray}
where summation over the repeated index $g$ is implicit and $\Xia{abj}{k}$ is identical to that of \EQ{e:Xia4}. Equations~(\ref{e:eom_P}-\ref{e:eom_chi}) and (\ref{e:eom_I}-\ref{e:def_Xita}) are the main results of this Section.

Quantifying the accuracy of \EQ{e:eom_I} relative to a direct integration of \EQ{e:eom_B} is prohibitively computationally expensive.  Discretizing $k$, $q$, and $p$ with $\sim 100$ values each implies $\sim 10^6$ combinations of $k$, $q$, and $p$.  Further, truncating $\ell$ at $\sim 10$ in each summation of \EQ{e:eom_B} implies $\sim 200$ quantities $B^{kqp}_{abc,\ell m, (k) j}$, for each of eight possible combinations of the indices $a$, $b$, and $c$, at each $(k,q,p)$ combination.  Solving this system of $\sim 1.6$~billion coupled evolution equations is beyond the scope of this article.  Instead, Sec.~\ref{sec:nu_NL_enhancement} tests the resulting power spectrum  by comparison with an N-body particle simulation.

\subsection{Mode-coupling integrals}
\label{subsec:trg_mode-coupling}

Extension of the mode-coupling integrals to the case of a massive neutrino flow is straightforward; we need only substitute the angle-dependent power spectrum of \EQ{e:power_Legendre2_expansion} into the definition \EQ{e:def_A} of $\Aa{acd,bef}{\vec k}$.  Splitting the result into terms $\Aa{acd,bef}{(2,2)\vec k}$ of $\Ptt$-type and $\Aa{acd,bef}{(1,3)\vec k}$ of $\Pot$-type, as discussed in Sec.~\ref{subsec:bkg_FFT_accel}, we have
\begin{eqnarray}
  \Aa{acd,bef}{\vec k}
  &=&
  \Aa{acd,bef}{(2,2)\vec k}
  + \Aa{adc,bfe}{(1,3)\vec k}
  + \Aa{acd,bef}{(1,3)\vec k}
  \label{e:A_22_13}
  \\
  \Aa{acd,bef}{(2,2)\vec k}
  &=&
  \sum_{\ell=0}^\infty \sum_{m=0}^\infty
  \intq \gamma_{acd}^{\vec k \vec q \vec p} \gamma_{bgh}^{\vec k \vec q \vec p} 
  \PLeg_\ell(\muaq)^2 \PLeg_m(\muap)^2 \Pa{ge\ell}{q} \Pa{hfm}{p}
  \label{e:def_A22}
  \\
  \Aa{acd,bef}{(1,3)\vec k}
  &=&
  \sum_{\ell=0}^\infty \sum_{m=0}^\infty
  \intq \gamma_{acd}^{\vec k \vec q \vec p} \gamma_{fgh}^{\vec p,\vec k,-\vec q}
  \PLeg_\ell(\mua)^2 \PLeg_m(\muaq)^2 \Pa{gb\ell}{k} \Pa{hem}{q}.
  \label{e:def_A13}
\end{eqnarray}
Equation~(\ref{e:eom_P}) for the power spectrum, \EQ{e:eom_I} for the bispectrum integrals, and \EQ{e:A_22_13} for the mode-coupling integrals, define our non-linear perturbation theory for flows of massive neutrinos, which we will call \FFTM{}.

Mode-coupling integrals represent the most computationally expensive part of this perturbation theory.  Fourteen combinations of $a$, $c$, $d$, $b$, $e$, and $f$ yield unique and non-zero $\Aa{acd,bef}{\vec k}$.  Truncating its expansion in polynomials in $\mua$ after  $\sim 10$ terms, and similarly truncating the $\ell$ and $m$ summations at $\sim 10$, implies $\sim 14000$ separate mode-coupling integrals, a thousandfold increase over ordinary Time-RG.  Thus the FFT acceleration techniques summarized in Sec.~\ref{subsec:bkg_FFT_accel} are necessary.  Use of the FAST-PT acceleration requires that the integrand of each $\Aa{acd,bef}{(2,2)\vec k}$ depend only upon $q$, $p$, and $\mupq$ alone, and that the integrand of each $\Aa{acd,bef}{(1,3)\vec k}$ depend only upon $q$ alone.
Deriving, implementing, and testing FFT-accelerated $\Aa{acd,bef}{(2,2)\vec k}$ and $\Aa{acd,bef}{(1,3)\vec k}$ will be carried out in the next two sections.  Readers not interested in the details of this acceleration may wish to skip to Sec.~\ref{subsec:p13_A_tests}, while readers looking for greater detail are inivited to study Appendices \ref{app:expansions_of_Lc_Mc} and \ref{app:divergence_cancellation} after the next two sections.

\section{Mode-coupling integrals of $P^{(2,2)}$-type}
\label{sec:mode-coupling_P22}

Evaluation of the $\Ptt$-type term, \EQ{e:def_A22}, is the goal of this Section.  Our starting point is the observation in Sec.~\ref{subsec:trg_lin_evol_I} that the bispectrum for a fluid of velocity $\vec \va$ is a function of $k$, $\mua$, and $\vec q$, and motivating a set of angular coordinates for $\vec q$ integration.  Since $\PLeg_\ell(\muaq)^2 \PLeg_m(\muap)^2$ is the only part of that integrand which cannot be expresed solely in terms of $q$, $p$, and $\mupq$, necessary for FAST-PT integration, we integrate directly over the remaining integration variable in order to reduce \EQ{e:def_A22} to a series of terms of the form of \EQ{e:def_J}. 

\subsection{Coordinate system and basis}
\label{subsec:p22_coordinates_basis}

We begin by defining a coordinate system for $\vec q$ integration of \EQ{e:def_A22} after the Dirac delta function has fixed $\vec p$ to $\vec k - \vec q$.  Let $\hat k$ be the vertical axis ($z_k$ axis) of the space of $\vec q$ vectors and $\vartheta$ be the angle between $\hat k$ and $\hat q$, so that $\cos(\vartheta) := \vtc := \hat k \cdot \hat q$.  Given a flow $\alpha$ with velocity $\vec \va$, which we assume not to be parallel to $\hat k$, define the $x_k$ axis so that $\vec \va$ lies in the $x_k z_k$ plane, and the $y_k$ axis by $\hat z_k \times \hat x_k = \hat y_k$.  Then, for a given $\vec q$, let $\varphi$ be the angle between the projection of $\vec q$ onto the $x_k y_k$ plane, $\vec q - (\vec q \cdot \hat k)\hat k$, and the $x_k$ axis.

In this coordinate system, and defining the shorthand $\nua := \sqrt{1-\mua^2}$, $\vts := \sin(\vartheta)$, $\vtct:=-(q\vtc-k)/p$, and $\vtst:=-q\vts/p$, we have the following identities:
\begin{eqnarray}
  \vec \va
  &=&
  \va \nua \hat x_k + \va \mua \hat z_k
  \\
  \vec q
  &=&
  q \vts \cos(\varphi) \hat x_k + q \vts \sin(\varphi) \hat y_k
  + q \vtc \hat z_k
  \\
  \vec p
  &=&
  -q \vts \cos(\varphi) \hat x_k - q \vts \sin(\varphi) \hat y_k
  - (q \vtc - k) \hat z_k
  \\
  \vec \va \cdot \vec q
  &=&
  q \va \nua \vts \cos(\varphi) + q \va \mua \vtc
  \\
  \vec \va \cdot \vec p
  &=&
  -q \va \nua \vts \cos(\varphi) - (q \vtc - k) \va \mua
  \\
  \mupq
  &:=&
  \hat q \cdot \hat p
  = \frac{k\vtc}{p} - \frac{q}{p}
  \Rightarrow
  \vtc = \frac{p}{k}\mupq + \frac{q}{k}
  \\
  \muaq
  &:=&
  \hat \va \cdot \hat q
  =
  \nua \vts \cos(\varphi) + \mua\vtc
  =
  \nua \vts \cos(\varphi)
  + \left(\frac{p}{k}\mupq + \frac{q}{k}\right)\mua
  \\
  \muap
  &:=&
  \hat \va \cdot \hat p
  =
  \nua \vtst \cos(\varphi) + \mua \vtct
  =
  \nua \vtst\cos(\varphi)
  + \left(\frac{q}{k} \mupq + \frac{p}{k}\right)\mua
\end{eqnarray}
The vertex functions $\gamma_{acd}$ can now be expressed as
\begin{eqnarray}
  \gamma_{001}^{\vec k \vec q \vec p}
  &=&
  \frac{1}{2} \left(1 + \frac{q}{p} \mupq\right)
  =
  \frac{k}{4q} \,
  \frac{\tfrac{q}{k}\vtc - 1}{c-\tfrac{1}{2}(\tfrac{q}{k}+\tfrac{k}{q})}
  \quad \textrm{and} \quad
  \gamma_{001}^{\vec p, \vec k, -\vec q}
  =
  \frac{1}{2} \left(1 - \frac{k}{q} \vtc \right)
  \label{e:gamma_001}
  \\
  \gamma_{010}^{\vec k \vec q \vec p}
  &=&
  \frac{1}{2} \left(1 + \frac{p}{q} \mupq\right)
  =
  \frac{k}{2q} \vtc
  \quad \textrm{and} \quad
  \gamma_{010}^{\vec p, \vec k, -\vec q}
  =
  \frac{1}{2} \left(1 - \frac{q}{k} \vtc \right)
  \label{e:gamma_010}
  \\
  \gamma_{111}^{\vec k \vec q \vec p}
  &=&
  \mupq^2
  + \frac{1}{2}\left( \frac{q}{p} + \frac{p}{q} \right) \mupq
  =
  \frac{k}{4q} \,
  \frac{1 - \tfrac{k}{q}\vtc}{c-\tfrac{1}{2}(\tfrac{q}{k}+\tfrac{k}{q})}
  \quad \textrm{and} \quad
  \gamma_{111}^{\vec p, \vec k, -\vec q}
  =
  \vtc^2 - \frac{1}{2}\left(\frac{q}{k} + \frac{k}{q}\right)\vtc. \qquad
  \label{e:gamma_111}
\end{eqnarray}

Since the power spectrum of \EQ{e:power_Legendre2_expansion} is expanded in the squares of Legendre polynomials, we summarize here the coefficients used to transform between this and other polynomial bases.  Legendre polynomials $\PLeg_\ell(\mu)$ can be expressed in terms of integer powers $\mu^L$ as:
\begin{eqnarray}
  \PLeg_\ell(\mu)
  &=&
  \sum_L \MLeg_{L \ell} \mu^L
  \quad \longleftrightarrow \quad
  \mu^L = \sum_\ell \MLeg^{-1}_{\ell L} \PLeg_\ell(\mu)
  \\
  \mathrm{where}\quad
  \MLeg_{L \ell}
  &=&
  \left\{
  \begin{array}{ll}
    \frac{2^\ell \Even_{\ell+L}}{L!(\ell-L)!}
    \left(\frac{L-\ell+1}{2}\right)_{\ell}
    &
    \textrm{if } 0 \leq L \leq \ell
    \\
    0 & \textrm{otherwise}
  \end{array}
  \right.
  \\
  \Even_n
  &=&
  \left\{
    \begin{array}{ll}
      1 & \textrm{if $n$ is an even integer}
      \\
      0 & \textrm{otherwise}
    \end{array}
    \right.
    \textrm{ and }
    \Odd_n =
    \left\{
    \begin{array}{ll}
      1 & \textrm{if $n$ is an odd integer}
      \\
      0 & \textrm{otherwise}
    \end{array}
    \right.\quad
    \label{e:def_E_O}
\end{eqnarray}
and $(x)_n := \Gamma(x+n)/\Gamma(x)$ is the Pocchammer symbol.  For squares of Legendre polynomials:
\begin{eqnarray}
  \PLeg_\ell(\mu)^2
  &=&
  \sum_L \sum_M \mu^{L+M} \MLeg_{L \ell} \MLeg_{M \ell}
  =:
  \sum_L \QLeg_{L \ell} \mu^{2L}
  \quad \longleftrightarrow \quad
  \mu^{2L}
  =
  \sum_\ell \QLeg^{-1}_{\ell L} \PLeg_\ell(\mu)^2\qquad
  \\
  \mathrm{where}\quad
  \QLeg_{L\ell}
  &:=&
  \sum_{M = \max(0,2L-\ell)}^{\min(\ell,2L)}
  \MLeg_{M,\ell} \MLeg_{2L-M,\ell}.
\end{eqnarray}
Each of $\MLeg_{L\ell}$, $\MLeg^{-1}_{\ell L}$, $\QLeg_{L\ell}$, and $\QLeg^{-1}_{\ell L}$ is an upper-triangular matrix whose entries vanish when the first (row) index is greater than the second (column) index.  Henceforth we use capital letters for the even-integer-power basis and lower-case letters for the Legendre-squared basis.

Coefficients of these basis functions transform in the opposite sense.  If some function $x(\mu) = \sum_\ell x_\ell \PLeg_\ell(\mu)^2 = \sum_L x_L \mu^{2L}$, then
\begin{eqnarray}
  x
  &=& \sum_\ell x_\ell \sum_L  \QLeg_{L\ell}\mu^{2L}
  = \sum_L \mu^{2L} \left(\sum_\ell \QLeg_{L\ell}x_\ell\right)
  \Rightarrow x_L = \sum_\ell \QLeg_{L\ell}x_\ell
  \\
  x
  &=& \sum_L x_L \sum_\ell \QLeg^{-1}_{\ell L} \PLeg_\ell(\mu)^2
  = \sum_\ell \PLeg_\ell(\mu)^2 \left(\sum_L  \QLeg^{-1}_{\ell L}  x_L \right)
  \Rightarrow x_\ell = \sum_L  \QLeg^{-1}_{\ell L}  x_L\qquad
\end{eqnarray}
and similarly, if $y(\mu) = \sum_\ell y_\ell \PLeg_\ell(\mu) = \sum_L y_L \mu^L$, then $y_L = \sum_\ell \MLeg_{L\ell} y_\ell \longleftrightarrow y_\ell = \sum_L \MLeg^{-1}_{\ell L} y_L$.

\subsection{Angular averages of Legendre polynomial products}
\label{subsec:p22_ang_avg}

Dependence of the factor $\PLeg_\ell(\muaq)^2 \PLeg_m(\muap)^2$ in \EQ{e:def_A22} upon $\varphi$ prevents its reduction to terms of the form of \EQ{e:def_J}, necessary for FFT-accelerated integration.  Our next step is to integrate explicitly over $\varphi$, allowing us to replace $\PLeg_\ell(\muaq)^2 \PLeg_m(\muap)^2$ by its angular average
\begin{equation}
  \left<\PLeg_\ell(\muaq)^2 \PLeg_m(\muap)^2\right>_\varphi
  :=
  \int_0^{2\pi} \frac{d\varphi}{2\pi} \PLeg_\ell(\muaq)^2 \PLeg_m(\muap)^2.
\end{equation}
Henceforth, we use triangular braces with one or more angles in the subscript, as above, to denote averaging over those angles.

Using the basis transformation coefficients of Sec.~\ref{subsec:p22_coordinates_basis}, we may expand this average.
\begin{eqnarray}
  \left<\PLeg_\ell(\muaq)^2 \PLeg_m(\muap)^2\right>_\varphi
  &=&
  \sum_{L=0}^\ell \sum_{M=0}^m \QLeg_{L\ell} \QLeg_{Mm}
  \left< \muaq^{2L} \muap^{2M} \right>_\varphi
  \label{e:PLegell2_PLegm2_expansion_0}
  \\
  \left< \muaq^{2L} \muap^{2M} \right>_\varphi
  &=:&
  \sum_{J=0}^\infty \mua^{2J} \rho_{JLM}(\vtc,\vts,\vtct,\vtst)
  \label{e:muaq2L_muap2M_expansion_0}
  \\
  \Rightarrow
  \left<\PLeg_\ell(\muaq)^2 \PLeg_m(\muap)^2\right>_\varphi
  &=&
  \sum_{j=0}^\infty \PLeg_j(\mua)^2
  \sum_{L=0}^\ell \sum_{M=0}^m \sum_{J=j}^\infty
  \QLeg_{L\ell} \QLeg_{Mm} \QLeg^{-1}_{jJ}
  \rho_{JLM}(\vtc,\vts,\vtct,\vtst).\qquad~
  \label{e:PLegell2_PLegm2_expansion_1}
\end{eqnarray}
The second line above defines the quantity $\rho_{JLM}$, which is independent of $\varphi$, after which \EQ{e:PLegell2_PLegm2_expansion_1} expresses the angular average in the desired $\PLeg_j(\mua)^2$ basis with $\varphi$-independent coefficients.  Thus we have reduced the goal of this subsection to a computation of $\rho_{JLM}$.

The product $\muaq^{2L} \muap^{2M}$ is expanded in binomial series as
\begin{eqnarray}
  \muaq^{2L} \muap^{2M}
  &=&
  \sum_{i=0}^{2L} \sum_{j=0}^{2M} \binom{2L}{i} \binom{2M}{j}
  (\mua \vtc)^{2L-i} (\mua \vtct)^{2M-j}
  (\vts \nua \cos\varphi)^i
  (\vtst \nua \cos\varphi)^j
  \nonumber\\
  &=&
  \sum_{i=0}^{2L} \sum_{j=0}^{2M} \binom{2L}{i} \binom{2M}{j}
  \vtc^{2L-i} \vtct^{2M-j} \vts^i \vtst^j \mua^{2L+2M-i-j}
  (\nua \cos\varphi)^{i+j}.\qquad
\end{eqnarray}
The angular average over odd powers of $\cos\varphi$ vanishes, while that over even powers is
\begin{equation}
  \left<\cos(\varphi)^{2r}\right>_\varphi = \frac{(2r)!}{4^r (r!)^2} =: \kappa_r~.
\end{equation}
Defining $r=(i+j)/2$, we recast the summation of the $\varphi$ average as
\begin{eqnarray}
  \left< \muaq^{2L} \muap^{2M} \right>_\varphi
  &=&
  \sum_{r=0}^{L+M} \sum_{i=0}^{2L} \binom{2L}{i} \binom{2M}{2r-i}
  \vtc^{2L-i} \vtct^{2M+i-2r} \vts^i \vtst^{2r-i} \mua^{2L+2M-2r}
  \nua^{2r} \kappa_r . \qquad
  \label{e:avg_muaq2L_muap2M}
\end{eqnarray}
In our binomial factor convention, $\binom{n}{i}$ vanishes for all integers $n$ and $i$ unless $0 \leq i \leq n$.  Thus the summand is non-zero only for $\max(0,2r-2M) \leq i \leq \min(2L,2r)$.

Next, we expand $\nua^{2r} = (1-\mua^2)^r$ in binomial series.
\begin{eqnarray}
  \left< \muaq^{2L} \muap^{2M} \right>_\varphi
  &=&
  \sum_{r=0}^{L+M} \! \sum_{i=0}^{2L}
  \binom{\!2L\!}{\!i\!} \binom{\!2M\!}{\!2r\!-\!i\!}
  \vtc^{2L-i} \vtct^{2M+i-2r} \vts^i \vtst^{2r-i} \mua^{2L+2M-2r} \kappa_r
  \sum_{s=0}^r \binom{r}{s} (-1)^{r-s} \mua^{2r-2s}
  \nonumber\\
  &=&
  \sum_{s=0}^{L+M} \sum_{r=s}^{L+M}\sum_{i=0}^{2L}
  \binom{2L}{i} \binom{2M}{2r-i} \binom{r}{s} (-1)^{r-s} \kappa_r \mua^{2L+2M-2s}
  \vtc^{2L-i} \vtct^{2M+i-2r} \vts^i \vtst^{2r-i}
  \nonumber\\
  &=&
  \sum_{J=0}^{L+M} \mua^{2J} \sum_{t=0}^J \sum_{i=0}^{2L}
  (-1)^{J-t} \binom{2L}{i} \binom{2M}{i+2t-2L} \binom{L+M-t}{J-t}
  \ldots\nonumber\\
  &~&\qquad\qquad\qquad
  \times \kappa_{L+M-t} \vtc^{2L-i} \vtct^{i+2t-2L} \vts^i \vtst^{2L+2M-2t-i}
\end{eqnarray}
where in the final equality we have changed summation variables to $J=L+M-s$ and $t=L+M-r$.  From this expression we may immediately write down $\rho_{JLM}$,
\begin{eqnarray}
  \rho_{JLM}(\vtc,\vts,\vtct,\vtst)
  &=&
  \sum_{t=0}^J \sum_{i=0}^{2L}
  (-1)^{J-t} \binom{2L}{i} \binom{2M}{i+2t-2L} \binom{L+M-t}{J-t}
  \ldots\nonumber\\
  &~&\qquad\qquad
  \times
  \kappa_{L+M-t} \vtc^{2L-i} \vtct^{i+2t-2L} \vts^i \vtst^{2L+2M-2t-i}
  \label{e:rho_JLM_expansion_0}
\end{eqnarray}
with $J$ restricted to the range $[0,L+M]$.

Using the identities $\vtc^2+\vts^2=1$, $\vtct^2+\vtst^2=1$, and $\vtc\vtct + \vts\vtst = \mupq$, we eliminate $\vtc$ and $\vtct$ in favor of $\vts$ and $\vtst$.  The $i$ summation may be split into low values $0 \leq i \leq 2L-t$ and high values $2L-t+1 \leq i \leq 2L$, with the exponent of $\vtct$ being greater than that of $\vtc$ for high $i$.  After some algebra we arrive at the expression
\begin{eqnarray}
  \left< \muaq^{2L} \muap^{2M} \right>_\varphi
  &=&
  \sum_{J=0}^{L+M} \mua^{2J} \sum_{g=L+M-J}^{L+M} \rho_{JLMg}
  \\
  \rho_{JLMg}
  &=&
  \sum_{e=0}^J {\hat h}^{JLM}_{0eg}  \vts^{2g-2M+e}\vtst^{2M-e} \mupq^e
  + \sum_{e=0}^{J-1} {\hat h}^{JLM}_{1eg}  \vts^{2L-e}\vtst^{2g-2L+e} \mupq^e\qquad
  \\
  {\hat h}^{JLM}_{0eg}
  &=&
  \sum_{r=L+M-J}^g \sum_{n=0}^{L+r-M} (-1)^{J+L+M+g+e} \kappa_r
  \binom{2L}{n} \binom{2M}{2r-n}
  \ldots\nonumber\\
  &~&\qquad\times
  \binom{r}{L+M-J} \binom{2M-2r+n}{e}\binom{L+r-M-n}{g+e+r-n-2M}
  \\
  {\hat h}^{JLM}_{1eg}
  &=&
  \sum_{r=L+M-J}^{\min(g,L+M-e-1)} \sum_{n=L+r-M+1}^{2L} (-1)^{J+L+M+g+e} \kappa_r
  \binom{2L}{n} \binom{2M}{2r-n}
  \ldots\nonumber\\
  &~&\qquad\times
  \binom{r}{L+M-J} \binom{2L-n}{e}\binom{M+n-L-r}{g+n+e-2L-r}
\end{eqnarray}

FAST-PT computation of $\Ptt$-type integrals requires that the integrand be expanded in terms proportional to $(q/k)^A (p/k)^B \mupq^C \Pa{ab\ell}{q} \Pa{cdm}{p}$ for integers $A$, $B$, and $C \geq 0$, as in \EQ{e:def_J}. Thus we substitute $\vts = (p/k)\sqrt{1-\mupq^2}$ and $\vtst = -(q/k)\sqrt{1-\mupq^2}$.  Defining ${\hat h}^{JLM}_{1Jg}$ to be zero, we may combine the two $e$ sums as
\begin{equation}
  \rho_{JLMg}
  =
  (1-\mupq^2)^g \left(\frac{p}{k}\right)^{2g}
  \sum_{e=0}^J \left[{\hat h}^{JLM}_{0eg} \left(\frac{q}{p}\right)^{2M-e}
    \!\!\!+{\hat h}^{JLM}_{1eg} \left(\frac{q}{p}\right)^{2g-2L+e} \right] \mupq^e.
  \quad
\end{equation}
Finally, we change bases to the squared Legendre polynomials:
\begin{eqnarray}
  \left< \PLeg_\ell(\muaq)^2 \PLeg_m(\muap)^2 \right>_\varphi
  &=&
  \sum_{j=0}^{\ell+m} \PLeg_j(\mua)^2 \sum_{g=0}^{\ell+m} \sum_{e=0}^{\ell+m-g}
  \sum_{f=0}^{g-\Odd_e} h^{jlm}_{efg} \left(\frac{p}{k}\right)^{2g}
  (1-\mupq^2)^g \mupq^e \left(\frac{q}{p}\right)^{2f+\Odd_e} \quad
  \label{e:PLegell2_Plegm2_expansion}
  \\
  h^{jlm}_{efg}
  &:=&
  \sum_{L=0}^\ell \sum_{J=j}^{L+f+\frac{e+\Odd_e}{2}}
  \QLeg^{-1}_{jJ} \QLeg_{L\ell}
  \QLeg_{f+\frac{e+\Odd_e}{2},m} {\hat h}^{J,L,f+\frac{e+\Odd_e}{2}}_{0eg}
  \ldots\nonumber\\
  &~&
  +
  \sum_{M=0}^m \sum_{J=j}^{g+M-f+\frac{e-\Odd_e}{2}}
  \QLeg^{-1}_{jJ} \QLeg_{Mm}
  \QLeg_{g-f+\frac{e-\Odd_e}{2},\ell} {\hat h}^{J,g-f+\frac{e-\Odd_e}{2},M}_{1eg}
\end{eqnarray}
which is the goal of this subsection.

Before proceeding, we note a symmetry in the $h^{jlm}_{efg}$ coefficients.  The angular average $\left<\PLeg_\ell(\muaq)^2 \PLeg_m(\muap)^2\right>_\varphi$ is invariant under the transformation $\ell \leftrightarrow m$ and $\vec q \leftrightarrow \vec p$.  Applying this transformation to the right hand side of Eq.~(\ref{e:PLegell2_Plegm2_expansion}), and changing from the summation variable $f$ to $f' := g - \Odd_e - f$, shows that the identity
\begin{equation}
  h^{jm\ell}_{e, g - \Odd_e - f, g} = h^{j\ell m}_{efg}
\end{equation}
is implied by this invariance.

\subsection{FAST-PT computation of $\Ptt$-type terms}
\label{subsec:p22_fast-pt}

Each $\Aa{acd,bef}{(2,2)\vec k}$ integrand is the product of two power spectra with two vertex functions.  Since this product of vertex functions can be decomposed into terms containing integer powers of $p/q$ and $\mupq$, we may  decompose each $\Aa{acd,bef}{(2,2) \vec k}$ into integrals whose integrands are proportional to $(p/q)^B \mupq^N$ times the product of two power spectra:
\begin{eqnarray}
  \Sa{BN,ab,cd}{\vec k}
  &:=&
  \intq \left(\frac{p}{q}\right)^B \mupq^N \Pa{ab}{\vec q} \Pa{cd}{\vec q}
  =
  \sum_{\ell=0}^\infty \sum_{m=0}^\infty
  \intq \left< \PLeg_\ell(\muaq)^2 \PLeg_m(\muap)^2\right>_\varphi
  \Pa{ab\ell}{q} \Pa{cdm}{p}\quad\quad
  \label{e:def_S_BN_ab_cd}
  \\
  &=:&
  \sum_{j=0}^\infty \PLeg_j(\mu)^2 \Sa{BNj,ab,cd}{k}
  \label{e:S_BN_ab_cd_Leg2_exp}
  \\
  \Sa{BNj,ab,cd}{k}
  &=:&
  \sum_{\ell=0}^\infty \sum_{m=0}^\infty \Sa{BNj,ab\ell,cdm}{k}
  \\
  \Sa{BNj,ab\ell,cdm}{k}
  &=&
  \sum_{g=0}^{\ell+m} \sum_{e=0}^{\ell+m-g} \sum_{f=0}^{g-\Odd_e} h^{jlm}_{efg}
  \intq \left(\frac{p}{k}\right)^{2g} \left(\frac{q}{p}\right)^{2f\!+\!\Odd_e\!-\!B}
  \!\!\!\!\!\!
  (1-\mupq^2)^g \mupq^{e+N} \Pa{ab\ell}{q} \Pa{cdm}{p}.
  \label{e:def_S_BNj_lm}
\end{eqnarray}

Expanding the $(1-\mupq^2)^g \mupq^{e+N}$ factor in \EQ{e:def_S_BNj_lm} in binomial series and changing bases to Legendre polynomials, we rewrite the integral in that equation as
\begin{equation*}
  k^{-2g} \sum_{n=0}^g \sum_{\lambda=0}^{e+N+2n}
  (-1)^n \binom{g}{n} \MLeg^{-1}_{\lambda,e+N+2n} 
  \intq q^{2f+\Odd_e-B} p^{2g-2f-\Odd_e+B} \PLeg_\lambda(\mupq)
  \Pa{ab\ell}{q} \Pa{cdm}{p}.
\end{equation*}
Defining $\Ja{AB\lambda,ab\ell,cdm}{k}$ as in Ref.~\cite{McEwen:2016fjn}
\begin{equation}
  \Ja{AB\lambda,ab\ell,cdm}{k}
  =
  \intq q^A p^B \PLeg_\lambda(\mupq) \Pa{ab\ell}{q} \Pa{cdm}{p}
\end{equation}
we see that $\Sa{BNj,ab\ell,cdm}{k}$ may be written as a linear combination of $\Ja{AB\lambda,ab\ell,cdm}{k}$ integrals.

Because our subsequent treatment of $\Sa{BNj,ab\ell,cdm}{k}$ closely parallels that of Ref.~\cite{McEwen:2016fjn} we will be brief.  The Fourier transform of $\Ja{AB\lambda,ab\ell,cdm}{k}$ is
\begin{eqnarray*}
  \Ja{AB\lambda,ab\ell,cdm}{}(\vec r)
  &=&
  \int \!\!\!\!\frac{d^3 k}{(2\pi)^3} e^{i \vec k \cdot \vec r}
  \Ja{AB\lambda,ab\ell,cdm}{k}
  =
  \int \!\!\!\frac{d^3 q}{(2\pi)^3} \frac{d^3 p}{(2\pi)^3}
  e^{i \vec q \cdot \vec r + i \vec p \cdot \vec r}
  q^\alpha p^\beta \PLeg_\lambda(\mupq) \Pa{ab\ell}{q} \Pa{cdm}{p}.
\end{eqnarray*}
Using the spherical harmonic expansions
\begin{equation*}
  \PLeg(\hat q \cdot \hat p)
  =
  \frac{4\pi}{2\ell+1} \sum_{m=-\ell}^\ell
  Y_{lm}(\hat q)^* Y_{lm}(\hat p)
  \quad \textrm{and} \quad
  e^{i\vec q \cdot \vec r}
  =
  4\pi \sum_\ell i^\ell j_\ell(qr)
  \sum_{m=-\ell}^\ell Y_{lm}(\hat q)^* Y_{lm}(\hat r)
\end{equation*}
we may factor the $\vec q$ and $\vec p$ integrals.  Integrating directly over both $\hat q$ and $\hat p$ results in
\begin{equation*}
  \Ja{AB\lambda,ab\ell,cdm}{}(\vec r)
  =
  \frac{(-1)^\lambda}{4\pi^4}
  \!\!\!\int_0^\infty \!\!\!\!\! dq q^{A+2} j_\lambda(qr) \Pa{ab\ell}{q}
  \!\!\int_0^\infty \!\!\!\!\!dp p^{B+2} j_\lambda(pr) \Pa{cdm}{p}
  =:
  \frac{(-1)^\lambda}{4\pi^4} I_{A\lambda}(r) I_{B\lambda}(r)
\end{equation*}
reducing the problem to the computation of the single-dimensional integral $I_{A\lambda}$.

FFT-based acceleration of this integral proceeds by extending the $k$ range of the power spectrum in both directions.  Assume a power spectrum defined at $\Nk$ points, with an even spacing $\Dk$ in $\log(k)$ such that $k_{i+1} = e^{\Dk} k_i$, and with $\Nk$ equal to $2^n$ for some positive integer $n$ as required by the FFT algorithm. We extrapolate it by $3\Nk/2$ in either direction, as in Ref.~\cite{Upadhye:2017hdl}.  Extrapolation uses a power law and then tapers off the power spectrum to zero in both directions.  Since discrete Fourier transforms such as the FFT assume a periodic input function, this zero-padding is necessary to avoid jump discontinuities in the power.  The total number of points in this extrapolated, zero-padded power spectrum is now $\Nkp = 4\Nk$.  Further, as described in Ref.~\cite{McEwen:2016fjn} the power spectrum is tilted by a factor $k^{-\nufpt}$, with $\nufpt=-2$, as a means of controlling divergences.  Its discrete Fourier transform is
\begin{eqnarray}
  \ca{ab\ell,n}
  &=&
  \sum_{h=0}^{\Nkp} \Pa{ab\ell}{}(k_h) k_h^{-\nufpt} e^{-2\pi i h n / \Nkp}
  =
  {\mathscr F}\left\{ \Pa{ab\ell}{}(k_h) k_h^{-\nufpt} \right\}_n^h
  \label{e:def_fft}
  \\
  \Rightarrow
  \Pa{ab\ell}{}(k_h)
  &=&
  \sum_n \ca{ab\ell,n} k_h^{\nufpt + 2\pi i n / (\Nkp \Dk)}.
  \label{e:def_ifft}
\end{eqnarray}
Here ${\mathscr F}$ represents the discrete Fourier transform, for which we may use an FFT.

With this theoretical machinery and the definitions
\begin{eqnarray}
  g(x,y)
  &:=&
  \int_0^\infty dz \, {\mathcal J}_x(z) z^y
  =
  \frac{\Gamma[(x+y+1)/2]}{\Gamma[(x-y+1)/2]}
  \\
  g_{\lambda A n}
  &:=&
  g\left(\lambda+\frac{1}{2},
  \nufpt + A + \frac{3}{2} + \frac{2\pi i n}{\Nk \Dk}\right).
\end{eqnarray}
in terms of Bessel function ${\mathcal J}$, we reduce $I_{A\lambda}$ to
\begin{equation*}
  I_{A\lambda}(r)
  =
  \sqrt{\frac{\pi}{2}}
  \sum_n \ca{ab\ell,n} g_{\lambda A n}
  r^{-3 - \nufpt - A - 2\pi i n / (\Nkp \Dk)}
  2^{\nufpt+A+3/2+2\pi i n/(\Nkp \Dk)}.
\end{equation*}
Multiplying two such integrals, we recover $\Ja{AB\lambda,ab\ell,cdm}{}(\vec r)$:
\begin{equation*}
  \Ja{AB\lambda,ab\ell,cdm}{}(\vec r)
  =
  \frac{(-1)^\lambda}{(2\pi)^3}
  \sum_n \sum_s \ca{ab\ell,n} \ca{cdm,s} g_{\lambda A n} g_{\lambda B s}
  2^{\frac{2\pi i (n+s) }{ \Nkp \Dk}}
  r^{-6 - 2\nufpt - A - B - \frac{2\pi i (n+s) }{ \Nkp \Dk} }.
\end{equation*}

Changing the inner summation index from $s$ to $t = s + n$ reveals a discrete convolution:
\begin{equation}
  \Ca{AB\lambda t,ab\ell,cdm}
  =
  \sum_{n} c_{\alpha,ab\ell,n} g_{\lambda An} c_{\alpha,cdm,t-n} g_{\lambda,B,t-n}.
\end{equation}
Though a brute-force evaluation of this quantity for all $2\Nkp$ values of $t$ would require ${\mathcal O}(\Nkp^2)$, steps, it may be computed in only ${\mathcal O}(\Nkp \log(\Nkp))$ by multiplying the FFTs of the two quantities $c_{\alpha,ab\ell,n} g_{\lambda A n}$ and $c_{\alpha,cdm,n} g_{\lambda,B,n}$ to be convolved, then taking the inverse FFT of the result.  Fourier transforming $\Ja{AB\lambda,ab\ell,cdm}{}(\vec r)$ back to $\vec k$-space and defining $F_t := \int_0^\infty dz \sin(z) z^{-5 - 2\nufpt - A - B - 2\pi i t / ( \Nkp \Dk ) }$ results in
\begin{eqnarray}
  \Ja{AB\lambda,ab\ell,cdm}{k_u}
  &=&
  \int_0^\infty dr 4\pi r^2 j_0(k_u r)  \Ja{AB\lambda,ab\ell,cdm}{}(r)
  \nonumber\\
  &=&
  \frac{(-1)^\lambda}{\pi^2} 2^{2 + 2\nufpt + A + B}
  \sum_t \Ca{AB\lambda t,ab\ell,cdm} F_t 2^{\frac{2\pi i t }{\Nkp \Dk}}
  k_u^{3 + 2\nufpt + A + B + \frac{2\pi i t }{ \Nkp \Dk }}
  \nonumber\\
  &=&
  \frac{(-1)^\lambda}{2\pi^2} (2 k_u)^{3 + 2\nufpt + A + B }
       {\mathscr F}^{-1}\left\{
       \Ca{AB\lambda t,ab\ell,cdm} F_t 2^{\frac{2\pi i t }{\Nkp \Dk}}
       \right\}^t_u
\end{eqnarray}
Thus the elementary $\Ptt$-type integral $\Ja{AB\lambda,ab\ell,cdm}{k}$ has been expressed as an inverse FFT of a discrete convolution, allowing its computation in ${\mathcal O}(\Nkp \log(\Nkp))$ steps.

We began the FFT acceleration procedure with a decomposition of $\Sa{BNj,ab\ell,cdm}{k}$ into $\Ja{AB\lambda,ab\ell,cdm}{k}$ integrals.  Reassembling these parts into the whole, we find after some bookkeeping of summation indices that
\begin{eqnarray}
  \Sa{BNj,ab\ell,cdm}{k_u}
  &=&
  \sum_{g=0}^{\ell+m}
  \frac{4^g (2k_u)^{3+2\nu}}{2\pi^2} {\mathscr F}^{-1} \left\{
  \sum_{\lambda=0}^{\ell+m+g+N} \sum_{f=0}^{f_\mathrm{max}}
  F_t 2^{\frac{2\pi i t}{\Nkp \Dk}} {\tilde C}_{\alpha, BNgt, ab\ell, cdm}
  \right\}^t_u\qquad
  \label{e:S_FFT}
  \\
  {\tilde C}_{\alpha, BNgt, ab\ell, cdm}
  &:=&
  \Ca{2f+\Odd_{\lambda+N}-B,2g-2f-\Odd_{\lambda+N}+B,\lambda,t, ab\ell, cdm}\ldots
  \nonumber\\
  &~&
  \times
  \sum_{w=w_\mathrm{min}}^{w_\mathrm{max}} \sum_{v=v_\mathrm{min}}^g
  (-1)^{\lambda+v} \binom{g}{v} \MLeg^{-1}_{\lambda, N+2v+2w+\Odd_{\lambda+N}}
  h^{j\ell m}_{2w+\Odd_{\lambda+N}, f, g}
  \label{e:def_Ct}
  \\
  w_\mathrm{min} &:=& \max(0,\tfrac{\lambda-N-\Odd_{\lambda+N}}{2}-g),\quad
  w_\mathrm{max}  :=  (\ell + m - g - \Odd_{\ell+m+g})/2
  \nonumber\\
  f_\mathrm{max} &:=& \min(m,g)-\Odd_{\lambda+N},\qquad\quad
  v_\mathrm{min}  :=  (\lambda-N-\Odd_{\lambda+N})/2 - w
  \nonumber
\end{eqnarray}
Since the summations in \EQ{e:def_Ct} are independent of $t$, they need only be computed once for each $\lambda$ and $f$. The final result, \EQ{e:S_FFT}, is the FFT-accelerated mode-coupling integral from which we will construct the $\Aa{acd,bef}{\vec k}$ integrals.  Its computational cost scales with $\Nkp$ as ${\mathcal O}(\Nkp \log(\Nkp))$ as was the case with $\Ja{AB\lambda,ab\ell,cdm}{k}$.

We will see that the $\Ptt$-type mode-coupling integrals dominate the total computational cost.  Suppose that we truncate the summations over angular modes, $\ell,~m < \Nmunl$, in the non-linear integrals of \EQ{e:def_S_BNj_lm}.  Since $j$, $g$, $e$, and $f$ can each take $\sim \Nmunl$ values, the total number of FFT-accelerated integrals making up all of the $\Sa{BNj,ab\ell,cdm}{k}$, hence the total computational cost of the $\Ptt$-type mode-coupling integrals, scales as ${\mathcal O}(\Nmunl^6)$.

\subsection{Decomposition of $\Aa{}{(2,2)}$ integrals}
\label{subsec:p22_A22_catalog}

The goal of this Section is to decompose the $(2,2)$-type components of the mode-coupling integral $\Aa{acd,bef}{\vec k}$ of \EQ{e:def_A22}.  Products of the vertex functions can be expanded as
\begin{eqnarray}
  \gamma_{001}^{\vec k \vec q \vec p}\gamma_{001}^{\vec k \vec q \vec p}
  &=&
  \frac{1}{4} + \frac{1}{2} \frac{q}{p} \mupq
  + \frac{1}{4} \frac{q^2}{p^2}\mupq^2
  \\
  \gamma_{001}^{\vec k \vec q \vec p}\gamma_{010}^{\vec k \vec q \vec p}
  &=&
  \frac{1}{4} + \frac{1}{4}\left(\frac{q}{p} + \frac{p}{q}\right)\mupq
  + \frac{1}{4} \mupq^2
  \\
  \gamma_{001}^{\vec k \vec q \vec p}\gamma_{111}^{\vec k \vec q \vec p}
  &=&
  \frac{1}{4}\left(\frac{q}{p}+\frac{p}{q}\right)\mupq
  + \frac{1}{4}\left(3 + \frac{q^2}{p^2}\right)\mupq^2
  + \frac{1}{2} \frac{q}{p} \mupq^3
\end{eqnarray}
\begin{eqnarray}
  \gamma_{010}^{\vec k \vec q \vec p}\gamma_{010}^{\vec k \vec q \vec p}
  &=&
  \frac{1}{4} + \frac{1}{2}\frac{p}{q}\mupq
  + \frac{1}{4} \frac{p^2}{q^2}\mupq^2
  \\
  \gamma_{010}^{\vec k \vec q \vec p}\gamma_{111}^{\vec k \vec q \vec p}
  &=&
    \frac{1}{4}\left(\frac{q}{p}+\frac{p}{q}\right)\mupq
  + \frac{1}{4}\left(3 + \frac{p^2}{q^2}\right)\mupq^2
  + \frac{1}{2} \frac{p}{q} \mupq^3
  \\
  \gamma_{111}^{\vec k \vec q \vec p}\gamma_{111}^{\vec k \vec q \vec p}
  &=&
  \frac{1}{4}\left( 2 + \frac{q^2}{p^2} + \frac{p^2}{q^2}\right)\mupq^2
  + \left(\frac{q}{p} + \frac{p}{q}\right)\mupq^3 + \mupq^4.
\end{eqnarray}

Substituting these into the expression for $\Aa{acd,bef}{(2,2)\vec k}$, we may immediately expand these mode-coupling integrals in terms of the $\Sa{BN,ab,cd}{\vec k}$ integrals:
\begin{eqnarray}
  \Aa{001,0ef}{(2,2)\vec k}
  &=&
  \frac{1}{4} \Sa{00,e0,f1}{\vec k}
  + \frac{1}{2} \Sa{-1,1,e0,f1}{\vec k}
  + \frac{1}{4} \Sa{-2,2,e0,f1}{\vec k}
  + \frac{1}{4} \Sa{00,e1,f0}{\vec k}
  \ldots\nonumber\\
  &~&
  + \frac{1}{4} \Sa{-1,1,e1,f0}{\vec k}
  + \frac{1}{4} \Sa{11,e1,f0}{\vec k}
  + \frac{1}{4} \Sa{02,e1,f0}{\vec k}
  \label{e:A22_001_0ef}
  \\
  \Aa{001,1ef}{(2,2)\vec k}
  &=&
  \frac{1}{4} \Sa{-1,1,e1,f1}{\vec k}
  + \frac{1}{4} \Sa{11,e1,f1}{\vec k}
  + \frac{3}{4} \Sa{02,e1,f1}{\vec k}
  + \frac{1}{4} \Sa{-2,2,e1,f1}{\vec k}
  + \frac{1}{2} \Sa{-1,3,e1,f1}{\vec k}\qquad
  \label{e:A22_001_1ef}
  \\
  \Aa{111,0ef}{(2,2)\vec k}
  &=&
  \frac{1}{4} \Sa{-1,1,e0,f1}{\vec k}
  + \frac{1}{4} \Sa{11,e0,f1}{\vec k}
  + \frac{3}{4} \Sa{02,e0,f1}{\vec k}
  + \frac{1}{4} \Sa{-2,2,e0,f1}{\vec k}
  + \frac{1}{2} \Sa{-1,3,e0,f1}{\vec k}
  \ldots\nonumber\\
  &~&
  + \frac{1}{4} \Sa{11,e1,f0}{\vec k}
  + \frac{1}{4} \Sa{-1,1,e1,f0}{\vec k}
  + \frac{3}{4} \Sa{02,e1,f0}{\vec k}
  + \frac{1}{4} \Sa{22,e1,f0}{\vec k}
  + \frac{1}{2} \Sa{13,e1,f0}{\vec k}
  \label{e:A22_111_0ef}
  \\
  \Aa{111,1ef}{(2,2)\vec k}
  &=&
  \frac{1}{2} \Sa{02,e1,f1}{\vec k}
  + \frac{1}{4} \Sa{-2,2,e1,f1}{\vec k}
  + \frac{1}{4} \Sa{22,e1,f1}{\vec k}
  + \Sa{-1,3,e1,f1}{\vec k}
  \ldots\nonumber\\
  &~&
  + \Sa{13,e1,f1}{\vec k}
  + \Sa{04,e1,f1}{\vec k}.
  \label{e:A22_111_1ef}
\end{eqnarray}
The above is a ``catalog'' expanding $\Aa{acd,bef}{(2,2)\vec k}$ in terms of the $S$ integrals of \EQ{e:def_S_BNj_lm}, allowing for the rapid computation of the mode-coupling integrals using FFT techniques.

\section{Mode-coupling integrals of $P^{(1,3)}$-type}
\label{sec:mode-coupling_P13}

Computation of $\Aa{acd,bef}{(1,3)\vec k}$, as defined in \EQ{e:def_A13}, is the goal of the present Section.  Reduction of this integral to a convolution, necessary for its acceleration, requires that its integrand depend only upon the single integration variable $q$.  Thus we must explicitly integrate over both $\vartheta$ and $\varphi$ in order to proceed.  We begin by averaging over $\vartheta$ and $\varphi$ factors containing integer powers of $\vtc$, $\vts$, and $\muaq$.  By analogy with the $\Sa{BN,ab,cd}{\vec k}$ integrals of Sec.~\ref{sec:mode-coupling_P22}, we then define $\Pot$-type mode-coupling terms from which we create a catalog of $\Aa{acd,bef}{(1,3)\vec k}$ values.  Concluding this Section is a numerical test of the full $\Aa{acd,bef}{\vec k}$ integrals.

\subsection{Angular averages of elementary terms}
\label{subsec:p13_ang_avg}

Our first goal is to show that the integrand of $\Aa{acd,bef}{(1,3)\vec k}$ may be explicitly averaged over $\vartheta$ and $\varphi$, leaving only one integral to be carried out numerically.  The chief difficulty is that the vertex factors $\gamma_{001}^{\vec k \vec q \vec p}$ and $\gamma_{111}^{\vec k \vec q \vec p}$ have $p^2 = k^2 + q^2 - 2 k q \vtc$ in their denominators, giving rise to badly-behaved terms which must be handled carefully.  Thus we begin by performing $\vartheta$ averages over elementary terms into which we will later decompose the integrals $\Aa{acd,bef}{(1,3)\vec k}$.  As a shorthand, we define the dimensionless variables $x := q/k$, $y(x) = (x + x^{-1})/2$, and $z(x) := (x - x^{-1})/2$. After $\vartheta$ averaging, all that remains is an integral over $x>0$.

Terms without $p^2$ in their denominators are trivial.  Once again using the convention that angles in the subscript of triangular brackets denote averages over those angles, we have
\begin{eqnarray}
  \left< \vts^{2r} \vtc^n \right>_\vartheta
  &=&
  \Even_n
  \frac{\Gamma(r+1) \Gamma(\tfrac{n+1}{2})
  }{2 \Gamma(r + \tfrac{n}{2} + \tfrac{3}{2})}
  \\
  \left< \vtc^m \muaq^n \right>_{\vartheta, \varphi}
  &=&
  \Even_{m+n}
  \sum_{r=0}^{n/2} \binom{n}{2r} \kappa_r \mua^{n-2r} \nua^{2r}
  \frac{\Gamma(r+1) \Gamma(\tfrac{n+1}{2})
  }{2 \Gamma(r + \tfrac{n}{2} + \tfrac{3}{2})}
\end{eqnarray}
(Recall the definitions of $\Even_n$ and $\Odd_n$ from \EQ{e:def_E_O}.)  Our convention for summations is that $\sum_{i=A}^B$ denotes a summation over all integers in the closed interval $[A,B]$, even if $A$ and $B$ are not integers.  Thus the actual upper limit of $s$ summation, for example, is $(n-\Odd_n)/2$.

Terms with denominators proportional to $p^2$, when integrated over $\vartheta$, give rise to terms proportional to $\Lx := \ln\left|\frac{1+x}{1-x}\right|$ which must then be integrated over $x$.  However, this factor has a logarithmic singularity at $x=1$.  Its Taylor series is
\begin{equation}
  \Lx(x)
  :=
  \ln\left|\frac{1+x}{1-x}\right|
  =
  \left\{
  \begin{array}{cl}
    2 \sum_{j=0}^\infty \frac{x^{2j+1}}{2j+1}
    & \textrm{ for $x<1$}
    \\
    2\sum_{j=0}^\infty \frac{x^{-(2j+1)}}{2j+1}
    & \textrm{ for $x>1$}
  \end{array}
  \right. .
\end{equation}
For each integer $N$, we define a function 
\begin{equation}
  \lambda_N(x)
  :=
  \left\{
  \begin{array}{cl}
    0
    & \textrm{ for $N=0$}
    \\
    \frac{1}{2} (1-x^N) \ln\left|\frac{1+x}{1-x}\right|
    + \sum_{j=0}^{(N-1)/2} \frac{x^N-2j-1}{2j+1}
    & \textrm{ for $N>0$}
    \\
    \lambda_{-N}(x^{-1})
    =
    \frac{1}{2} (1-x^N) \ln\left|\frac{1+x}{1-x}\right|
    + \sum_{j=0}^{(-N-1)/2} \frac{x^N+2j+1}{2j+1}
    & \textrm{ for $N<0$}
  \end{array}
  \right.
\end{equation}
that is finite and bounded for all positive real $x$. In the remainder of this subsection, we will compute all necessary angular averages with $p^2$ in the denominator, expressed as sums of $\lambda_N$ functions and integer powers of $x$.

The simplest such angular average is
\begin{equation}
  \left<  \frac{z \vtc^n}{\vtc - y} \right>_\vartheta
  =
  -z y^n \Lx + z\sum_{j=0}^{(n-1)/2} \frac{y^{n-2j-1}}{2j+1}
  \label{e:P13_avgTheta_elem1_intermediate1}
\end{equation}
for integer $n \geq 0$.  After some algebra, we expand the first term on the right hand side as
\begin{eqnarray}
  -z y^n \Lx
  &=&
  \frac{1}{2^n} \sum_{i=0}^{n+1} \bNi{n}{i} \lambda_{n+1-2i}(x)
  - \frac{1}{2^n} \sum_{j=0}^{(n-1)/2} U^{(n)}_j (x^{n-2j} - x^{-n+2j})
  \\
  \mathrm{where}\quad
  \bNi{n}{i}
  &:=&
  \binom{n}{i} - \binom{n}{i-1}
  \quad\mathrm{and}\quad
  U^{(n)}_j
  :=
  \sum_{i=0}^j \frac{\bNi{n}{i}}{2j-2i+1}.
\end{eqnarray}
The second term on the right hand side of Eq.~(\ref{e:P13_avgTheta_elem1_intermediate1}) can be recast as
\begin{eqnarray}
  z\sum_{j=0}^{(n-1)/2} \frac{y^{n-2j-1}}{2j+1}
  &=&
  \frac{1}{2^n} \sum_{j=0}^{(n-1)/2} V^{(n)}_j (x^{n-2j} - x^{-n+2j})
  \quad\mathrm{where}\quad
  V^{(n)}_j
  :=
  \sum_{i=0}^j \frac{4^i \bNi{n-2i-1}{j-i}}{2i+1}.\qquad
\end{eqnarray}

The final step is to show that $U^{(n)}_j$ and $V^{(n)}_j$ are equal.  Defining $u_n(\xi) := \sum_{j=0}^\infty U^{(n)}_j \xi^{2j}$ and $v_n(\xi) := \sum_{j=0}^\infty V^{(n)}_j \xi^{2j}$, we will show that $u_n(\xi) = v_n(\xi) = (\xi^{-1}-\xi)(1+\xi^2)^n \arctanh(\xi)$.  Since the $U^{(n)}_j$ and $V^{(n)}_j$ are Taylor series coefficients of this function, they must also be equal.

For $u_n$, after reversing the summation order of $j$ and $i$, we have
\begin{eqnarray}
  u_n(\xi)
  &=&
  \sum_{i=0}^\infty \sum_{j=i}^\infty
  \binom{n}{i} \frac{\xi^{2j}}{2j-2i+1}
  -
  \sum_{i=1}^\infty \sum_{j=i}^\infty
  \binom{n}{i-1} \frac{\xi^{2j}}{2j-2i+1}
  \nonumber\\
  &=&
  \sum_{i=0}^\infty \sum_{j'=0}^\infty
  \binom{n}{i} \frac{\xi^{2j' + 2i}}{2j'+1}
  -
  \sum_{i=1}^\infty \sum_{j'=0}^\infty
  \binom{n}{i-1} \frac{\xi^{2j'+2i}}{2j'+1}
  \nonumber\\
  &=&
  \xi^{-1} \arctanh(\xi) \sum_{i=0}^\infty \binom{n}{i} \xi^{2i}
  -
  \xi^{-1} \arctanh(\xi) \sum_{i'=0}^\infty \binom{n}{i'} \xi^{2i'+2}
  \nonumber\\
  &=&
  (\xi^{-1}-\xi) (1+\xi^2)^n \arctanh(\xi)
  \nonumber
\end{eqnarray}
with $j':=j-i$ and $i':=i-1$ and the Taylor series $\arctanh(\xi) = \sum_{j=0}^\infty \tfrac{\xi^{2j+1}}{2j+1}$.

The complication for $v_n(\xi)$ is that, for $2j > n-1$, the upper arguments of the binomial coefficients become negative.  Using the standard generalization $\binom{n}{i} = (-1)^i \binom{i-n-1}{i}$ to extend the binomial coefficients to negative $n$, we have
\begin{eqnarray}
  v_n(\xi)
  &=&
  \sum_{i=0}^\infty \sum_{j'=0}^\infty
  \binom{n-2i-1}{j'} \frac{4^i \xi^{2j'+2i}}{2i+1}
  -
  \sum_{i=0}^\infty \sum_{j'=1}^\infty
  \binom{n-2i-1}{j'-1} \frac{4^i \xi^{2j'+2i}}{2i+1}
  \nonumber\\
  &=&
  \sum_{i=0}^\infty \frac{(2\xi)^{2i}}{2i+1}
  \sum_{j'=0}^\infty \binom{n-2i-1}{j'} \xi^{2j'}
  -
  \xi^2 \sum_{i=0}^\infty \frac{(2\xi)^{2i}}{2i+1}
  \sum_{j''=0}^\infty \binom{n-2i-1}{j''} \xi^{2j''}
  \nonumber\\
  &=&
  \sum_{i=0}^\infty \frac{(2\xi)^{2i}}{2i+1} (1+\xi^2)^{n-2i-1}
  -
  \xi^2 \sum_{i=0}^\infty \frac{(2\xi)^{2i}}{2i+1} (1+\xi^2)^{n-2i-1}
  \nonumber\\
  &=&
  \frac{1}{2} (\xi^{-1}-\xi) (1+\xi^2)^n
  \arctanh\left(\frac{2\xi}{1+\xi^2}\right)
  \nonumber
\end{eqnarray}
with $j'':= j'-1$.  Note that these extended binomial coefficients are used only in this particular proof.  In practice we only need $V^{(n)}_j$ for $2j \leq n-1$, for which the upper argument of each binomial coefficient is non-negative.  The hyperbolic arctangent can be simplified as:
\begin{equation}
  \arctanh\left(\frac{2\xi}{1+\xi^2}\right)
  =
  \frac{1}{2}
  \ln \left|
  \frac{1+\tfrac{2\xi}{1+\xi^2}}{1-\tfrac{2\xi}{1+\xi^2}}
  \right|
  =
  \ln\left|\frac{1+\xi}{1-\xi}\right|
  =
  2 \, \arctanh(\xi).
  \nonumber
\end{equation}
Thus $U^{(n)}_j$ and $V^{(n)}_j$ are equal, and
\begin{equation}
  \left<  \frac{z \vtc^n}{\vtc - y} \right>_\vartheta
  =
  \frac{1}{2^n} \sum_{i=0}^{n+1} \bNi{n}{i} \lambda_{n+1-2i}(x).
  \label{e:P13_avgTheta_elem1}
\end{equation}

A more general term multiplies the above by $x^m$ for integer $m$.  Note that
\begin{equation}
  x \lambda_N(x)
  =
  \lambda_{N+1}(x) - \lambda_1(x) + 1 +
  \left\{
  \begin{array}{cl}
    \frac{1}{N+1} & \textrm{if $N\geq 0$ is even;}\\
    \frac{x}{N}   & \textrm{if $N\leq -1$ is odd;}\\
    0             & \textrm{otherwise.}
  \end{array}
  \right.
\end{equation}
Applying this repeatedly, we find that for any positive integer $m$,
\begin{eqnarray}
  \left<\frac{z x^m \vtc^n}{\vtc-y}\right>_\vartheta
  &=&
  \frac{1}{2^n} \sum_{i=0}^{n+1} \bNi{n}{i} \lambda_{m+n+1-2i}(x)
  - \frac{1}{2^n} \sum_{j=0}^m \sigmamnj{m}{n}{j} x^j
  \\
  \left<\frac{z x^{-m} \vtc^n}{\vtc-y}\right>_\vartheta
  &=&
  -\frac{1}{2^n} \sum_{i=0}^{n+1} \bNi{n}{i} \lambda_{2i-m-n-1}(x)
  + \frac{1}{2^n} \sum_{j=0}^m \sigmamnj{m}{n}{j} x^{-j}
\end{eqnarray}
where the coefficients $\sigmamnj{m}{n}{j}$ are defined recursively for non-negative integers $m$ and $n$ by
\begin{eqnarray}
  \sigmamnj{0}{n}{0} &=& 0
  \quad\mathrm{and}\quad
  \sigmamnj{m}{n}{0}
  =
  \Even_{m+n} \sum_{i=0}^{(m+n)/2} \frac{\bNi{n}{i}}{m+n-2i+1}
  \textrm{ for $m \geq 1$}
  \label{e:sigmamn0}
  \\
  \sigmamnj{m}{n}{1}
  &=&
  \sigmamnj{m-1}{n}{0}
  +
  \Odd_{m+n}
  \sum_{i=0}^{(n-m+1)/2} \frac{\bNi{n}{i}}{n-m-2i+2}
  \textrm{ for $m \geq 1$}
  \label{e:sigmamn1}
  \\
  \sigmamnj{m}{n}{j}
  &=&
  \sigmamnj{m-1}{n}{j-1} \textrm{ for $m\geq 1$ and $j \geq 2$}.
  \label{e:sigmamnj_recursion}
\end{eqnarray}

The most general elementary $\varphi$ average with $p^2$ in the denominator is
\begin{eqnarray}
  \left< \frac{z \muaq^\ell x^m \vtc^n}{\vtc-y} \right>_{\!\!\vartheta,\varphi}
  &=&
  \sum_{r=0}^{\ell/2} \binom{\ell}{2r} \kappa_r \mua^{\ell-2r} \nua^{2r}
  \sum_{s=0}^r (-1)^{r-s} \binom{r}{s}
  \left< \frac{z x^m \vtc^{\ell+n-2s}}{\vtc-y} \right>_\vartheta
  \\
  &=&
  \sum_{t=0}^{\ell/2} \mua^{\Odd_\ell + 2t}
  \sum_{s=0}^{\ell/2} (-1)^{s+t}
  \left< \frac{z x^m \vtc^{\ell+n-2s}}{\vtc-y} \right>_{\!\!\vartheta}
  \ldots\nonumber\\
  &~&\times
  \sum_{r=0}^{\min(s,t)}  \kappa_{\frac{\ell-\Odd_\ell}{2}-r}
  \binom{\ell}{2r+\Odd_\ell} \binom{\frac{\ell-\Odd_\ell}{2}-r}{s-r}
  \binom{\frac{\ell-\Odd_\ell}{2}-r}{t-r}.
\end{eqnarray}
This may be expressed more concisely by defining the functions
\begin{eqnarray}
  \Lc_{\ell m n}(\mua,x)
  &:=&
  \sum_{r=0}^{\ell/2} \!\binom{\!\ell\!}{\!2r\!} \kappa_r \mua^{\ell-2r} \nua^{2r}
  \!\!\sum_{s=0}^r \!\binom{r}{s} (-1)^{r-s}
  \!\!\!\! \sum_{i=0}^{\ell+n-2s+1}
  \frac{\bNi{\ell+n-2s}{i} \lambda_{\ell+m+n+1-2s-2i}(x)}{2^{\ell+n-2s}}\qquad
  \label{e:def_Lc}
  \\
  \Mc^{(j)}_{\ell m n}(\mua)
  &:=&
  \sum_{r=0}^{\ell/2} \binom{\ell}{2r} \kappa_r \mua^{\ell-2r} \nua^{2r}
  \sum_{s=0}^r \binom{r}{s} (-1)^{r-s}
  \frac{\sigmamnj{|m|}{\ell+n-2s}{j}}{2^{\ell+n-2s}}.
  \label{e:def_Mc}
\end{eqnarray}
Appendix~\ref{app:expansions_of_Lc_Mc} expands and simplifies these polynomials in $\mua$ for cases of interest here,
\begin{eqnarray}
  \Lc_{2L,m,n}(\mua,x)
  &=&
  \sum_{t=0}^L (-1)^t \binom{L}{t} \mua^{2t}
  \sum_{v=0}^{2L+n+1} {\hat g}_{Lntv} \, \lambda_{m-2L-n-1+2v}(x)
  \label{e:Lc_from_appA}
  \\
  \Mc^{(j)}_{2L,m,2I+\Odd_j-m}(\mua)
  &=&
  \sum_{t=0}^L {\hat d}^{(j)}_{LmIt} \mu^{2t}
  \label{e:Mc_from_appA}
\end{eqnarray}
where ${\hat g}_{Lntv}$ and ${\hat d}^{(j)}_{LmIt}$ are respectively given by \EQ{e:def_hat_g} and \EQS{e:def_hat_d0}{e:def_hat_d1}.  In terms of these, we rewrite the expressions for the general elementary terms as
\begin{eqnarray}
  \left< \muaq^\ell x^m \vtc^n \right>_{\!\!\vartheta,\varphi}
  &=&
  x^m \Mc^{(1)}_{\ell,1,n}(\mua)
  \label{e:p13_fac0}
  \\
  \left< \frac{z \muaq^\ell x^m \vtc^n}{\vtc-y} \right>_{\!\!\vartheta,\varphi}
  &=&
  \Lc_{\ell m n}(\mua,x)
  - \sign(m) \sum_{j=0}^{|m|} x^{j\cdot \sign(m)} \Mc^{(j)}_{\ell,|m|,n}(\mua).\qquad
  \label{e:p13_fac1}
\end{eqnarray}
Equations~(\ref{e:p13_fac0},~\ref{e:p13_fac1}) are the key results of this subsection.

\subsection{$\Pot$-type power spectrum integrals}
\label{subsec:p13_integrals}

Since the vertex factors $\gamma_{abc}$ of Eqs.~(\ref{e:gamma_001}-\ref{e:gamma_111}) can all be expanded in terms containing only integer powers of $x$ and $\vtc$, the $\Aa{acd,bef}{(1,3)\vec k}$ of \EQ{e:def_A13} can themselves be expanded in
\begin{equation}
  \mua^{2I} \intq \left< \muaq^{2L} x^m \vtc^n \right>_{\!\!\vartheta,\varphi}
  \Pa{abi}{k} \Pa{cd\ell}{q}
  \quad\textrm{ and }\quad
  \mua^{2I}
  \intq \left< \frac{z \muaq^{2L} x^m \vtc^n}{\vtc-y} \right>_{\!\!\vartheta,\varphi}
  \Pa{abi}{k} \Pa{cd\ell}{q}.
  \nonumber
\end{equation}
The evenness of the exponents of $\mua$ and $\muaq$ is due to the fact that power spectra are expanded in terms of the squares of Legendre polynomials, which are necessarily even in their respective arguments.  Thus the above may be expressed using $\Lc_{2L, m, n}$ and $\Mc^{(r)}_{2L, m, n}$.

Consider a term $z x^m \vtc^n / (\vtc-y)$ arising from a pair of vertex factors.  Multiplying by the power spectra and integrating, we have
\begin{eqnarray}
  \intq \frac{z x^m \vtc^n}{\vtc-y} \Pa{ab}{\vec k} \Pa{cd}{\vec q}
  &=&
  \intq \frac{z x^m \vtc^n}{\vtc-y} \sum_{i,\ell}
  \PLeg_i(\mua)^2 \PLeg_\ell(\muaq)^2 \Pa{abi}{k} \Pa{cd\ell}{q}
  \nonumber\\
  &=&
  \sum_{i,\ell} \sum_{I,L}  \intq \frac{z x^m \vtc^n}{\vtc-y} \mua^{2I} \muaq^{2L}
  \QLeg_{Ii} \QLeg_{L\ell} \Pa{abi}{k} \Pa{cd\ell}{q}
  \nonumber\\
  &=&
  \sum_{i,\ell}\sum_{I,L} \QLeg_{Ii} \QLeg_{L\ell} \mua^{2I}
  \Bigg[ \intq \Lc_{2L,m,n}(\mua,x)\Pa{abi}{k} \Pa{cd\ell}{q}
    \ldots\nonumber\\
    &~&\quad
    - \sign(m)\sum_{r=0}^{|m|}  \Mc^{(r)}_{2L, m, n}(\mua) 
     \intq x^{r\,\sign(m)} \Pa{abi}{k} \Pa{cd\ell}{q}
     \Bigg].\qquad
\end{eqnarray}
Similarly, considering a term $x^m \vtc^n$, we have
\begin{eqnarray}
  \intq x^m \vtc^n \Pa{ab}{\vec k} \Pa{cd}{\vec q}
  &=&
  \sum_{i,\ell}\sum_{I,L} \QLeg_{Ii} \QLeg_{L\ell} \mua^{2I}
  \intq  x^m \vtc^n \muaq^{2L} \Pa{abi}{k} \Pa{cd\ell}{q}
  \nonumber\\
  &=&
  \sum_{i,\ell}\sum_{I,L} \QLeg_{Ii} \QLeg_{L\ell} \mua^{2I}
  \Mc^{(1)}_{2L,1,n}(\mua) \intq x^m \Pa{abi}{k} \Pa{cd\ell}{q}.
\end{eqnarray}
These motivate the definitions
\begin{eqnarray}
  \Wa{abcd,tnr}{\vec k}
  &=&
  \sum_{i,\ell}\sum_{I,L} \QLeg_{Ii} \QLeg_{L\ell} \mua^{2I}
  \Mc^{(r)}_{2L,t,n}(\mua) \intq \Pa{abi}{k} \Pa{cd\ell}{q}
  \\
  \Za{abcd,tn}{\vec k}
  &=&
  \sum_{i,\ell}\sum_{I,L} \QLeg_{Ii} \QLeg_{L\ell} \mua^{2I}
  \intq \Lc_{2L,t,n}(\mua,x) \Pa{abi}{k} \Pa{cd\ell}{q}
\end{eqnarray}
where $i$ and $\ell$ are summed from $0$ to $\infty$; $I$ from $0$ to $i$; and $L$ from $0$ to $\ell$.  Each of these integrals, in turn, may be further decomposed.

We simplify $\Wa{abcd,tnr}{\vec k}$ by noting that, for given $i$ and $\ell$, the summand is an even polynomial in $\mua$ multiplying a single integral over $\vec q$.  Thus we may define $d^{j\ell m}_{tIr}$ by
\begin{eqnarray}
  \sum_{j=0}^\infty \PLeg_j(\mua)^2 d^{j\ell m}_{tIr}
  &=&
  \sum_{L=0}^\ell \sum_{M=0}^m \QLeg_{L\ell} \QLeg_{Mm} \mua^{2L}
  \Mc^{(r)}_{2M,t,2I+\Odd_r-t}(\mua)
  \nonumber\\
  \Rightarrow
  d^{j\ell m}_{tIr}
  &=&
  \sum_{M=0}^m \sum_{L=0}^M \sum_{J=\max(L,j)}^{\ell+L}
  \QLeg^{-1}_{jJ} \QLeg_{J-L,\ell} \QLeg_{Mm} \hat d^{(r)}_{MtIL}.
\end{eqnarray}
This allows the expression of $\Wa{abcd,tnr}{\vec k}$ in squared Legendre polynomials:
\begin{eqnarray}
  \Wa{abcd,tnr}{\vec k}
  &=&
  \sum_{j=0}^\infty \PLeg_j(\mua)^2 \Wa{abcd,tnr,j}{k}
  \\
  \Wa{abcd,tnr,j}{k}
  &=&
  \sum_{\ell=0}^\infty \sum_{m=0}^\infty d^{j\ell m}_{tIr} \Wa{ab\ell,cdm}{k}
  \quad\textrm{ with } I:=\frac{t+n-r}{2}
  \\
  \Wa{ab\ell,cdm}{k}
  &=&
  \intq \Pa{ab\ell}{k} \Pa{cdm}{q}.
\end{eqnarray}

In the same manner, by applying the appropriate basis transformations to ${\hat g}_{Lntv}$ of Eq.~(\ref{e:def_hat_g}), we may expand $\Za{abcd,tn}{\vec k}$ in squared Legendre polynomials:
\begin{eqnarray}
  g^{j\ell m}_{nv}
  &=& \!\!\!\!
  \sum_{M=M_{\min}}^m \sum_{L=0}^M \sum_{J=\max(L,j)}^{\ell+L} \!\!\!\!\!\!\!\!\!
  \QLeg^{-1}_{jJ}\QLeg_{J-L,\ell}\QLeg_{Mm} (\!-1\!)^L
  \binom{\!M\!}{\!L\!}{\hat g}_{M,n,L,v\!+\!M\!-\!m} \qquad
  \\
  M_{\min}
  &=&
  \max(0, m-v, v-m-n-1)
  \\
  \Za{abcd,tn}{\vec k}
  &=&
  \sum_{j=0}^\infty \PLeg_j(\mua)^2 \Za{abcd,tn,j}{k}
  \\
  \Za{abcd,tn,j}{k}
  &=&
  \sum_{\ell=0}^\infty \sum_{m=0}^\infty \sum_{v=0}^{2m+n+1}
  g^{j\ell m}_{nv} \Za{ab\ell,cdm,t-2m-n-1+2v}{k}
  \\
  \Za{ab\ell,cdm,N}{k}
  &=&
  \intq \lambda_{N}(q/k) \Pa{ab\ell}{k} \Pa{cdm}{q}
\end{eqnarray}
In the next subsection, we will use these integrals, $\Wa{abcd,tnr,j}{k}$ and $\Za{abcd,tn,j}{k}$, as the building blocks of the $P^{(1,3)}$-type mode-coupling integrals.

\subsection{Decomposition of $\Aa{acd,bef}{(1,3)}$ integrals}
\label{subsec:p13_A13_catalog}

Section~\ref{subsec:trg_mode-coupling} split $\Aa{acd,bef}{\vec k}$ of \EQ{e:A_22_13} into a single $\Ptt$-type term as well as two $\Pot$-type terms, $\Aa{acd,bef}{(1,3)\vec k}$ and $\Aa{adc,bfe}{(1,3)\vec k}$.  In the subsequent Section, \EQS{e:gamma_001}{e:gamma_111} expanded the vertex factors in terms of $x=q/k$ and $\vtc$.  Proceeding as in Sec.~\ref{subsec:p22_A22_catalog}, we construct a catalog of $\Pot$-type contributions to $\Aa{acd,bef}{\vec k}$.

Including both $\Pot$-type terms from \EQ{e:A_22_13}, the catalog of terms is
\begin{eqnarray}
  \Aa{001,bef}{(1,3)\vec k} + \Aa{010,bfe}{(1,3)\vec k}
  &=&
  -\frac{1}{4} \Wa{1bef,121}{\vec k}
  + \delK_{f,1} \frac{1}{4} \Wa{1b1e,121}{\vec k}
  \ldots\nonumber\\
  &~&
  + \frac{\delK_{f,0}}{8}
  \Big[ \Wa{0b1e,101}{\vec k} - \Wa{0b1e,110}{\vec k} + \Za{0b1e,00}{\vec k}
    - \Za{0b1e,-1,1}{\vec k} 
    \ldots\nonumber\\
    &~&~~~~~~~
    + \Wa{1b0e,101}{\vec k} + \Wa{1b0e,110}{\vec k}
    + \Za{1b0e,00}{\vec k} - \Za{1b0e,11}{\vec k} \Big]
  \label{e:A13_001}
  \\
  \Aa{111,bef}{(1,3)\vec k} + \Aa{111,bfe}{(1,3)\vec k}
  &=&
  \frac{\delK_{e,0}}{8}
  \Big[ \Wa{0b1f,200}{\vec k} - \Wa{0b1f,310}{\vec k}
    + \Za{0b1f,-2,0}{\vec k} - \Za{0b1f,-3,1}{\vec k}
    \ldots\nonumber\\
    &~&~~~~~~~
    + \Wa{1b0f,200}{\vec k} - \Wa{1b0f,110}{\vec k}
    + \Za{1b0f,-2,0}{\vec k} - \Za{1b0f,-1,1}{\vec k} \Big]
  \ldots\nonumber\\
  &~&
  +\frac{\delK_{f,0}}{8}
  \Big[ \Wa{0b1e,200}{\vec k} - \Wa{0b1e,310}{\vec k}
    + \Za{0b1e,-2,0}{\vec k} - \Za{0b1e,-3,1}{\vec k}
    \ldots\nonumber\\
    &~&~~~~~~~
    + \Wa{1b0e,200}{\vec k} - \Wa{1b0e,110}{\vec k}
    + \Za{1b0e,-2,0}{\vec k} - \Za{1b0e,-1,1}{\vec k} \Big].
  \label{e:A13_111}
\end{eqnarray}
FFT acceleration results in several additional terms which are divergent, but which precisely cancel divergent terms arising from $\Ptt$-type mode-coupling integrals, as discussed in Appendix~\ref{app:divergence_cancellation}.  These have not been included in the above catalog.

Our FFT-accelerated computation of all mode-coupling integrals is now complete.  We will conclude this section with numerical tests of the mode-coupling integrals computed above.

\subsection{Tests of $\Aa{acd,bef}{\vec k}$ computation}
\label{subsec:p13_A_tests}

\begin{figure}[tbp]
  \includegraphics[width=77mm]{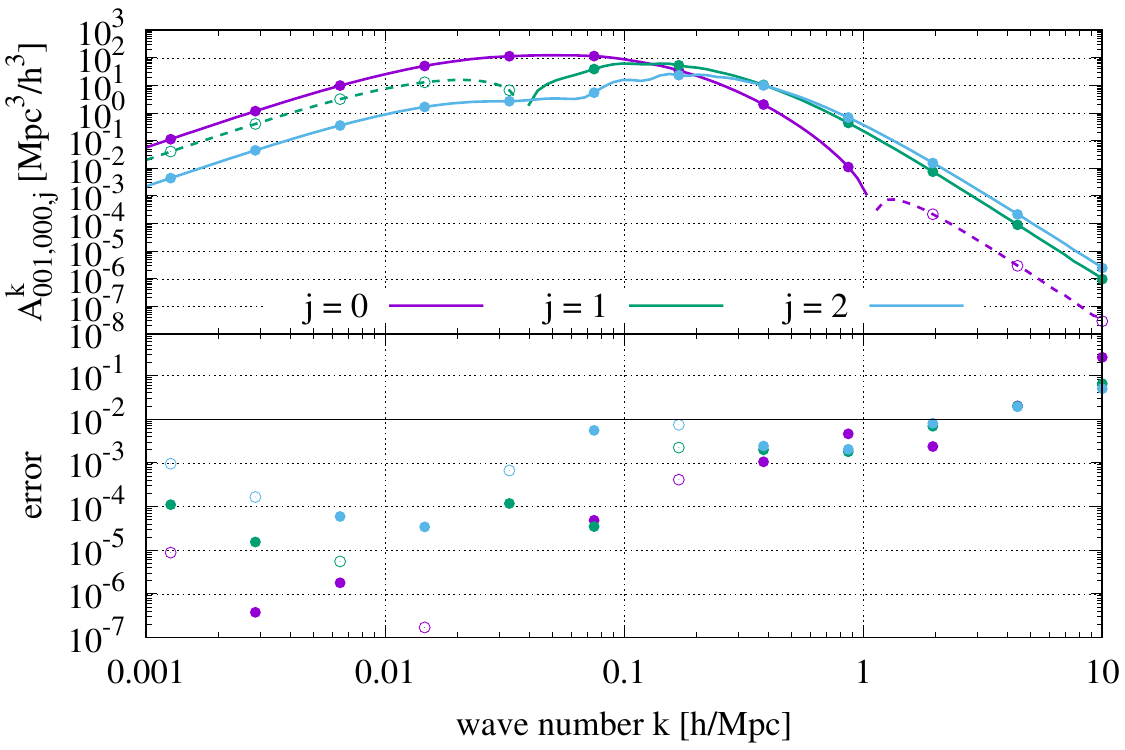}
  \includegraphics[width=77mm]{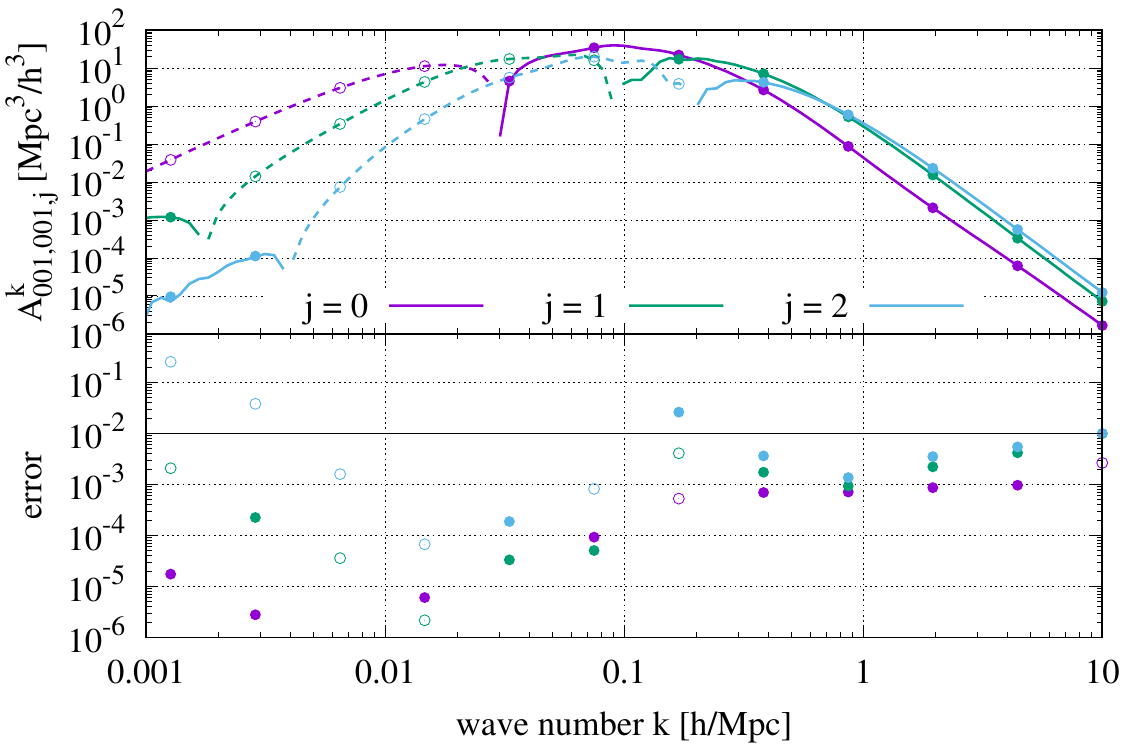}%

  \includegraphics[width=77mm]{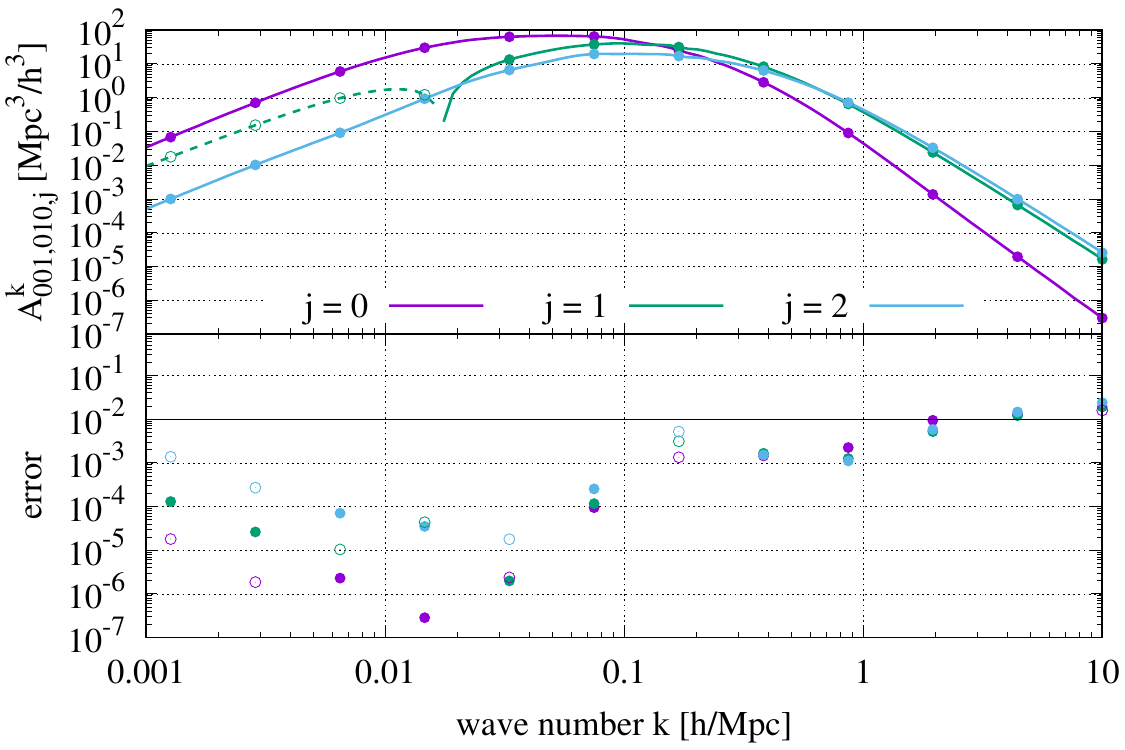}
  \includegraphics[width=77mm]{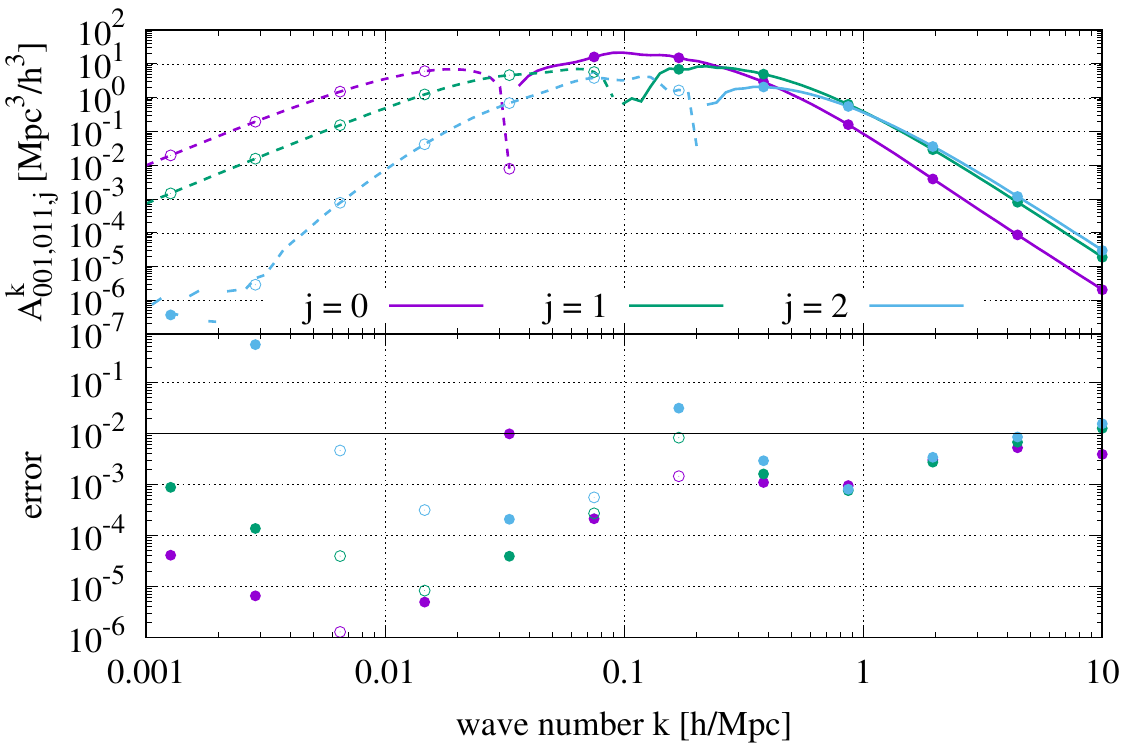}%

  \includegraphics[width=77mm]{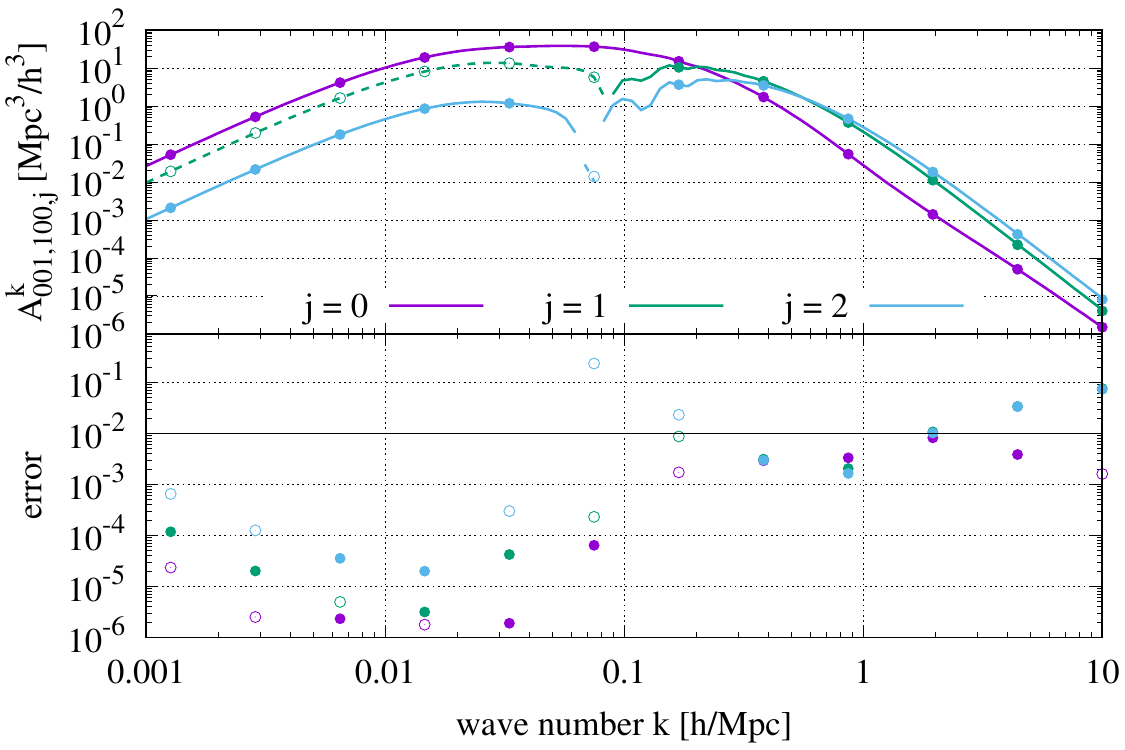}
  \includegraphics[width=77mm]{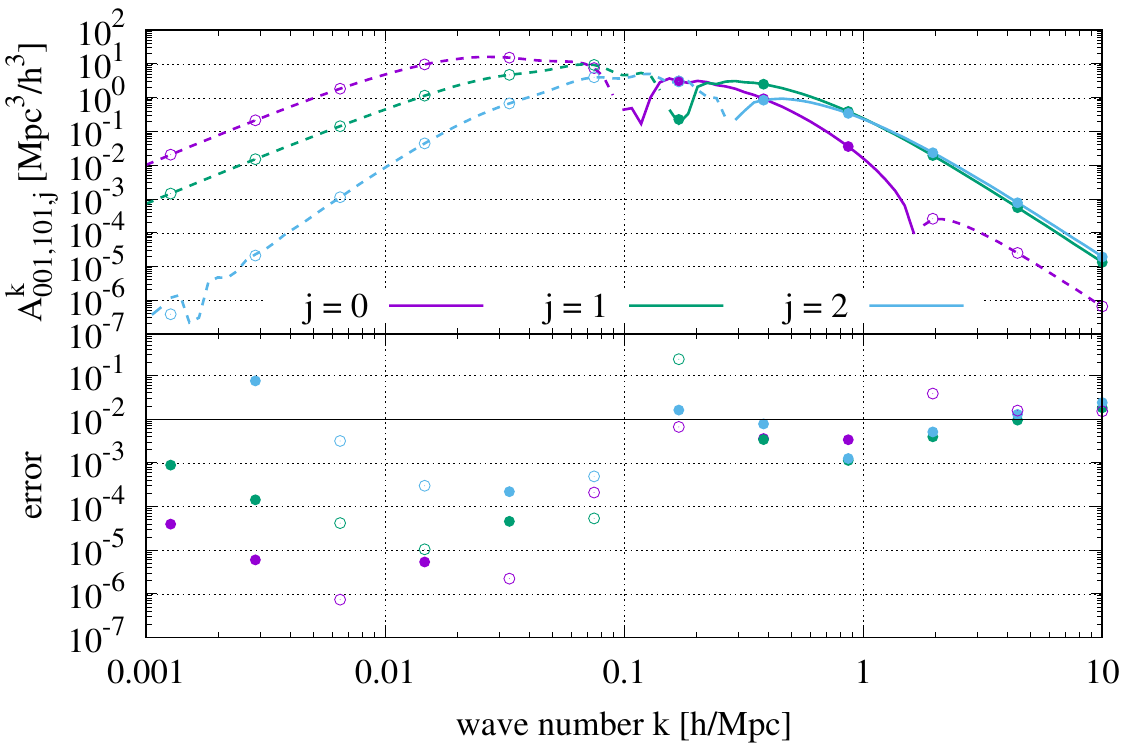}%

  \includegraphics[width=77mm]{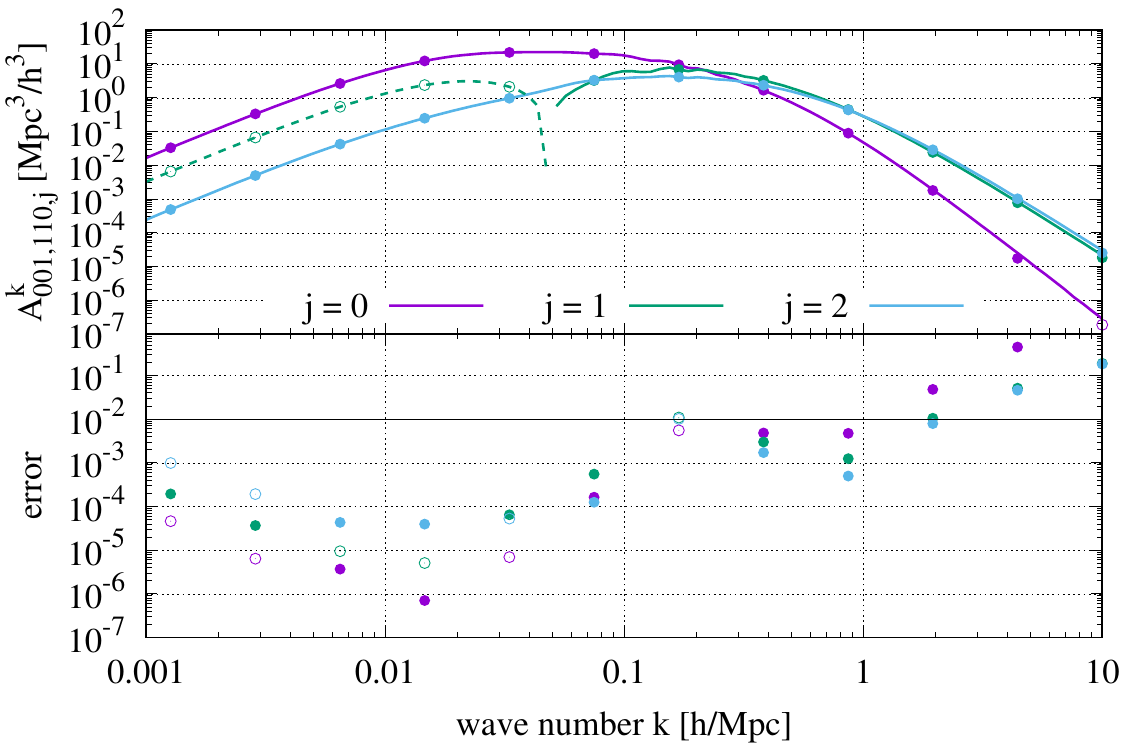}
  \includegraphics[width=77mm]{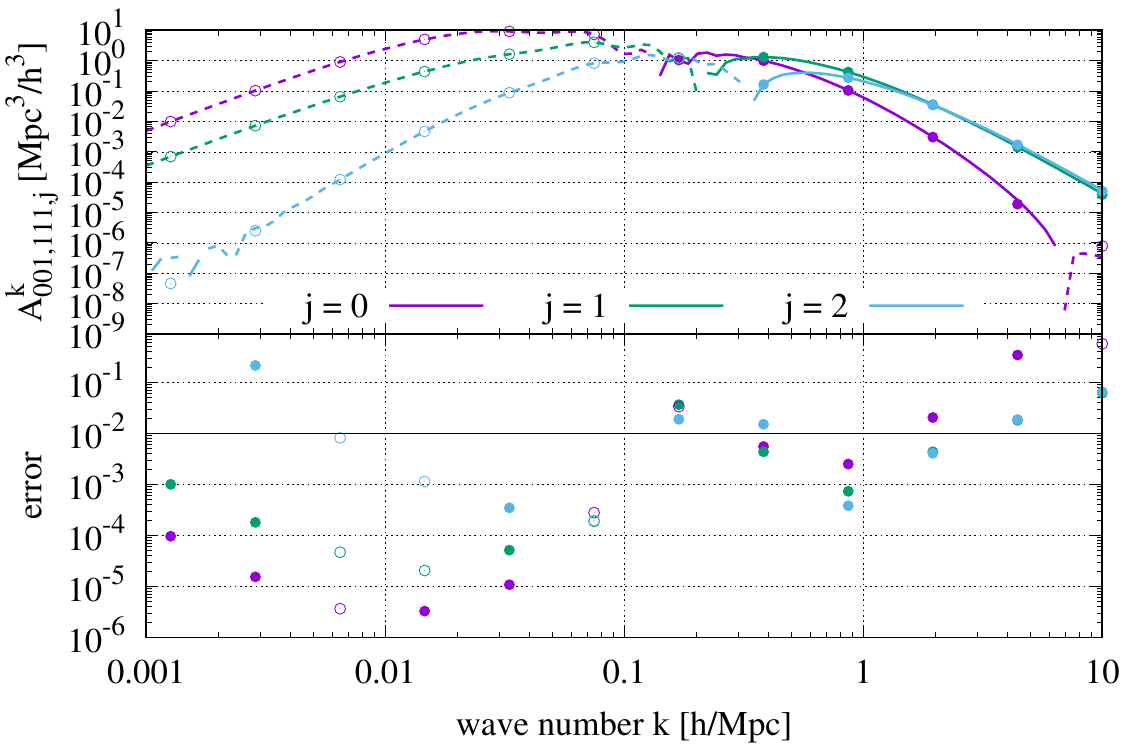}%

  \caption{
    Mode-coupling integrals $\Aa{001,bef,j}{k}$ of \EQS{e:A_22_13}{e:def_A13}
    computed using the FFT techniques of
    Sections~\ref{sec:mode-coupling_P22}-\ref{sec:mode-coupling_P13} (lines)
    compared against direct numerical integration of \EQ{e:def_A} (points).
    Dashed lines and open points denote negative values.  Within each plot,
    the upper panel shows $\Aa{001,bef,j}{k}$ and the lower panel the
    fractional error in the FFT computation.  The $\Aa{001,bef,j}{k}$
    are plotted using $z=0$ linear power spectra and $\Nmunl = 3$ angular modes
    for a neutrino flow with $\va = 0.00154 = 461$~km/s.
    \label{f:test_A_001}
  }
\end{figure}

\begin{figure}[tbp]
  \includegraphics[width=77mm]{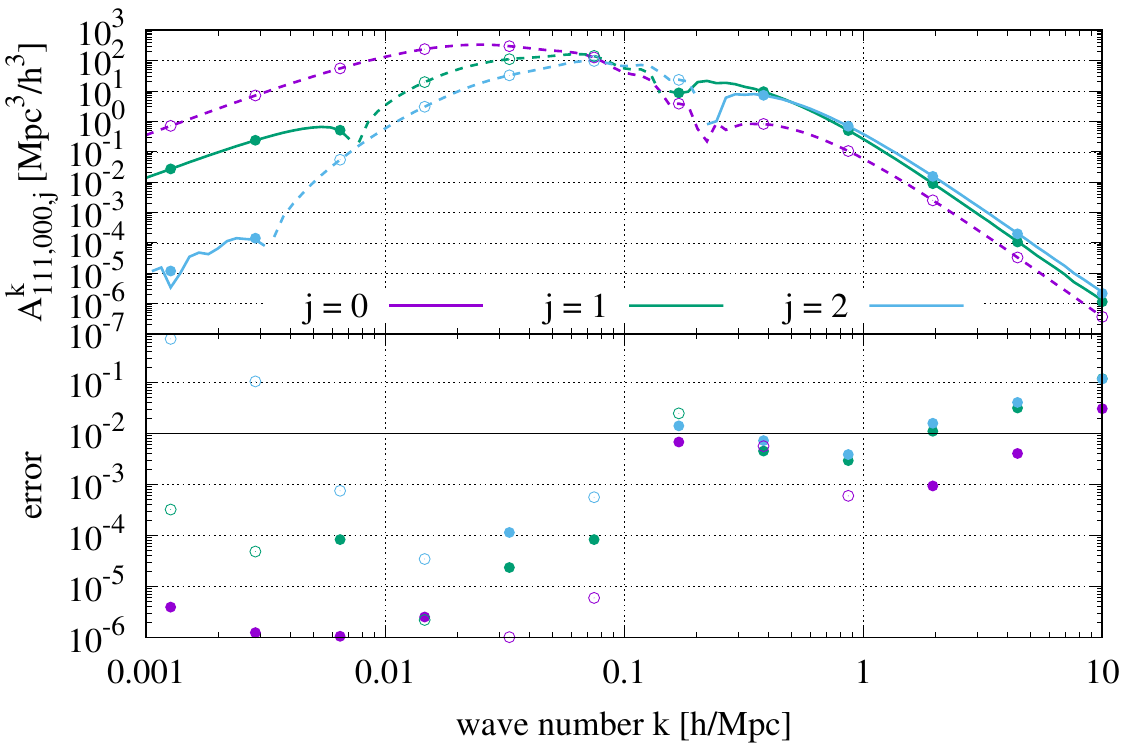}
  \includegraphics[width=77mm]{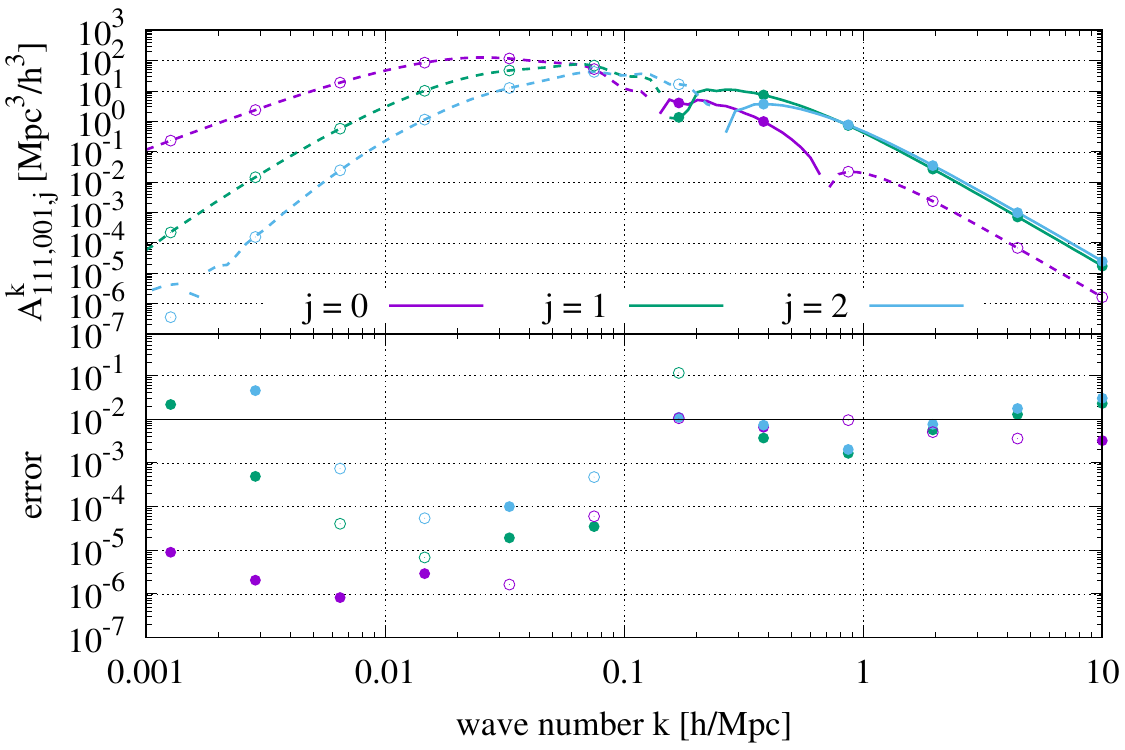}%

  \includegraphics[width=77mm]{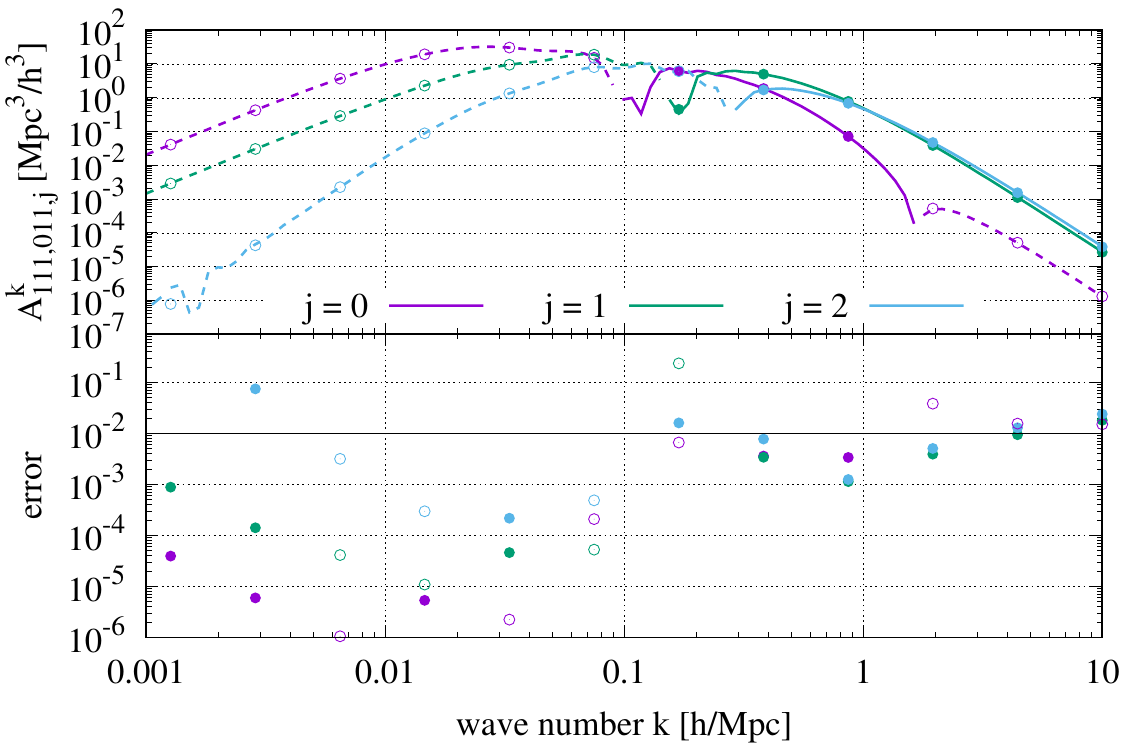}
  \includegraphics[width=77mm]{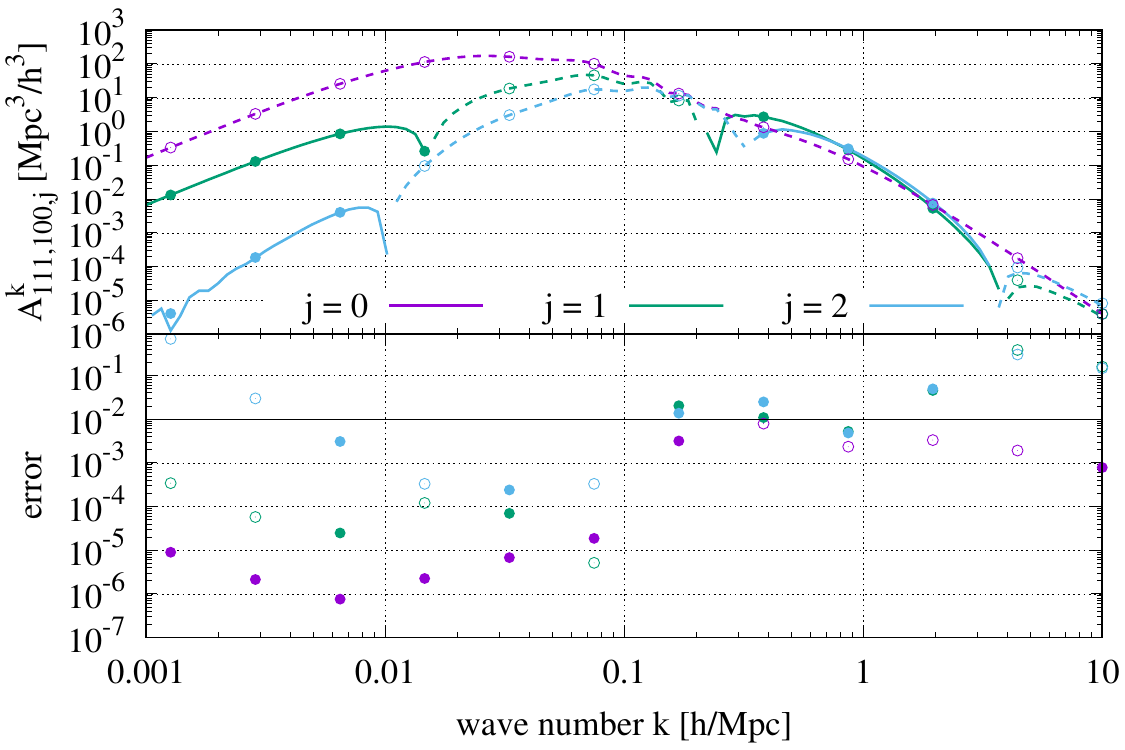}%

  \includegraphics[width=77mm]{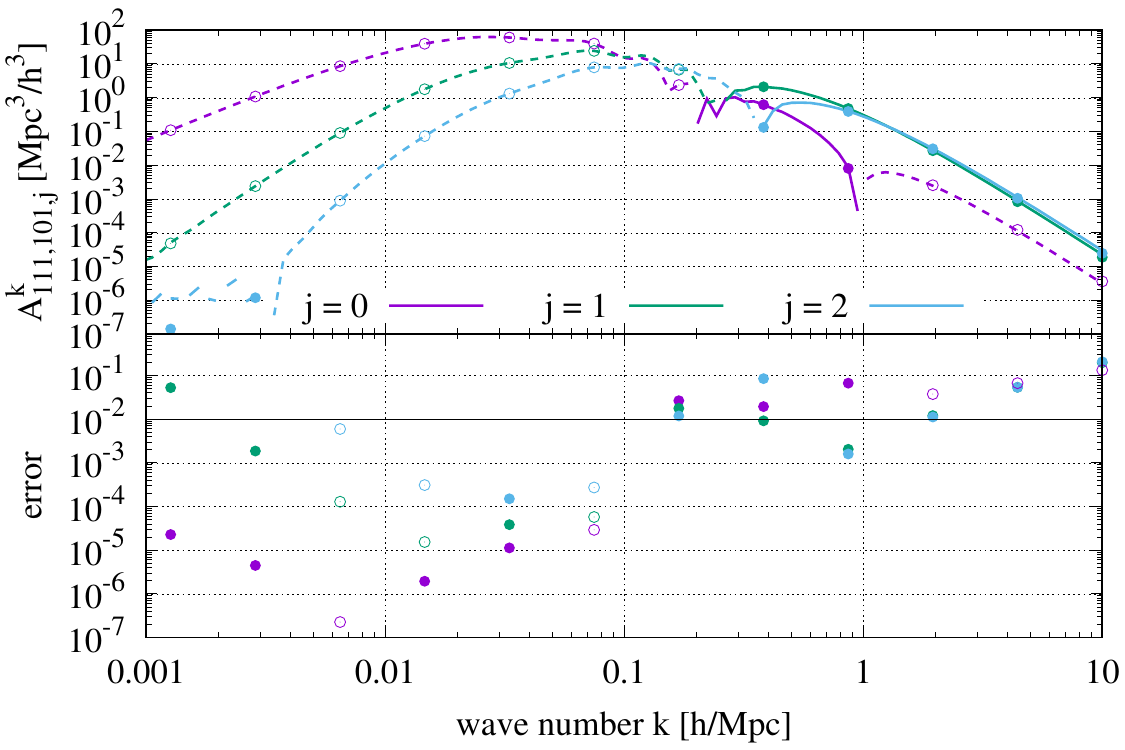}
  \includegraphics[width=77mm]{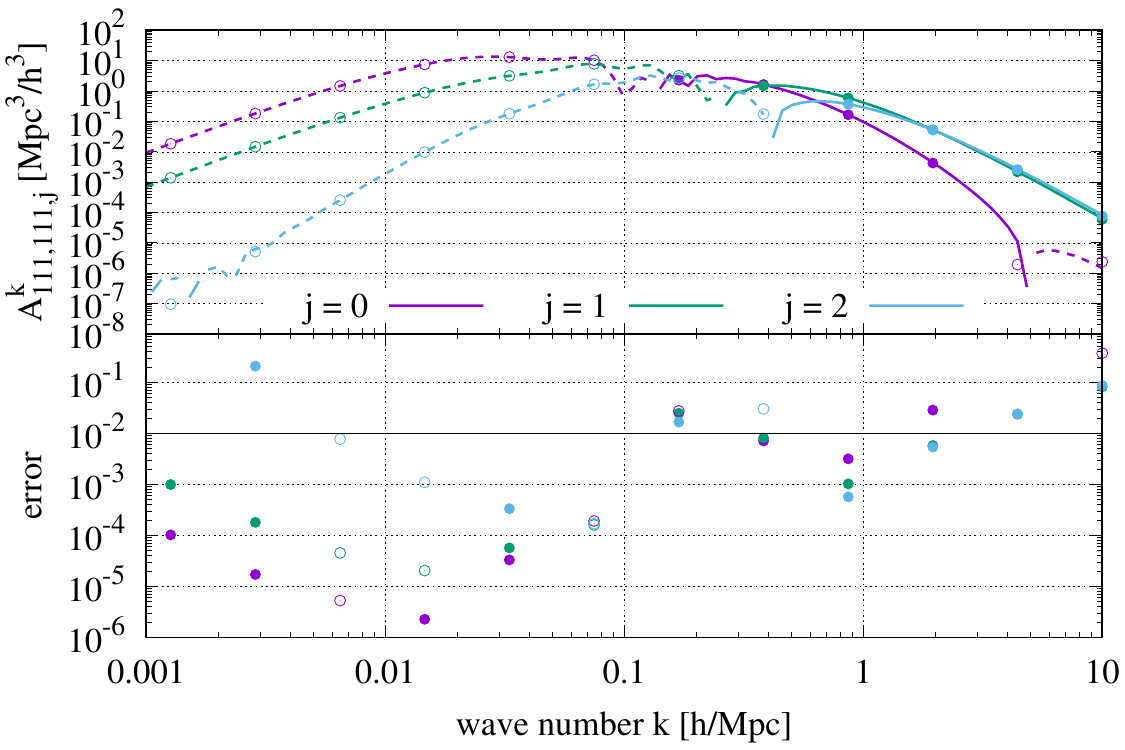}%

  \caption{
    Similar to Fig.~\ref{f:test_A_001} but for $\Aa{111,bef,j}{k}$.
    The symmetry $\Aa{111,bef,j}{k} = \Aa{111,bfe,j}{k}$ means that there
    are only six unique integrals of this type.
    \label{f:test_A_111}
  }
\end{figure}

The culmination of Sections~\ref{sec:mode-coupling_P22}-\ref{sec:mode-coupling_P13} is a complete FFT-accelerated calculation of the mode-coupling  integrals $\Aa{acd,bef,j}{k}$ of \EQS{e:A_22_13}{e:def_A13}, with $\Ptt$-type terms given by \EQS{e:A22_001_0ef}{e:A22_111_1ef} and $\Pot$-type terms by \EQS{e:A13_001}{e:A13_111}.  Before applying these to a Time-RG computation of neutrino clustering, we quantify their accuracy by comparing them to a direct, brute-force numerical integration of \EQ{e:def_A} in Figs.~\ref{f:test_A_001}-\ref{f:test_A_111}.

Mode-coupling integrals were calculated using linear $z=0$ neutrino power spectra for a cosmology with $\Ono h^2 = 0.01$, and for a neutrino flow with $\va = 0.00154$, which were themselves computed using the methods of Ref.~\cite{Chen:2020bdf}. Direct integration used absolute and relative error tolerances of $\epsilon_\mathrm{abs} = 10^{-16}$ and $\epsilon_\mathrm{rel} = 10^{-7}$.  Both integration methods input $\Nmunl = 3$ angular modes into the non-linear mode-coupling integrals.  Figures~\ref{f:test_A_001}-\ref{f:test_A_111} show both the value of each $\Aa{acd,bef,j}{k}$ and the error in the FFT-accelerated computation.

\afterpage{\clearpage}

Errors in the FFT-accelerated mode-coupling integrals are less than $1\%$ over a broad range of wave numbers, with a few important exceptions:
\begin{enumerate}
\item regions near a zero of a given mode-coupling integral;
\item small scales, corresponding to wave numbers $k \gtrsim 2~h/$Mpc;
\item large scales, corresponding to wave numbers $k \lesssim 0.005~h/$Mpc,
  particularly for $j > 1$.
\end{enumerate}
Fourier transforms mix numerical errors across the entire $k$ range.  Each $\Aa{acd,bef,j}{k}$ has a large dynamic range, so FFT acceleration turns small relative errors around the peak of $\Aa{acd,bef,j}{k}$ into large relative errors where $\Aa{acd,bef,j}{k}$ is small.  Since the power spectra for higher Legendre moments are suppressed by more factors of $k\va/\Hc$, which is small at low $k$, this translates into larger low-$k$ errors for larger $j$.

Note, however, that within a large range of wave numbers around the free-streaming scale, $\kfs \approx 0.2~h/$Mpc for this model, the error in a particular FFT-accelerated $\Aa{acd,bef,j}{k}$ exceeds $1\%$ only where that integral is small, and therefore subdominant to other mode-coupling integrals.  Moreover, $j=0$ mode-coupling integrals, which have the most direct impact on the monopole power spectrum, tend to be the most accurate.

Acceleration substantially improves the speed of the mode-coupling integral computation.  Direct numerical integration with tolerances of $\epsilon_\mathrm{abs}=10^{-6}$ and $\epsilon_\mathrm{rel} = 10^{-2}$ required $1800$~sec on a quad-core desktop computer.  FFT acceleration reduced the running time to $4.6$~sec, speeding the calculation by a factor of about $400$.

\begin{figure}[tbp]
  \includegraphics[width=77mm]{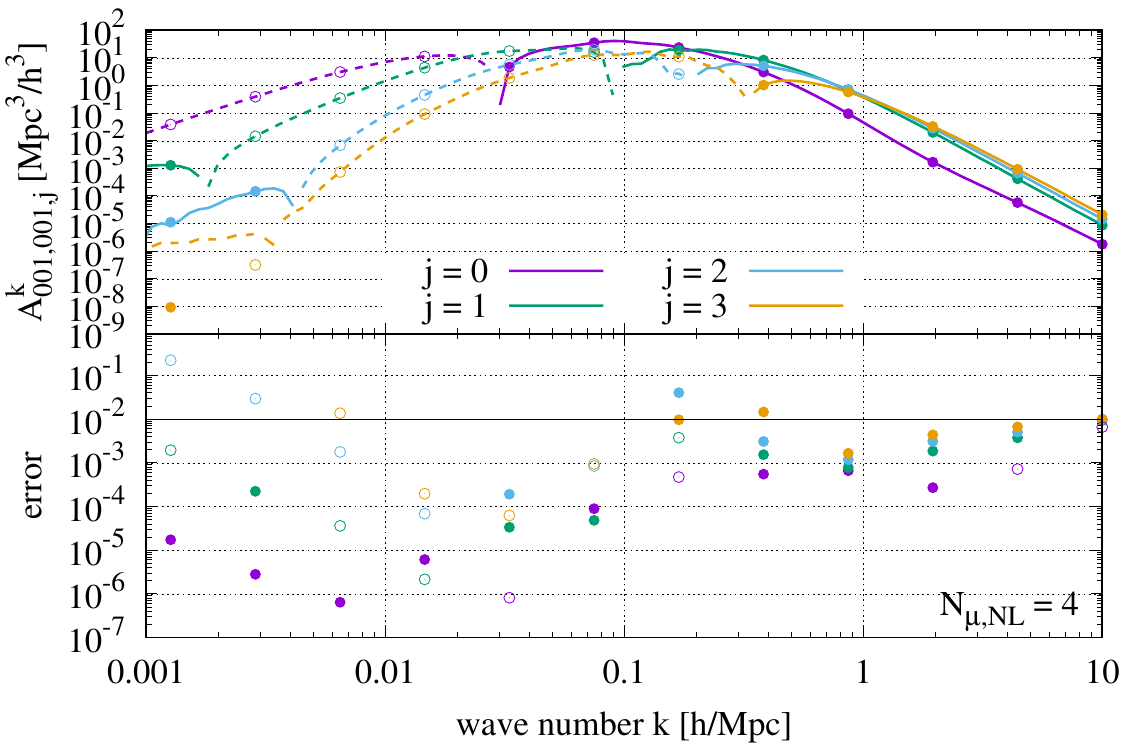}
  \includegraphics[width=77mm]{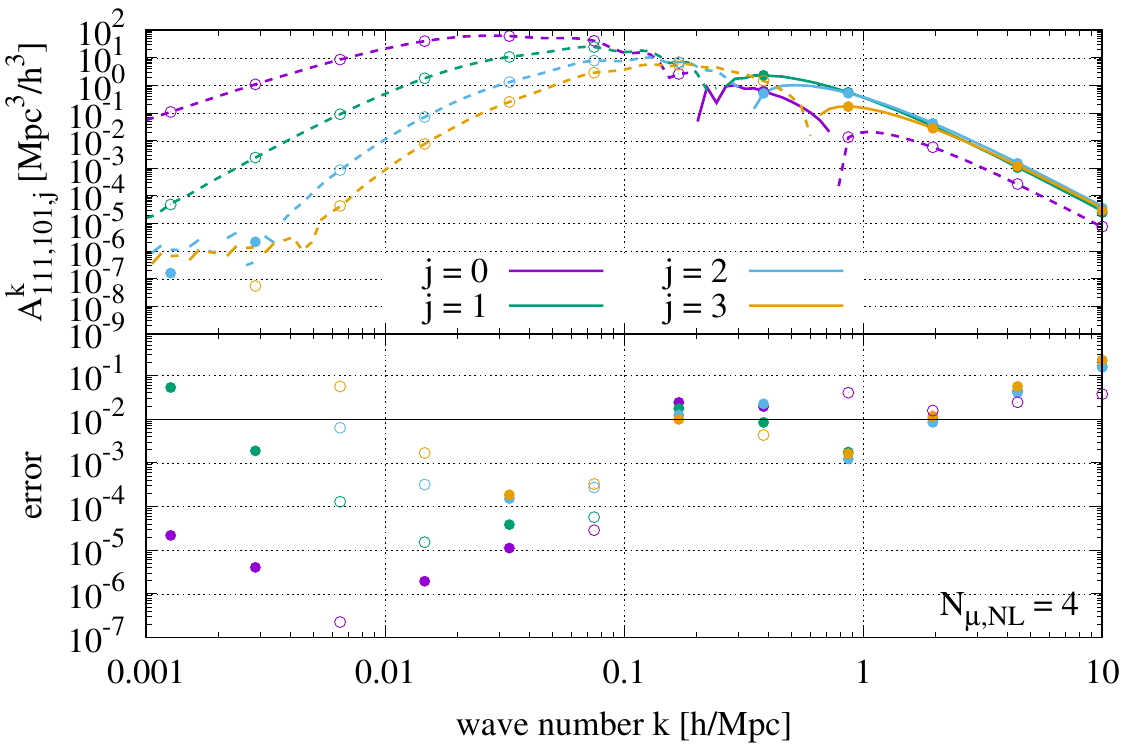}%

  \caption{
    Similar to Fig.~\ref{f:test_A_001} but for $\Nmunl=4$.  Only the
    mode-coupling integrals $\Aa{001,001,j}{k}$ (left)
    and $\Aa{111,101,j}{k}$ (right) are shown.
    \label{f:test_A_Nmu4}
  }
\end{figure}

\begin{figure}[tbp]
  \includegraphics[width=77mm]{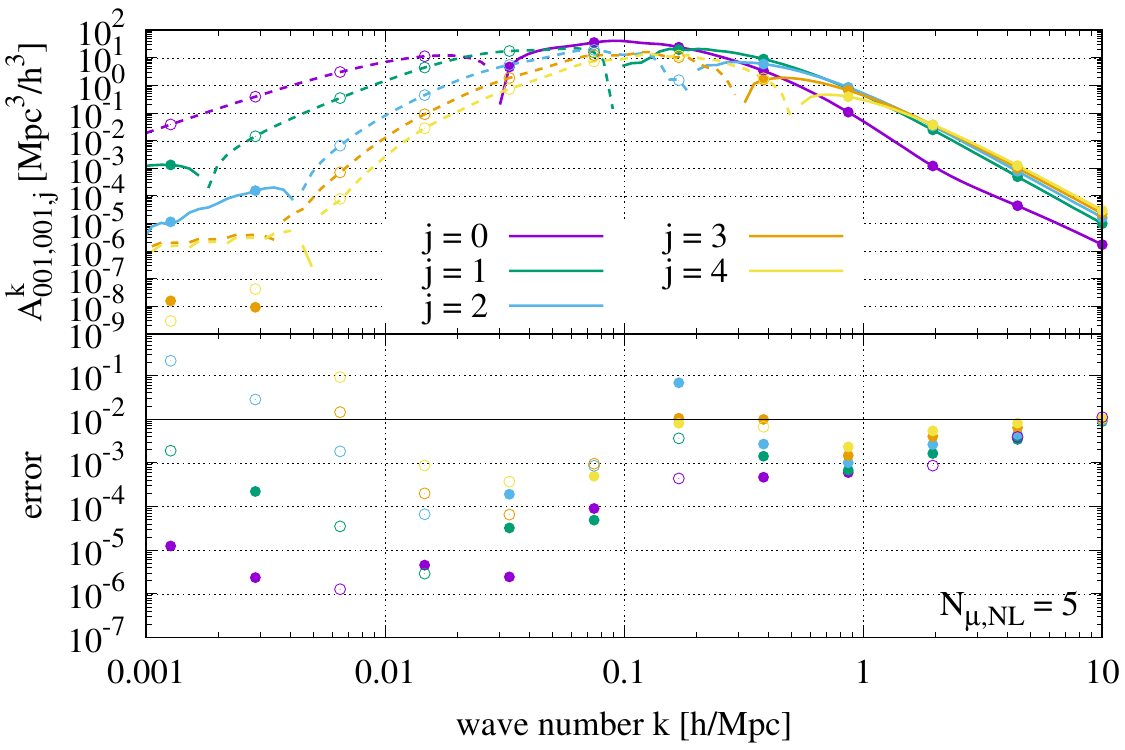}
  \includegraphics[width=77mm]{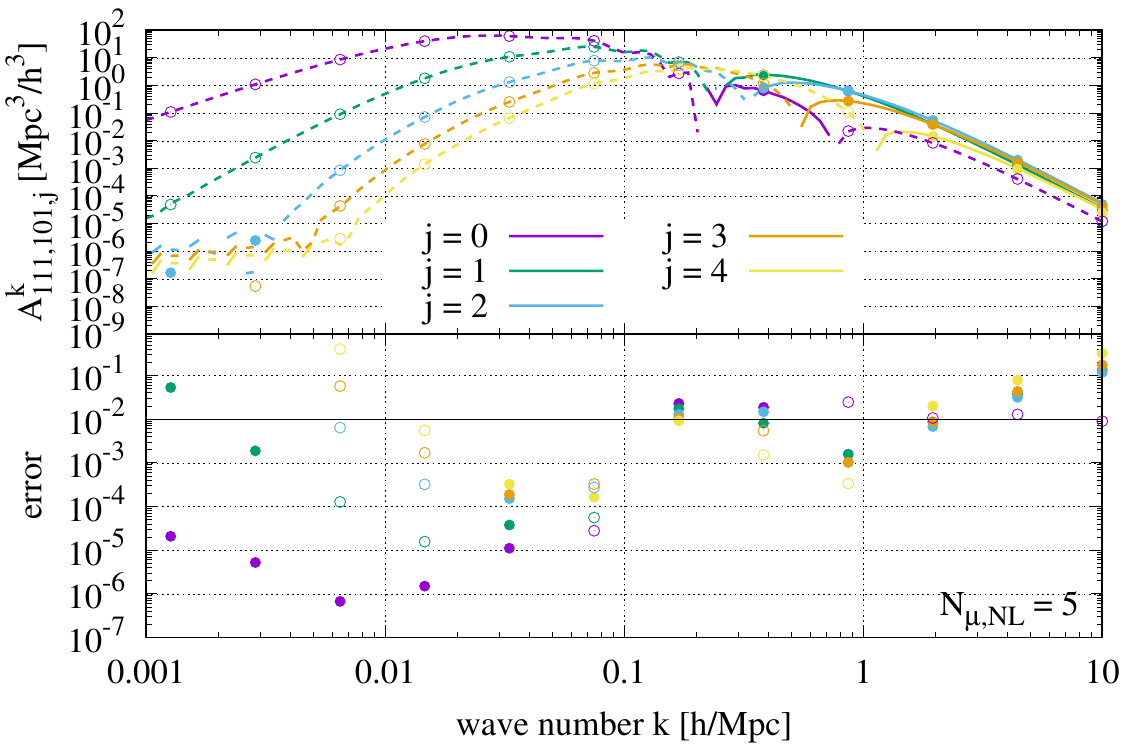}%

  \caption{
    Similar to Fig.~\ref{f:test_A_001} but for $\Nmunl=5$.  Only the
    mode-coupling integrals $\Aa{001,001,j}{k}$ (left)
    and $\Aa{111,101,j}{k}$ (right) are shown.
    \label{f:test_A_Nmu5}
  }
\end{figure}

Next, we consider the accuracy and computational expense of mode-coupling integrals with more angular modes $\Nmunl$.  We focus on two representative mode-coupling integrals: $\Aa{001,001,j}{k}$, the non-linear source for $\Ia{001,001,j}{k}$, which enhances density clustering; and $\Aa{111,101,j}{k}$, because it has the largest errors in the $0.1~h/{\rm Mpc} < k < 1~h/{\rm Mpc}$ range.

Figures~\ref{f:test_A_Nmu4} and \ref{f:test_A_Nmu5} plot these two mode-coupling integrals for $\Nmunl$ of $4$ and $5$, respectively.  Most encouragingly, $\Aa{001,001,j}{k}$ is accurate to $\leq 1\%$ for $k$ within an order of magnitude of the free-streaming scale $k \sim 0.2~h/$Mpc except for  $\Aa{001,001,2}{k}$ and  $\Aa{001,001,3}{k}$ near their zeros.  Furthermore, increasing $\Nmunl$ does not appear to worsen errors in $\Aa{111,101,j}{k}$, which for $\Nmunl=5$ remain under $3\%$ over the same $k$ range.

Examination of the higher-$j$ mode-coupling integrals shows a floor $\sim 10^{-6}~{\rm Mpc}^3/h^3$ below which $\Aa{acd,bef,j}{k}$ becomes dominated by noise.  This floor is about seven or eight orders of magnitude below the peak of each mode-coupling integral.  For all $j$ shown, mode-coupling integrals become noise-dominated well below $k=0.01~h/$Mpc.  Fortunately, neutrino non-linearities are negligible in that regime.  In order to prevent this numerical noise from contaminating our calculation, we neglect non-linear corrections below some $k_{\rm low} \sim 0.01~h/$Mpc, following the suggestion of Ref.~\cite{Vollmer:2014pma} for ordinary Time-RG.

Increasing $\Nmunl$ from $3$ to $4$ and $5$ increases the computation time for the full set of mode-coupling integrals from $4.6$~sec to $23.5$~sec and $87$~sec, respectively on a four-processor desktop computer.  FFT acceleration therefore provides non-linear mode-coupling integrals at a reasonable computational cost, and with $\sim 1\%$-level accuracy over a broad range of scales on either side of the free-streaming scale.  In the next Section we include these mode-coupling integrals in a complete non-linear perturbation theory for the massive neutrinos.

\section{\FFTM{} and non-linear enhancement of neutrino power}
\label{sec:nu_NL_enhancement}

\subsection{Procedure, convergence, and computational cost}
\label{subsec:procedure}

At last we assemble the multi-fluid linear theory of Sec.~\ref{subsec:bkg_mflr} and Refs.~\cite{Dupuy:2013jaa,Dupuy:2014vea,Chen:2020bdf}, the non-linear equations of motion from Sec.~\ref{sec:time-rg_for_massive_nu}, and the non-linear mode-coupling integrals of Secs.~\ref{sec:mode-coupling_P22} and \ref{sec:mode-coupling_P13}, into a complete non-linear perturbative code using multiple flows to represent the massive neutrinos, which we call \FFTM{}.

Throughout this Section, unless otherwise identified, we assume a $\nu\Lambda$CDM cosmology with the following parameters:
\begin{eqnarray}
  h&=&0.6766; \quad
  n_\mathrm{s}=0.9665; \quad
  A_\mathrm{s}=2.135 \times 10^{-9};
  \nonumber\\
  \Omo h^2&=&0.14175; \quad
  \Obo h^2=0.02242; \quad
  \Ono h^2=0.005.
  \label{e:nu05}
\end{eqnarray}
This neutrino density is near the upper limit allowed by recent analyses such as Refs.~\cite{Upadhye:2017hdl,Sgier:2021bzf} provided that the dark energy equation of state and its first derivative are also allowed to vary.  We treat the three neutrino species as degenerate, corresponding to a single neutrino mass $m_\nu = 0.155$~eV and a free-streaming scale $\kfs = 0.13 (1+z)^{-1/2}~h/$Mpc~\cite{Ringwald:2004np,Wong:2008ws}.

Section~\ref{subsec:p13_A_tests} noted a numerical noise arising at $k \lesssim 0.01~h/$Mpc for mode-coupling integrals $\Aa{acd,bef,j}{k}$ with $j \gtrsim 3$.  Since non-linear corrections to neutrino clustering are negligible in that region, \FFTM{} smoothly turns off low-$k$ non-linear corrections by multiplying each mode-coupling integral in \EQ{e:eom_I} by a low-$k$ suppression factor:
\begin{equation}
  \Aa{acd,bef,j}{k} \longrightarrow
  \Aa{acd,bef,j}{k} \times \frac{2 }{ [1 + e^{(k_{\rm low}/k)^4}]}
  \quad\mathrm{with }~k_{\rm low} = 0.01~h/\mathrm{Mpc.}
  \label{e:f_A_low-k} 
\end{equation}

\begin{table}[tbp]
  \begin{center}
    \begin{footnotesize}
      \begin{tabular}{r|ccccccccc}
        $\Nmunl$ & $1$ & $2$ & $3$ & $4$ & $5$ & $6$ & $7$ & $8$ & $9$ \\
        \hline
        Cost [sec]
        & $0.079$ & $0.58$ & $4.6$ & $23.5$ & $87$
        & $250$ & $630$ & $1380$ & $2780$
        \\
        $k_\mathrm{stab}$ [$h/$Mpc]
        & $\geq 10$ & $\geq 10$ & $\geq 10$ & $4.8$ & $4.4$
        & $4.4$ & $3.7$ & $3.7$ & $0$
      \end{tabular}
    \end{footnotesize}
  \end{center}
  \caption{
    Computational cost and stability region of \FFTM{} integration vs.~number
    of angular modes $\Nmunl$ used in non-linear mode-coupling integrals.
    The cost is the time required to run a full set of mode-coupling integrals
    $\Aa{acd,bef,j}{k}$ for a single fluid at a single time step over $N_k=128$
    logarithmically-spaced wave numbers $k$ on a standard four-processor
    desktop machine.
    The numerical stability threshold $k_\mathrm{stab}$ is the maximum wave
    number to which the equations of motion, \EQS{e:eom_P}{e:eom_chi}, can be
    integrated for the model of \EQ{e:nu05}, as discussed in
    Sec.~\ref{subsec:enu_full_trg}.  For $\Nmunl=9$, instabilities prevented
    integration of the equations of motion below $z=0.7$.
    \label{t:A_cost_kNI}
  }
\end{table}

Table~\ref{t:A_cost_kNI} lists the computational cost of a mode-coupling integral computation for a range of $\Nmunl$.  Even including the $400$-fold speed improvement through FFT acceleration techniques in \FFTM{}, neutrino mode-coupling integrals for large $\Nmunl$ remain computationally expensive, due to the ${\mathcal O}(\Nmunl^6)$ scaling of the $\Ptt$-type integrals identified in Sec.~\ref{subsec:p22_fast-pt}.  The running times in the table, for a single fluid $\alpha$ at a single time step, assuming $N_k=128$ wave numbers, is approximated by ($0.005 \Nmunl^6 + 0.07 \Nmunl^2$) seconds.

We mitigate this computational cost in two ways.  Firstly, we do not include non-linear corrections for a given flow $\alpha$ until its maximum dimensionless density monopole power $\Delta_{\alpha 000}^2(k) = k^3 \Pa{000}{k} / (2\pi^2)$ exceeds a predetermined threshold of $\Delta^2_\mathrm{thresh} = 10^{-4}$.  Following Ref.~\cite{Audren:2011ne}, we initialize $\Ia{acd,bef,j}{k}$ to $2\Aa{acd,bef,j}{k}$ once this threshold is reached.

\begin{figure}[tbp]
  \includegraphics[width=155mm]{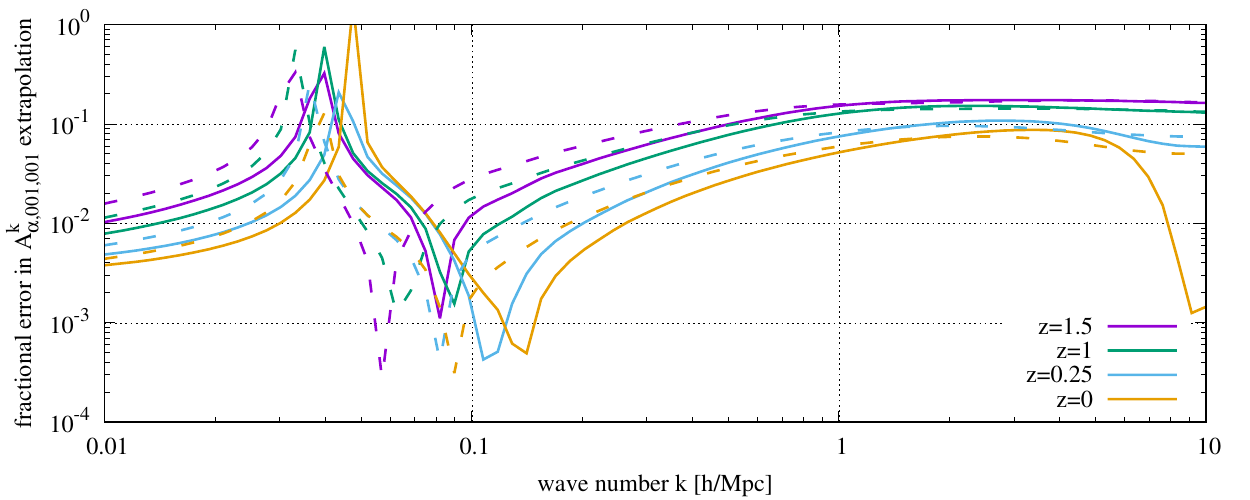}%
  \caption{
    Maximum extrapolation error in mode-coupling integrals $\Aa{001,001,0}{k}$
    for $\alpha=0$ (solid) and $\alpha=1$ (dashed). We fix $\Nmunl=3$.
    \label{f:A_extrap}
  }
\end{figure}

Secondly, even when this threshold is crossed, we restrict computation of neutrino mode-coupling integrals to the $N_a=50$ scale factors $a_n = 0.02n$ for $1 \leq n \leq N_a$.  Between these computation times, we approximate each $\Aa{acd,bef,j}{k}$ as scaling like the clustering-limit monopole power spectra in its integrand, that is, as $(\delta_{\alpha,0}^0)^{3-b-e-f} (\theta_{\alpha,0}^0)^{1+b+e+f}$.  The error associated with this approximation is greatest just before one of the $a_n$, when mode-coupling integrals from step $a_{n-1}$ are extrapolated all the way to $a_n$.

Figure~\ref{f:A_extrap} quantifies the maximum error in $\Aa{001,001,0}{k}$ resulting from this extrapolation for the case of $N_\tau=10$ flows.  Around $\kfs$ the error is no more than $4\%$ for $z \leq 1.5$.  At $z=0$, where non-linear corrections are most significant, the error is $< 1\%$ around $k=\kfs$ and no more than $6\%$ at any $\kfs \leq k \leq 1~h/$Mpc.  Aside from $k \approx 0.04~h/$Mpc, where the mode-coupling integrals pass through zero, the maximum error resulting from this extrapolation for  $z \leq 0.25$ is $\lesssim 10\%$.  This decrease in error with rising scale factor is due to our choice of $a_n$, since the fractional increase $(a_{n+1}-a_n)/a_n$ is lower at larger scale factors.

\begin{figure}[tbp]
  \includegraphics[width=155mm]{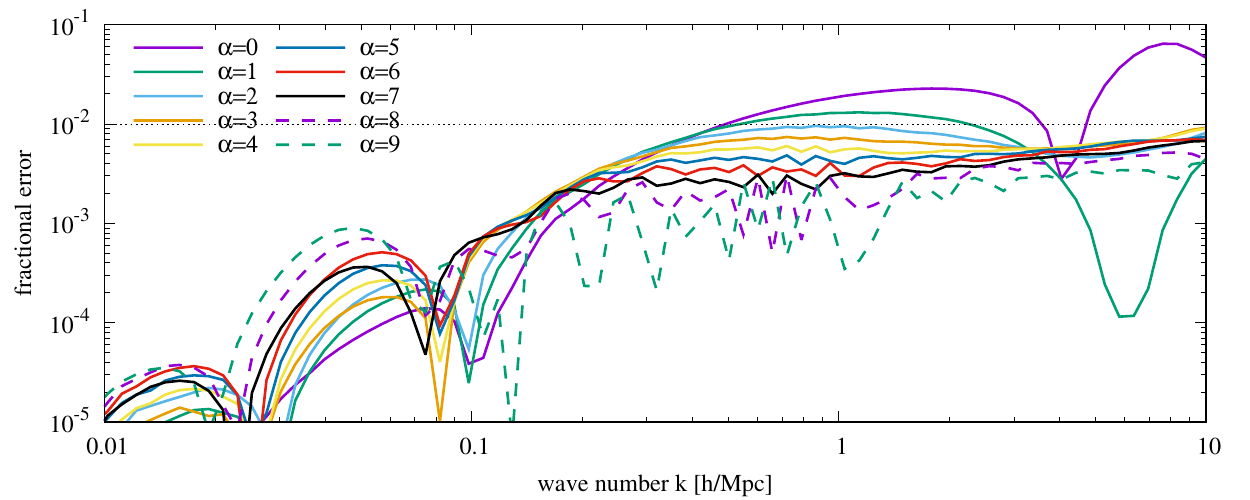}%
  \caption{
    Density monopole errors at $z=0$ associated with neglecting neutrino
    non-linearity below the threshold $\Delta^2_\mathrm{thresh} = 10^{-4}$ and
    computing mode-coupling integrals only at $a_n = 0.02n$. 
    \label{f:tol_errors}
  }
\end{figure}

Combined errors due to both of these approximations are quantified by reducing both the $\Delta^2$ threshold and the spacing between mode-coupling recomputations.  Figure~\ref{f:tol_errors} compares our default calculation above to one whose threshold is ten times smaller, $\Delta^2_\mathrm{thresh} = 10^{-5}$, and whose mode-coupling integrals are computed four times more frequently, $a_n = 0.005n$.  In the $k\leq 1~h/$Mpc range, the resulting error is about $2\%$ for $\alpha=0$.  Across the entire $k$ range, errors are under $1.5\%$ for $\alpha=1$, and under a percent for all higher-$\alpha$ flows.  Thus we regard these two approximations as well-founded.  Henceforth we use $\Delta^2_\mathrm{thresh} = 10^{-4}$ and $a_n = 0.02n$ in all of our \FFTM{} computations.

Also listed in Table~\ref{t:A_cost_kNI} is the maximum wave number $k_\mathrm{kstab}$ to which our integration of the equations of motion is stable.  In practice we integrate \EQS{e:eom_P}{e:eom_chi} using a $5$th-order Runge-Kutta method with adaptive step size control, as implemented in the GNU Scientific Library (GSL) of Ref.~\cite{Galassi_2009}.  We begin with $k_\mathrm{stab}$ set to the maximum wave number considered, $10~h/$Mpc.  Each time that the GSL integrator is unable to progress, we reduce $k_\mathrm{stab}$ by one step, or a factor of $\exp(\Dk)$, and resume integration.

Evidently, \FFTM{} is numerically stable for a broad range of wave numbers reaching $\approx 30$ times the free-streaming scale. For $\Nmunl \geq 6$, we have also raised $k_\mathrm{low}$ in \EQ{e:f_A_low-k} to $0.05~h/$Mpc to mitigate low-$k$ noise.   For $\Nmunl=9$, the highest value considered here, we were unable to integrate the equations of motion all the way to $z=0$.  Integration reached $z=0.7$ with a stability threshold of $0.5~h/$Mpc before failing.  Henceforth we only consider $\Nmunl \leq 8$.  We also caution that, for a large number $N_\tau$ of flows, the slowest will be much more non-linear than those considered here, hence will reach our non-linear threshold $\Delta^2_\mathrm{thresh} = 10^{-4}$ sooner, leaving more time for numerical instabilities to develop.  

\begin{figure}[tbp]
  \includegraphics[width=155mm]{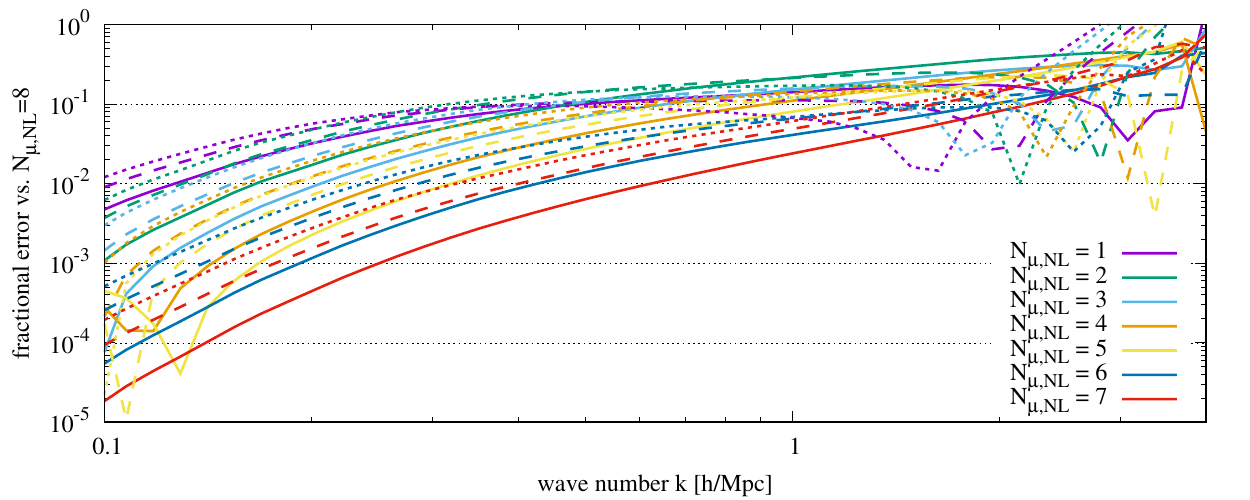}%
  \caption{
    Convergence of the $z=0$ density monopole perturbation $\delta_{\alpha 0}^k$
    for $\alpha=0$ (solid lines), $\alpha=1$ (dashed), and $\alpha=2$ (dotted),
    assuming $N_\tau=10$ flows in total.  The error in each perturbation is
    quantified by comparison to the $\Nmunl=8$ case.
    \label{f:convergence}
  }
\end{figure}

Finally, we discuss convergence of the code with increasing $\Nmunl$.  Again considering $N_\tau=10$ flows, Fig.~\ref{f:convergence} quantifies the error in each $\delta_{\alpha 0}^k$ for a given $\Nmunl$ by comparison to $\Nmunl=8$.  At the free-streaming scale, perturbation theory is accurate at the percent level for $\Nmunl \geq 3$.  In the range $k \lesssim 2~h/$Mpc, perturbation theory is converging, in the sense that its errors continue declining as $\Nmunl$ is increased.  For  $k \leq 1~h/$Mpc, the $\alpha=0$ density monopoles are $8\%$ accurate for $\Nmunl=5$, $4\%$ for $\Nmunl=6$, and $2.5\%$ for $\Nmunl=7$.  Thus \FFTM{} is  convergent and accurate for computing the neutrino power spectrum all the way up to $k=1~h/$Mpc.  In the remainder of this Section, we will: (i) couple neutrinos to a Time-RG CDM+baryon fluid, illustrating the high-$z$ and qualitative low-$z$ behavior of their non-linear corrections; (ii) couple neutrinos to a more accurate calculation of CDM+baryon clustering emulated from N-body simulations, providing neutrino power computations that are reliable even at low redshifts; (iii) describe a set of simulations carried out by a companion article, Ref.~\cite{Chen:2022dsv}; (iv) demonstrate the accuracy of \FFTM{} using these simulations; and (v) explore the dependency of non-linearity upon neutrino mass.

\subsection{\FFTM{} with Time-RG perturbations for CDM, baryons}
\label{subsec:enu_full_trg}

\begin{figure}[tbp]
  \includegraphics[width=155mm]{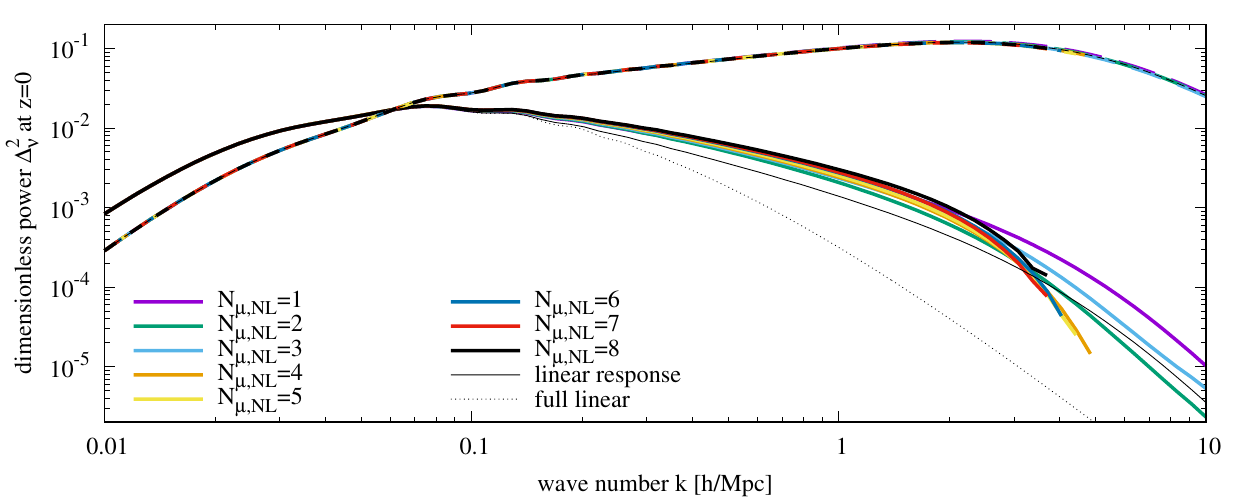}%

  \caption{
    Dimensionless $z=0$ neutrino monopole power spectra
    $\Delta^2_{aa0}(k) = \frac{k^3}{2\pi^2} P_{aa0}(k)$ for a range of $\Nmunl$,
    for the model of \EQ{e:nu05}.  Solid lines show the $\delta$ power spectrum
    ($a=0$) and dashed lines the $\theta$ power spectrum ($a=1$).  The
    multi-fluid linear response of Ref.~\cite{Chen:2020bdf} is shown for
    comparison.
    \label{f:D2nu_all}
  }
\end{figure}

As a first application of \FFTM{}, we pair it with a Time-RG perturbative evolution of the CDM+baryon fluid.  We begin by considering the neutrino power spectra themselves.  Working in the cosmological model of \EQ{e:nu05}, we discretize the massive neutrino population into $N_\tau = 10$ velocity bins, with $0 \leq \alpha \leq 9$, and $N_\mu = 10$ angular modes, with $0 \leq \ell \leq 9$.  Velocities corresponding to these bins are listed in Table~\ref{t:nu05_Ntau10_velocities}.  Figure~\ref{f:D2nu_all} shows our monopole ($\ell=0$) power spectra for several values of $\Nmunl$, the number of angular modes used to compute the mode-coupling integrals.  Note that $\Nmunl$ of one is equivalent to the ordinary Time-RG of Ref.~\cite{Pietroni:2008jx} with a modified linear evolution matrix $\Xia{ab0}{k}$, while increasing $\Nmunl$ includes more of the corrections developed in this work.

\begin{table}[tbp]
  \begin{center}
    \begin{footnotesize}
      \begin{tabular}{r|cccccccccc}
        $\alpha$ & $0$ & $1$ & $2$ & $3$ & $4$ & $5$ & $6$ & $7$ & $8$ & $9$ \\
        \hline
        $\va/c$ & $0.00101$ & $0.0016$ & $0.00204$ & $0.00245$ & $0.00286$
        & $0.0033$ & $0.0038$ & $0.00442$ & $0.00528$ & $0.00697$
      \end{tabular}
    \end{footnotesize}
  \end{center}
  \caption{
    Velocities $\va$ at $z=0$ for $N_\tau=10$ flows in the cosmological model
    of \EQ{e:nu05}.
    \label{t:nu05_Ntau10_velocities}
  }
\end{table}

Velocity power spectra are large at small scales due to free streaming.  Non-linear corrections to them are negligibly small, a few percent at most, for all $\Nmunl$.  Evidently, linear response accurately captures the behavior of the neutrino velocity dispersion power spectrum down to very small scales, $k = 10~h/$Mpc.  Thus the remainder of this Section will focus on the power spectra of density perturbations.

Regarding the density power spectra in Fig.~\ref{f:D2nu_all}, all of the non-linear power spectra $\Nmunl \geq 1$ in the region $k \sim 1~h/$Mpc are $\sim 2$ times larger than the linear response power spectrum.  At somewhat smaller scales, $k \gtrsim 2~h/{\rm Mpc}$, $\Nmunl$ of $1$ overpredicts clustering relative to larger values, while $\Nmunl$ of $2$ underpredicts it.  Evidently from \EQ{e:eom_delta}, free-streaming suppression of the monopole $\delta_{\alpha 0}^k$ is proportional to the dipole $\delta_{\alpha 1}^k$.  By including a non-linear enhancement to this dipole, $\Nmunl=2$ indirectly suppresses the monopole.  In turn, $\Nmunl=3$ includes an enhancement of the quadrupole $\delta_{\alpha 2}^k$, which indirectly enhances the monopole by suppressing the dipole.  Thus we recommend at least $\Nmunl=3$.

At still smaller scales, $2~h/{\rm Mpc} < k \lesssim 4~h/{\rm Mpc}$, power spectra using the higher $\Nmunl$ drop sharply, falling below the linear response power.  For $\Nmunl \geq 4$, we encounter the numerical instabilities listed in Table~\ref{t:A_cost_kNI}.  Integration of the equations of motion fails to converge for $k \gtrsim 4~h/$Mpc, suggesting that this drop below $4~h/$Mpc is the result of the same numerical instability.  References~\cite{Audren:2011ne,Vollmer:2014pma,Chen:2020bdf} have noted and  mitigated small-scale numerical instabilities in ordinary Time-RG for a cold fluid.  These are due in part to the scalarized equations of motion for $\delta$ and $\theta$ neglecting small-scale velocity vorticity.  Our perturbation theory appears to exacerbate these small-scale instabilities at high $\Nmunl$.

Neutrino clustering is dominated at very large scales, $k \ll \kfs$, by fully linear growth; at intermediate scales, $k \sim \kfs$, by linear response to non-linear CDM+baryon growth; and at small scales, $k \gg \kfs$, by non-linear growth.  We define the non-linear enhancement
\begin{equation}
  \Ea{a\ell}{k}
  :=
  \Pa{aa\ell}{k} / \Pa{aa\ell}{\mathrm{[MFLR]}k} - 1
  \label{e:def_Enu}
\end{equation}
where $\Pa{aa\ell}{\mathrm{[MFLR]}k}$ is the power spectrum computed using the Multi-Fluid Linear Response (MFLR) method of Ref.~\cite{Chen:2020bdf}.  We will devote the remainder of this Section to computing $\Ea{a\ell}{k}$ and comparing it against the enhancement seen in N-body simulations including massive neutrino particles.  We begin with the calculation described above, with \FFTM{} coupled to CDM and baryons evolved using Time-RG perturbation theory.

\begin{figure}[tbp]
  \includegraphics[width=155mm]{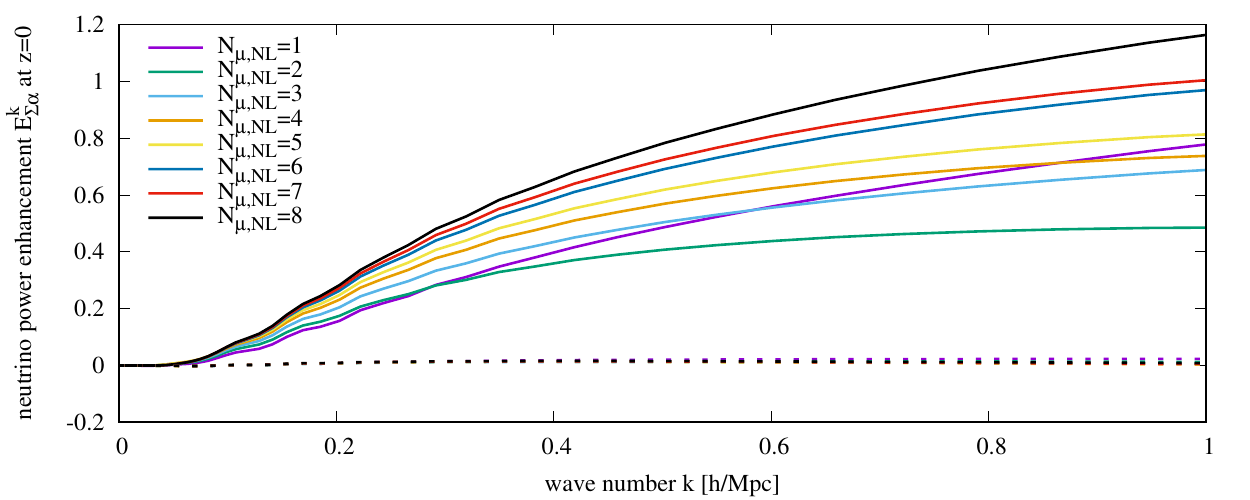}%
  \caption{
    Monopole non-linear enhancement of \EQ{e:def_Enu_total_mono}
    at $z=0$ for the model of \EQ{e:nu05}.  Solid lines show the density
    ($a=0$) enhancement and dashed lines the velocity ($a=1$) enhancement.
    \label{f:Enu_all}
  }
\end{figure}

Here we study the enhancement  in the region $k \leq 1~h/$Mpc, well below the unphysical power suppression at $k \sim 4~h/$Mpc.  We begin with the total monopole enhancement
\begin{equation}
  E_{\Sigma\alpha, a, 0}^k
  :=
  \left(\Sigma_\alpha \psia{a0}{k}/N_\tau\right)^2
  / \left(\Sigma_\alpha \psia{a0}{[\mathrm{MFLR}]k}/N_\tau\right)^2 - 1.
  \label{e:def_Enu_total_mono}
\end{equation}
Figure~\ref{f:Enu_all} shows the total enhancements of the density and velocity power spectra.  Density enhancement rises from a few percent at the free-streaming scale to $\approx 100\%$ at $k=1~h/$Mpc.  The enhancement for $\Nmunl=7$ is within $5\%$ of that for $\Nmunl=8$ up to $k=0.4~h/$Mpc; within $10\%$ up to $k=0.7~h/$Mpc, or about five times $\kfs$; and within $14\%$ up to $k=1~h/$Mpc.  Velocity enhancement is no more than a few percent for any $\Nmunl$.  

Considering the neutrino power, rather than the enhancement,  $\Nmunl=3$ agrees with  $\Nmunl=8$ to $10\%$ up to $k=0.3~h/$Mpc and to $20\%$ up to $k=0.8~h/$Mpc.  Meanwhile, $\Nmunl$ of $4$ and $5$ are accurate to $10\%$ up to $k$ of $0.4~h/$Mpc and $0.5~h/$Mpc, respectively, and to $<20\%$ across the entire range $k \leq 1~h/$Mpc, while $\Nmunl$ of $6$ and $7$ are accurate to $5\%$ up to $k$ of $0.4~h/$Mpc and $0.7~h/$Mpc, respectively, and to $<10\%$ for $k \leq 1~h/$Mpc.

\begin{figure}[tbp]
  \includegraphics[width=77mm]{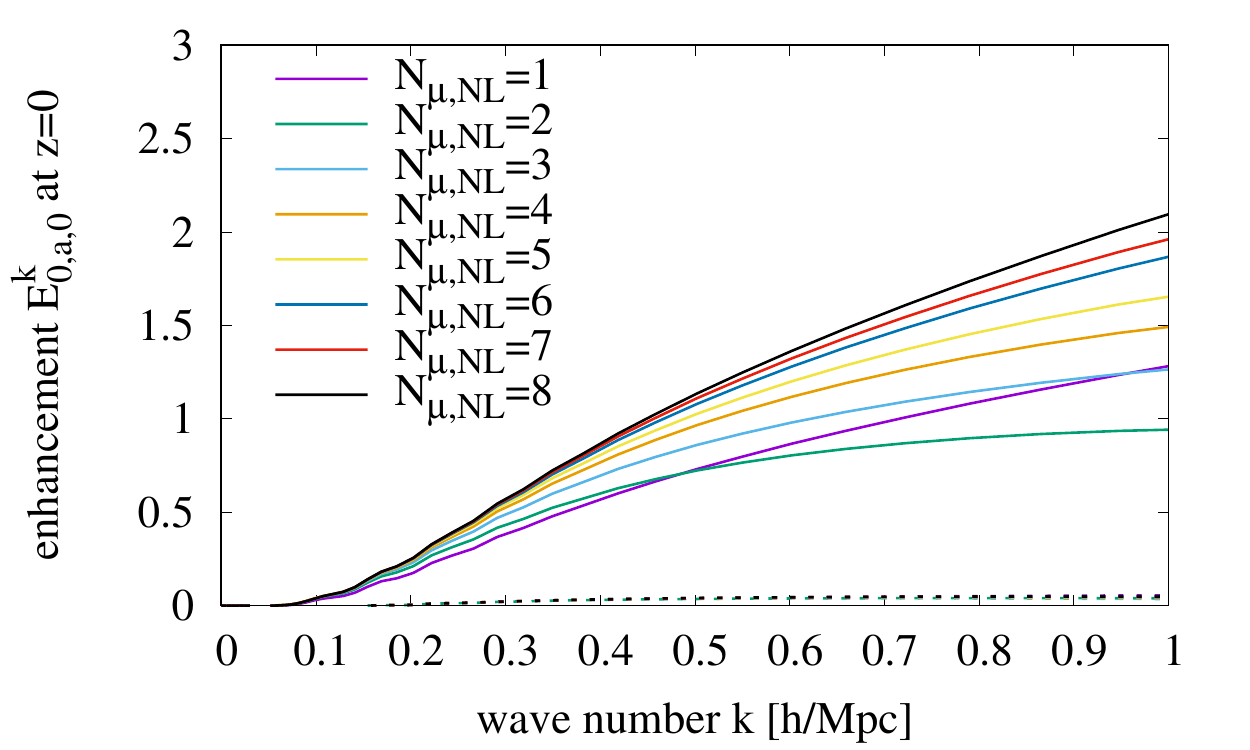}
  \includegraphics[width=77mm]{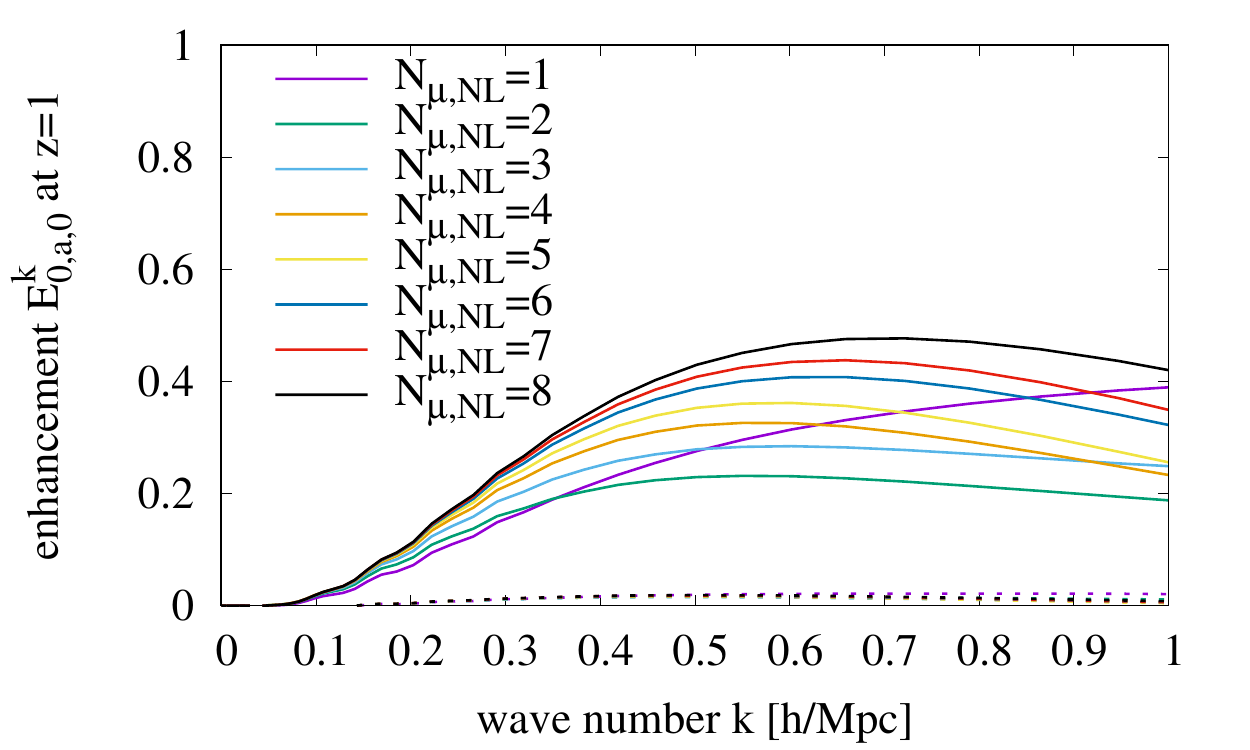}%

  \includegraphics[width=77mm]{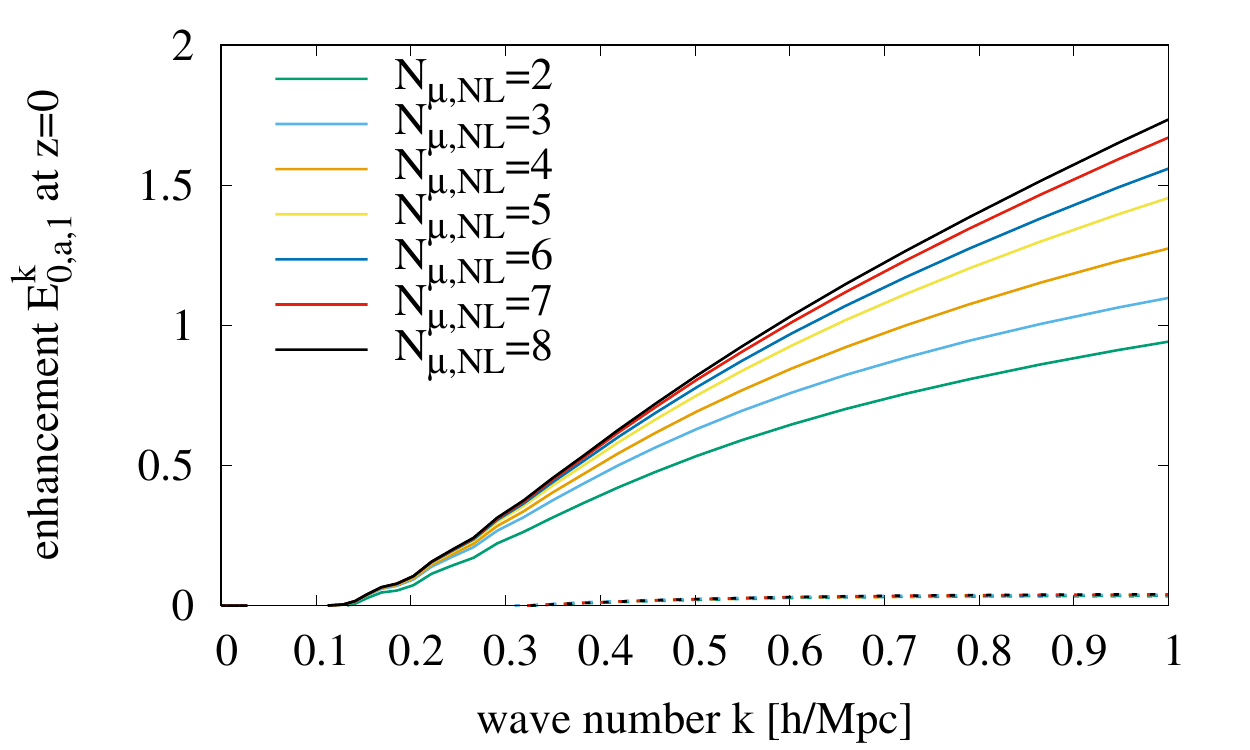}
  \includegraphics[width=77mm]{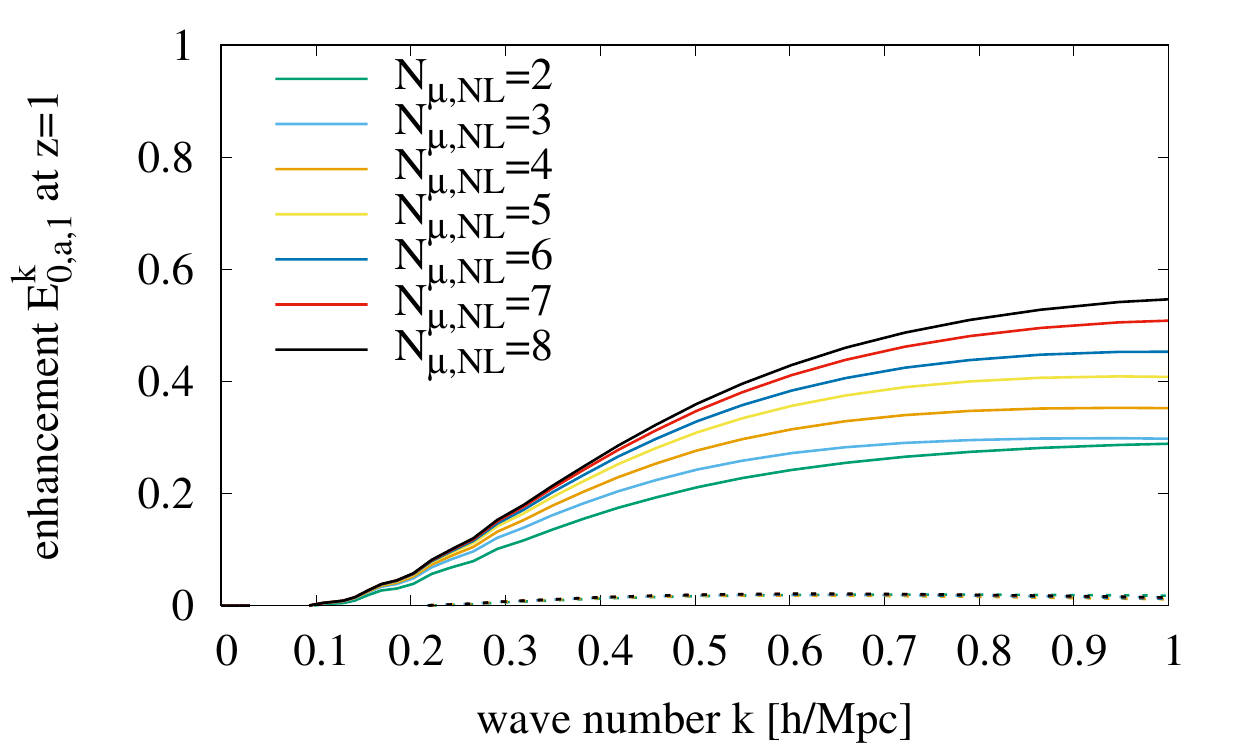}%

  \caption{
    Neutrino enhancement $\Ea{a\ell}{k}$ for $\alpha=0$;  $\ell$ of $0$
    (top row) and $1$ (bottom row); at $z=0$ (left column) and $z=1$
    (right column) for a range of $\Nmunl$. Enhancements to $\Pa{00\ell}{k}$
    are shown as solid lines and those to $\Pa{11\ell}{k}$ as dotted lines.
    \label{f:Enu_alpha01_ell01_Time-RG}
  }
\end{figure}

Next, we consider the power enhancement for the individual $\alpha=0$ flow using \EQ{e:def_Enu}.  Figure~\ref{f:Enu_alpha01_ell01_Time-RG} shows the density and velocity enhancements for $\ell=0$ and $1$, each at redshifts $0$ and $1$.  Dipole enhancements are not shown for $\Nmunl=1$, which only applies non-linear corrections to the monopole.  Since $\alpha=0$ corresponds to the slowest-moving, hence the most strongly-clustering, of the neutrino flows, its $z=0$ density enhancement is approximately twice the average value of Fig.~\ref{f:Enu_all}.  For $\Nmunl=8$ and $k=1~h/$Mpc, the enhancement is greater than two, implying that non-linear corrections more than triple the $\alpha=0$ linear response power spectrum at this scale.  Even at $z=1$, non-linear corrections represent an $\approx 50\%$ enhancement to the linear response power.  The dipole density power enhancements at both redshifts are comparable in magnitude to the monopole enhancements.  

\begin{figure}[tbp]
  \includegraphics[width=155mm]{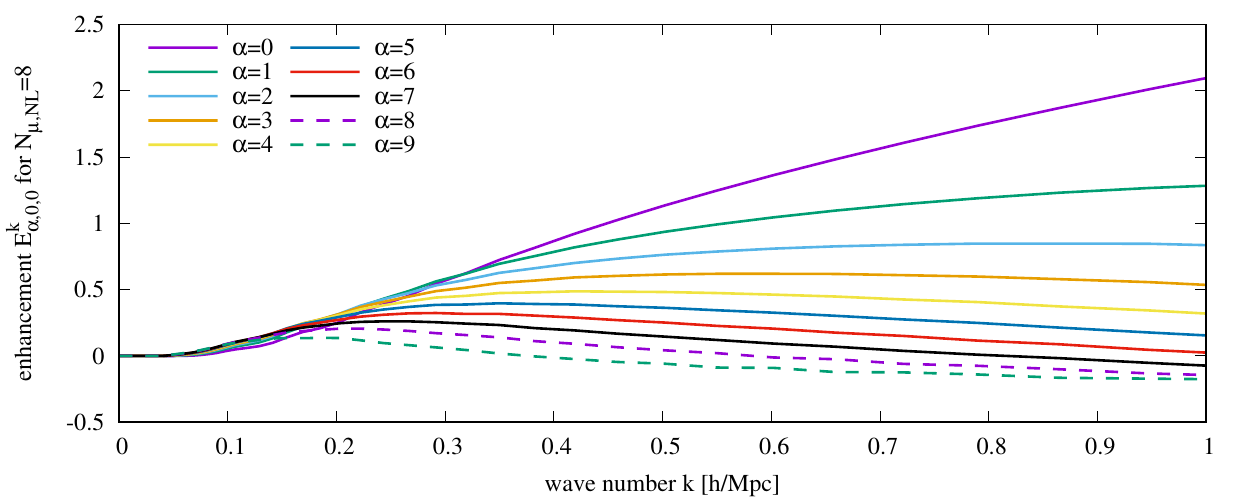}%
  \caption{
    Neutrino density monopole enhancement for all $\alpha$, at $z=0$,
    with $\Nmunl=8$ fixed.
    \label{f:Enu_alpha0-9_z0}
  }
\end{figure}

Figure~\ref{f:Enu_alpha0-9_z0} fixes $\Nmunl=8$ and displays the density monopole enhancements for all $0 \leq \alpha < N_\tau=10$.  The enhancement for $\alpha=1$ plateaus at about two-thirds the value of $E_{000}^{1~h/{\rm Mpc}}$, while $E_{200}^k$ peaks at less than half of this value.  The total density monopole enhancement for $5 \leq \alpha \leq 9$, or \EQ{e:def_Enu_total_mono} but with both summations limited to $5 \leq \alpha \leq 9$,  peaks at $28\%$.  However, these flows cluster so weakly that this enhancement always represents less than $1\%$ of the total power.  Thus a high-accuracy calculation of the neutrino power spectrum could neglect non-linearity in these flows altogether.  Similarly, neglecting the flows $3 \leq \alpha \leq 9$ would only introduce a $3.5\%$ error into a power spectrum calculation.

In summary, we have shown that our self-contained non-linear perturbative calculation for massive neutrinos and the CDM+baryon fluid has converged at the ${\mathcal O}(10\%)$ level across the range $k \lesssim 1~h/$Mpc.  Non-linear growth increases the density power in the model of \EQ{e:nu05} by a factor of more than $2$ at $k=1~h/$Mpc, meaning that it cannot be neglected in any computation of the neutrino power spectrum.

\subsection{\FFTM{} with N-body CDM+baryon source}
\label{subsec:enu_nbody_cb_source}

One source of error in the previous Section is its use of Time-RG perturbation theory for the CDM+baryon fluid, whose clustering is highly non-linear at small scales and late times.  References~\cite{Upadhye:2013ndm,Upadhye:2017hdl} show that Time-RG is roughly accurate down to $z \approx 1$ but breaks down for $z \lesssim 0.5$.  Since the CDM+baryon fluid makes up the significant majority of the total matter, and hence the gravitational source for neutrino clustering, its Time-RG perturbation theory errors impact our $\Ea{aj}{k}$ calculations.

\begin{figure}[tbp]
  \includegraphics[width=155mm]{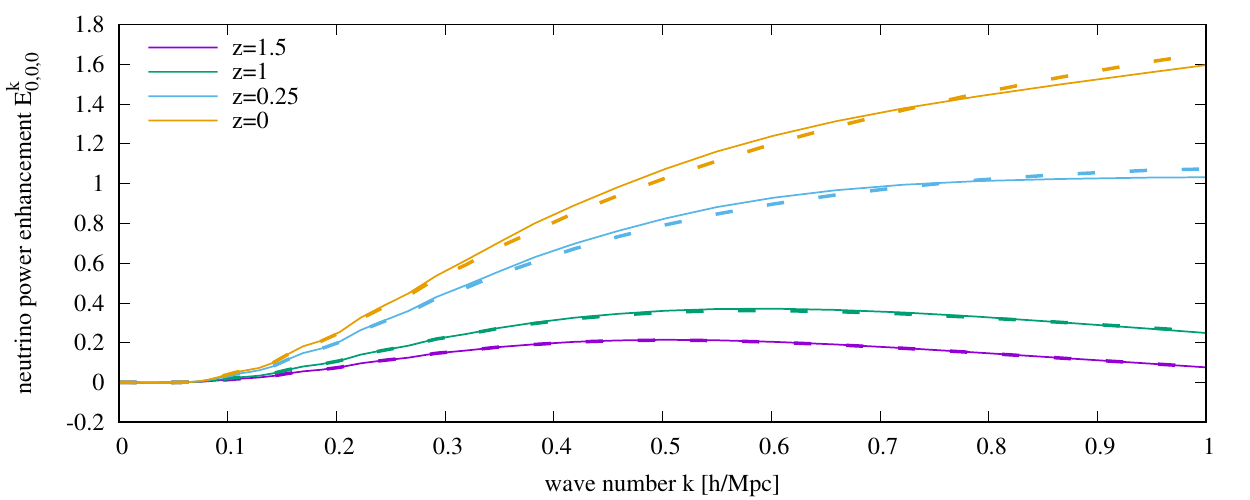}%
  \caption{
    Neutrino density monopole enhancement for $\alpha=0$, for a range of $z$,
    with $\Nmunl=5$ fixed.  Solid lines show the neutrino $E_{\alpha,0,0}^k$
    using the emulated CDM+baryon power spectrum of Ref.~\cite{Moran:2022iwe},
    while dashed lines use Time-RG for the CDM and baryons as in
    Sec.~\ref{subsec:enu_full_trg}.
    \label{f:Enu_MT4vsTRG}
  }
\end{figure}

In order to quantify this error, we include in \FFTM{} the CDM+baryon power spectrum computed by the Mira-Titan IV emulator of Ref.~\cite{Moran:2022iwe}.  This is calibrated against a suite of N-body simulations which account for neutrinos purely linearly, an approximation whose impact on the CDM+baryon power was limited to $\lesssim 0.1\%$ in  Ref.~\cite{Chen:2020bdf}.  Specifically, for $z\leq 2$ we replace the CDM+baryon density $\delta_\mathrm{cb}$ in $\Phi$ of \EQ{e:eom_Phi} by the square root of the corresponding emulator power.\footnote{Note the differing notation between this work and Ref.~\cite{Moran:2022iwe}, whose power spectrum is $(\Ocbo/\Omo)^2$ times the one defined here, and whose units are Mpc$^3$ rather than Mpc$^3/h^3$.}  Above $z=2$, the upper limit of the emulated power spectrum, we return to Time-RG for the CDM and baryons.

Neutrino power spectra resulting from this emulated CDM+baryon source are somewhat smaller than those in Fig.~\ref{f:D2nu_all}, for example about $10\%$ smaller at $z=0$ and $k=1~h/$Mpc.  Importantly, the breakdown in \FFTM{} around $k=4~h/$Mpc for $\Nmunl \geq 4$ persists.  Thus this instability affects neutrino perturbation theory directly, rather than through a Time-RG error in the CDM+baryon sector.

Figure~\ref{f:Enu_MT4vsTRG} compares the $\alpha=0$ neutrino density monopole enhancements for emulated and perturbative CDM+baryon sources.  Differences are small at high redshifts, amounting to just $0.9\%$ of the linear response power spectrum at $z=1$ and $k=1~h/$Mpc.  This difference rises to $6\%$ at $z=0$.  Thus Time-RG perturbation theory for the CDM+baryon fluid accurately quantifies the neutrino non-linear enhancement down to $z=1$, and is sufficiently accurate for estimates and convergence tests of \FFTM{} even at lower redshifts.

\subsection{Hybrid N-body neutrino simulations}
\label{subsec:enu_hybrid_sims}

Hybrid N-body simulations carried out in our companion paper, Ref.~\cite{Chen:2022dsv}, and run on the Katana cluster~\cite{KatanaCluster}, also divided the neutrino population into $N_\tau$ flows. Each of these was evolved using either the multi-fluid linear response perturbation theory of Ref.~\cite{Chen:2020bdf} or through a realization as dynamical particles whose masses, positions, and velocities quantified the density and velocity perturbations of that flow.  This hybrid simulation procedure was implemented by modifying the {\tt Gadget-4} code of Ref.~\cite{Springel:2020plp} based upon the earlier modifications of Ref.~\cite{Chen:2020bdf}.  In practice our simulations use $N_\tau=20$ flows but realize pairs of flows as particles, making the first pair similar to our $\alpha=0$ flow above, the next pair to $\alpha=1$, {\em etc}.

Reference~\cite{Chen:2022dsv} showed that non-linear clustering of neutrinos for the model of \EQ{e:nu05} represented only a $0.2\%$ correction to the total matter power spectrum, which sources the gravitational potential.  Thus we quantify the non-linear enhancement $\Ea{a\ell}{k}$ of a single flow $\alpha$ by realizing only that flow as simulation particles, while all others are tracked using multi-fluid linear response, with the resulting error well below a percent.  

Each of our hybrid N-body simulations uses $512^3$ particles for the CDM+baryon fluid and an additional $512^3$ for the neutrino flow being realized as particles.  All particles were initialized at $z=99$ using the Zel'dovich approximation for initial velocities, which Ref.~\cite{Elbers:2022tvb} showed to be $\approx 2\%$ accurate for $k \leq 1~h/$Mpc, and we included an additional randomly-directed comoving velocity of magnitude $\va$ for the neutrino particles.  Weak neutrino clustering at small scales makes the neutrino power spectrum especially prone to contamination by shot noise.  We mitigated this through the use of small simulation volumes, cubic boxes with edge lengths $L$ of either $256$~Mpc$/h$ or $128$~Mpc/h, and for which we assumed periodic boundary conditions.  The larger-box simulation used a force-softening length of $20$~kpc$/h$, while the smaller-box simulation used $10$~kpc$/h$.  From each simulated power spectrum we subtracted the approximate shot noise $L^3/N$, though residual noise remains.  

\subsection{Comparison to hybrid N-body simulations}
\label{subsec:enu_nbody_comparison}

\begin{figure}[tbp]
  \includegraphics[width=51.5mm]{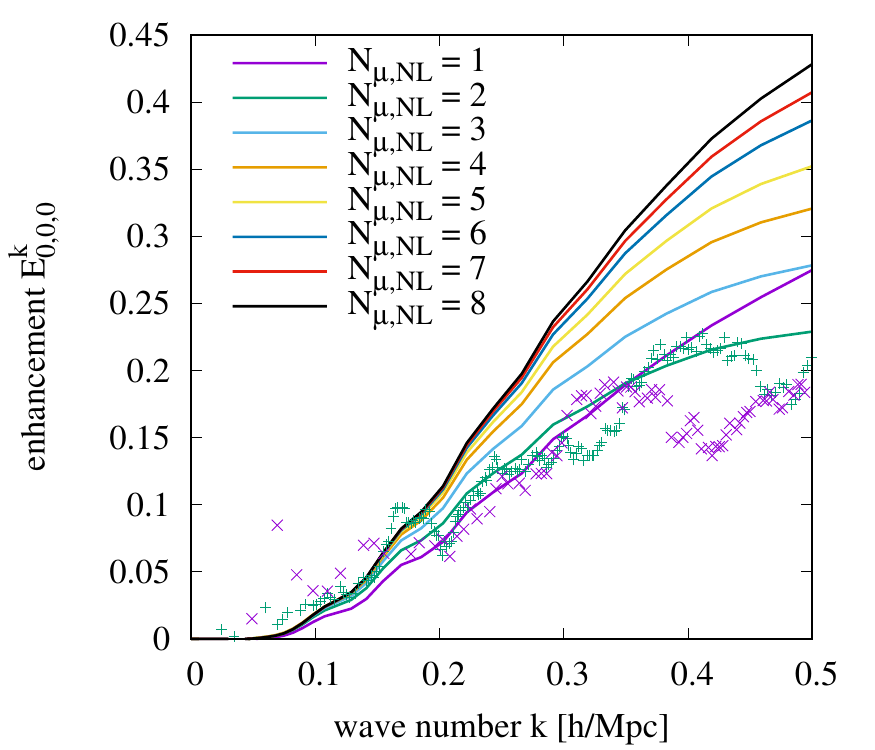}%
  \includegraphics[width=51.5mm]{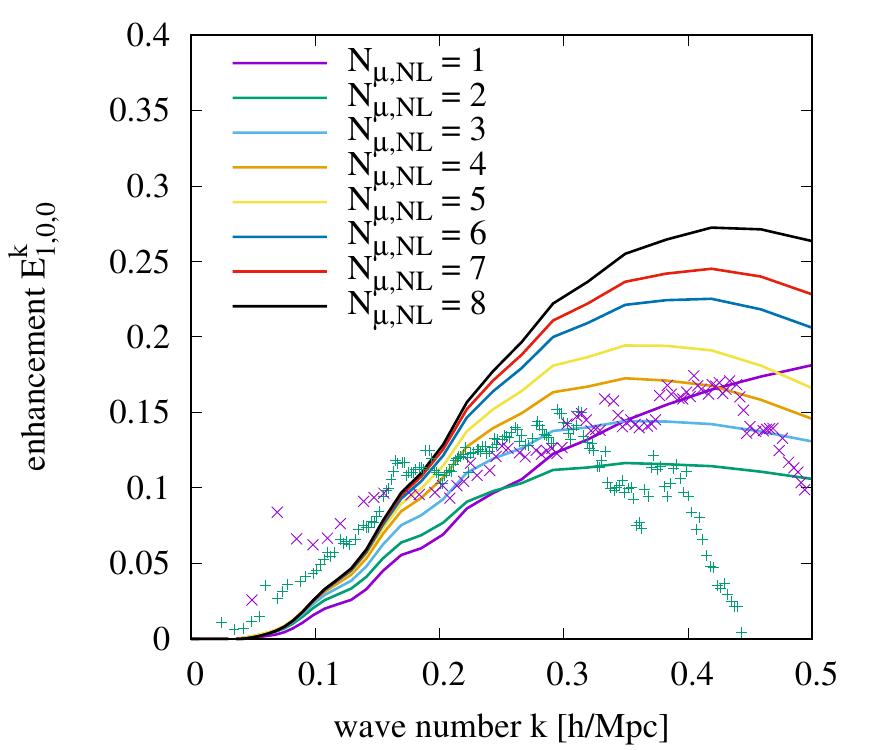}%
  \includegraphics[width=51.5mm]{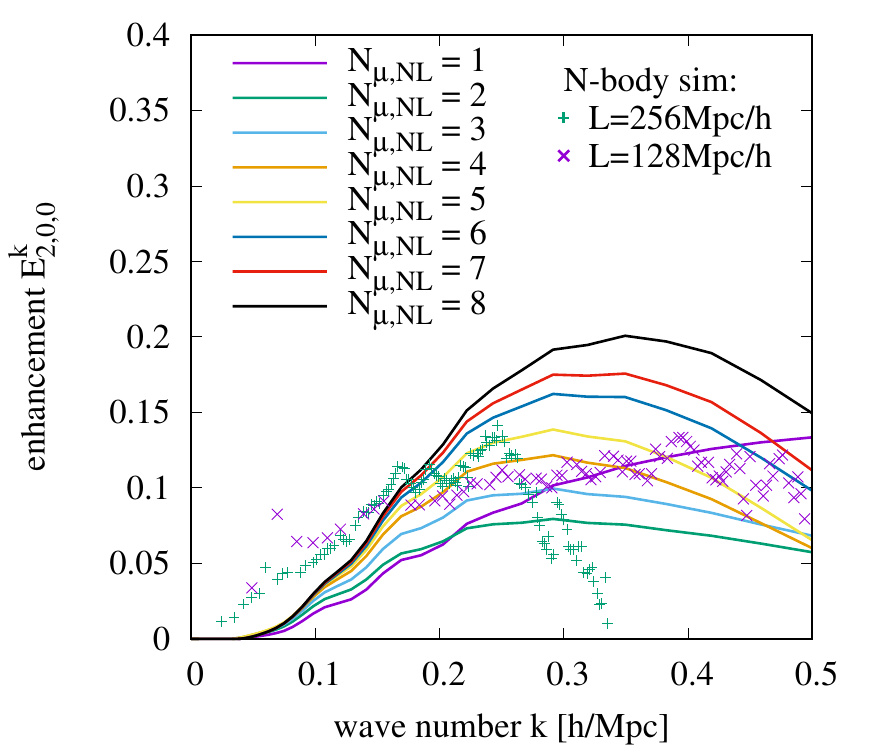}%
  \caption{
    Neutrino density monopole enhancement for $\alpha$ of $0$, $1$, and $2$
    at redshift $z=1$.  Shown are the non-linear perturbative calculations
    (thick lines) and hybrid N-body computations (points).
    \label{f:Enu_vs_Nbody_alpha0-2_z1}
  }
\end{figure}

We begin with a test of \FFTM{} in conjuction with Time-RG perturbation theory for the CDM+baryon fluid, as discussed in Sec.~\ref{subsec:enu_full_trg}.  Since the CDM and baryons cluster very non-linearly at late times, their perturbative treatment is unreliable for $z \lesssim 0.5$.  Thus in Fig.~\ref{f:Enu_vs_Nbody_alpha0-2_z1} we compare the N-body and perturbative enhancements only at $z=1$.  N-body enhancements are smoothed using a centered $20$-point moving average.

Neutrino clustering at $z=1$ is weak and its power spectrum is dominated by linear response.  Thus N-body measurements of the enhancement are quite noisy.  Figure~\ref{f:Enu_vs_Nbody_alpha0-2_z1} shows that the N-body enhancement drops below all of the perturbative calculations at $k=0.5~h/$Mpc for $\alpha=0$ and even falls below zero for larger $\alpha$.  Focusing on the region $k \lesssim 2\kfs \approx 0.25~h/$Mpc, we see close agreement between N-body and perturbative $\Ea{00}{k}$.  Agreement is closest for $\Nmunl \geq 3$, though noise in the N-body enhancements makes a more detailed comparison difficult.  At larger scales, $k \lesssim \kfs$, perturbation theory underpredicts power by a few percent, possibly due to the free-streaming closure approximation of Sec.~\ref{subsec:trg_lin_evol_I}.

\begin{figure}[tbp]
  \includegraphics[width=51.5mm]{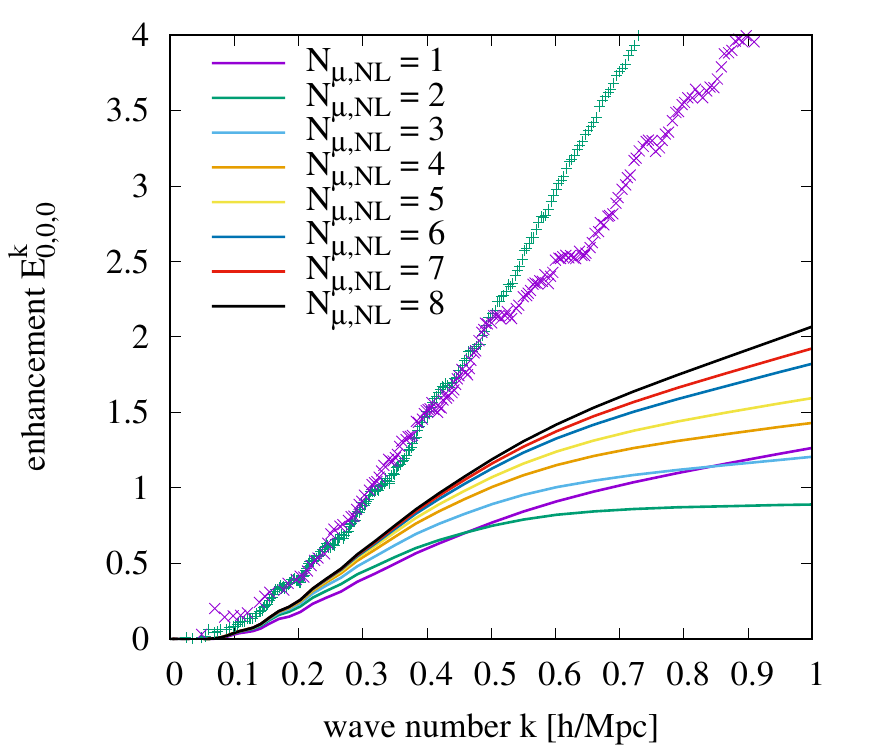}%
  \includegraphics[width=51.5mm]{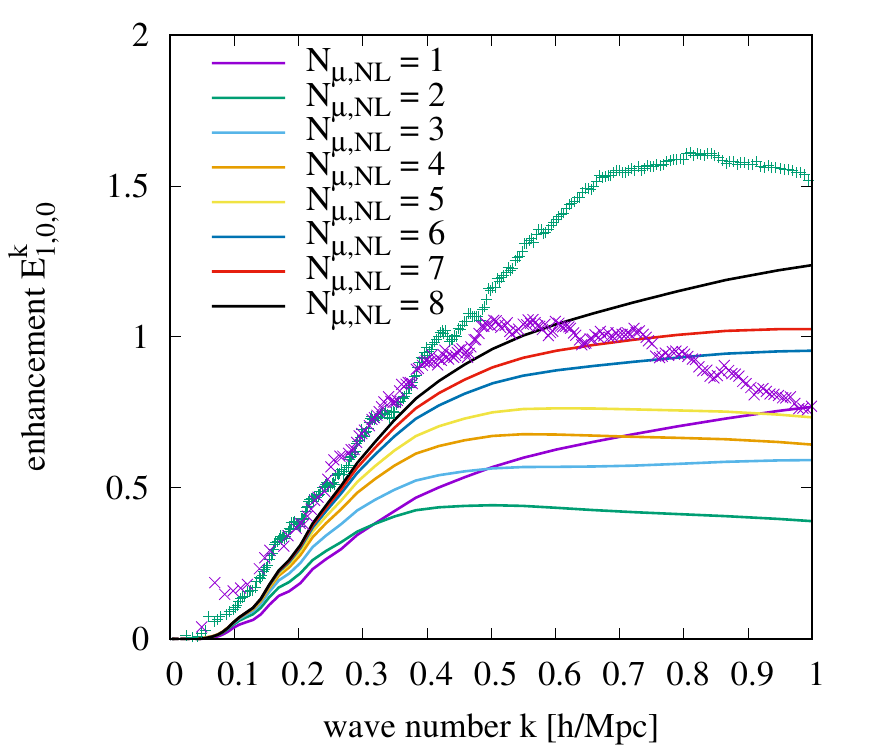}%
  \includegraphics[width=51.5mm]{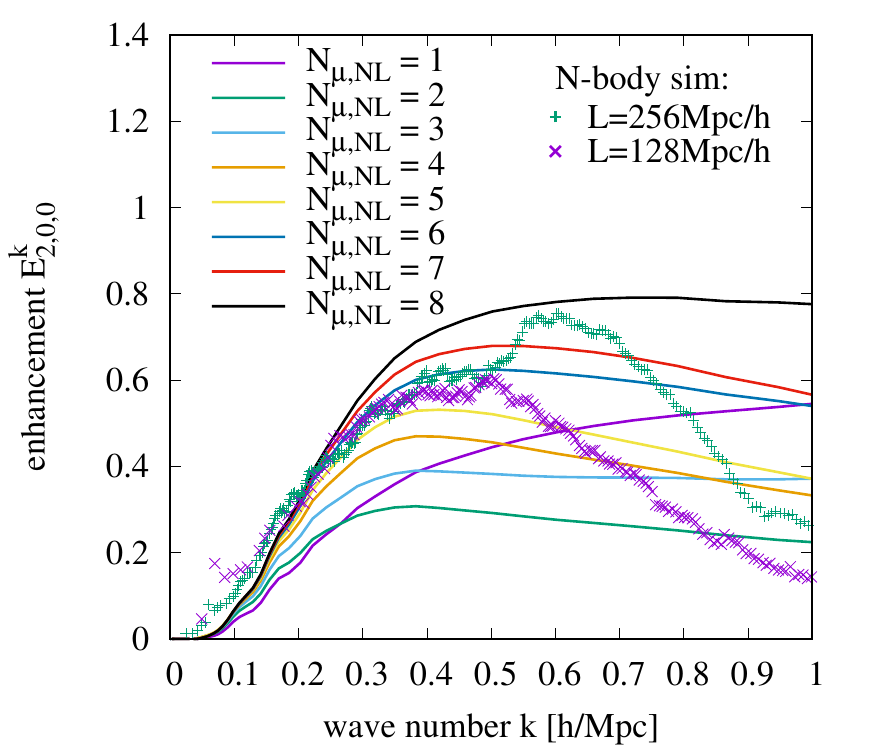}%
  \caption{
    Neutrino density monopole enhancement for $\alpha$ of $0$, $1$, and $2$
    at redshift $z=0$.  CDM+baryon power below a redshift of $2$ was computed
    using the emulator of Ref.~\cite{Moran:2022iwe}.  Shown are the non-linear
    perturbative calculations (thick lines) as well as
    hybrid N-body computations (points).
    \label{f:Enu_vs_Nbody_alpha0-2_z0}
  }
\end{figure}

Testing \FFTM{} at $z=0$ requires a more accurate calculation of the CDM+baryon power such as that emulated in Ref.~\cite{Moran:2022iwe}, as discussed in Sec.~\ref{subsec:enu_nbody_cb_source}.  Figure~\ref{f:Enu_vs_Nbody_alpha0-2_z0} compares our perturbation theory sourced by this emulated power to the hybrid simulation for $\alpha$ of $0$, $1$, and $2$ at $z=0$.  The N-body enhancements are substantially less noisy than those at $z=1$.  Since neutrinos cluster more strongly at late times, particularly on small scales, residual shot noise represents a smaller error at $z=0$.

Evidently perturbation theory for $\alpha=2$ agrees closely with the N-body enhancement in Fig.~\ref{f:Enu_vs_Nbody_alpha0-2_z0}~(right).  Agreement is best for $\Nmunl \geq 3$ in the region $k \leq 2 \kfs \approx 0.25~h/$Mpc and for $\Nmunl \geq 5$ in the region $k \leq 4 \kfs \approx 0.5~h/$Mpc.  Beyond this $k$ range, the N-body enhancement becomes noisy and its agreement with perturbation theory difficult to assess.  Moreover, at $k \gtrsim 0.5~h/$Mpc we see disagreements between our own simulations, with the $L=256$~Mpc$/h$ box likely suffering from larger residual shot noise, but the $L=128$~Mpc$/h$ box systematically underpredicting power due to neglecting perturbations larger than the box size.  Nevertheless, we may conclude that \FFTM{} accurately captures the non-linear enhancement for $\alpha \geq 2$, or $80\%$ of the total neutrino population for the model of \EQ{e:nu05}, down to $z=0$, over the range $k \lesssim 4\kfs$.

Perturbation theory is still impressive for the second-slowest flow, $\alpha=1$.  It agrees reasonably well with simulations up to $k/\kfs$ of $1.5-2$.  Even up to $k=1~h/$Mpc, the perturbation theory error in $E_{1,0,0}^k$ is limited to about $0.5$, representing an $\approx 25\%$ in the density monopole power spectrum for that flow.  Since $\alpha=1$ itself represents $10\%$ of the neutrino population, perturbation theory is useful for this flow in an accurate calculation of the total neutrino power spectrum.

N-body power for the slowest flow, shown in Fig.~\ref{f:Enu_vs_Nbody_alpha0-2_z0}~(left), is far in excess of any of the perturbative calculations for $k \gtrsim 3\kfs$.  Perturbation theory has clearly converged in this region; errors in $\delta_{00}^{3\kfs}$ for $\Nmunl$ of $1-7$, relative to $\Nmunl=8$, are respectively $8\%$, $7\%$, $4.5\%$, $2.6\%$, $1.6\%$, $0.79\%$, and $0.34\%$.  Thus we must conclude that perturbation theory itself is inaccurate for $\alpha=0$ at small scales.  This accords with our intuition that perturbation theory will break down as $\Ea{a\ell}{k}$ approaches unity for a given $\alpha$.  A percent-level-accurate computation of the small-scale neutrino power spectrum must use a particle treatment at least for flow $\alpha=0$ in the range $z < 1$.

Finally, we consider the accuracy of the \FFTM{} computation of the total neutrino power spectrum with an emulated CDM+baryon power.  Reference~\cite{Chen:2020bdf} demonstrated that $N_\tau=50$ and $N_\mu=20$ provides $3\%$ accuracy in the linear perturbations up to $k=1~h/$Mpc, so we adopt those values here.  We note a caveat for large $N_\mu$ such as $20$.  Mode-coupling integrals integrate over the product of two power spectra.  Thus truncating \EQ{e:power_Legendre2_expansion} for power spectra used in mode-couplings at $\ell = \Nmunl-1$ allows the $\Aa{acd,bef}{\vec k}$ themselves to depend upon $\mathcal{P}_j(\mua)^2$ for $j$ as high as $2\Nmunl-2$.  Our procedure necessarily neglects $\Aa{acd,bef,j}{k}$ for which $j \geq N_\mu$, so raising $N_\mu$ to $20$ includes more high-$j$ mode-coupling integrals, making the computation more vulnerable to low-$k$ noise.  Consequently, we raise $k_\mathrm{low}$ in \EQ{e:f_A_low-k} to $0.02 (1+j)~h/$Mpc for $\Nmunl=7$ and $0.01 (1+j^2)~h/$Mpc for $\Nmunl=8$.  For $\Nmunl=8$, we further reduce $\Nk$ to $64$ and $k_\mathrm{max}$ to $1~h/$Mpc in order to reduce numerical instability and computational expense.  Total running times using $32$ CPUs on the Katana Cluster~\cite{KatanaCluster} are $27$~hours for $\Nmunl=6$ and $\Nk=128$; $66$~hours for $\Nmunl=7$ and $\Nk=128$; and $75$~hours for $\Nmunl=8$ and $\Nk=64$.

\begin{figure}[tbp]
  \includegraphics[width=155mm]{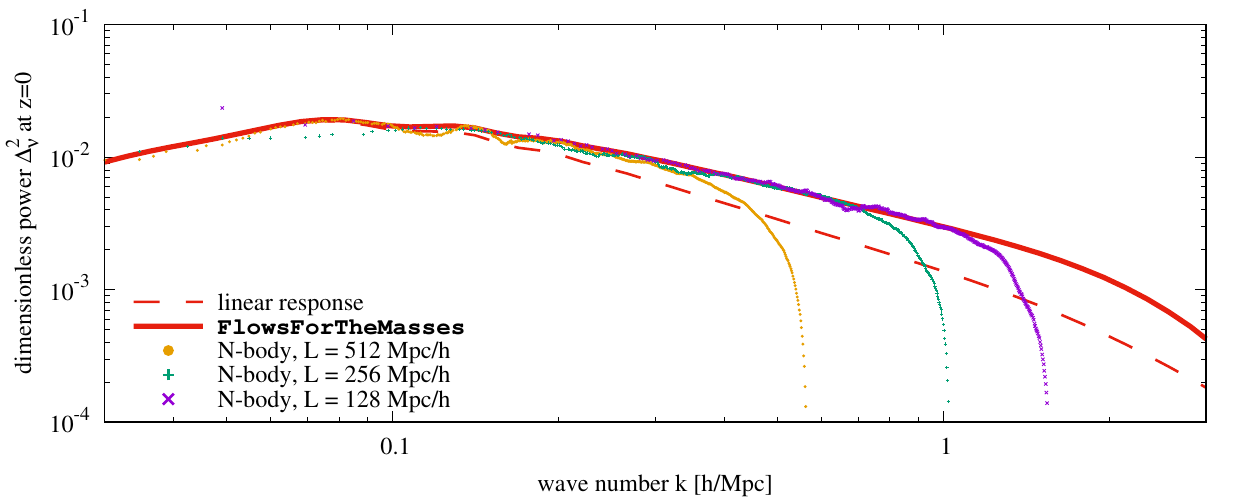}%
  \caption{
    Dimensionless neutrino density monopole power $\Delta^2_\nu(k)$ for the
    model of \EQ{e:nu05} at redshift $z=0$.  The solid line shows \FFTM{} with
    $N_\tau=50$ flows, $N_\mu=20$ angular modes, and $\Nmunl=7$; the dashed line
    shows multi-fluid linear response with the same $N_\tau$ and $N_\mu$; and
    points show the N-body power spectra with shot noise subtracted.
    \label{f:D2nu_all_vs_Nbody}
  }
\end{figure}

Figure~\ref{f:D2nu_all_vs_Nbody} compares the resulting total neutrino power spectrum for the model of \EQ{e:nu05} to our hybrid N-body simulation.  The N-body power realizes the slowest half of the neutrino population as particles while following the rest using multi-fluid linear response, an approximation shown in Sec.~\ref{subsec:enu_full_trg} to be accurate to within $1\%$.  Agreement between \FFTM{} and the N-body power is evident over a large range of wave numbers up to $k = 1~h/$Mpc, even as both rise to more than twice the linear response power.

\begin{figure}[tbp]
  \includegraphics[width=155mm]{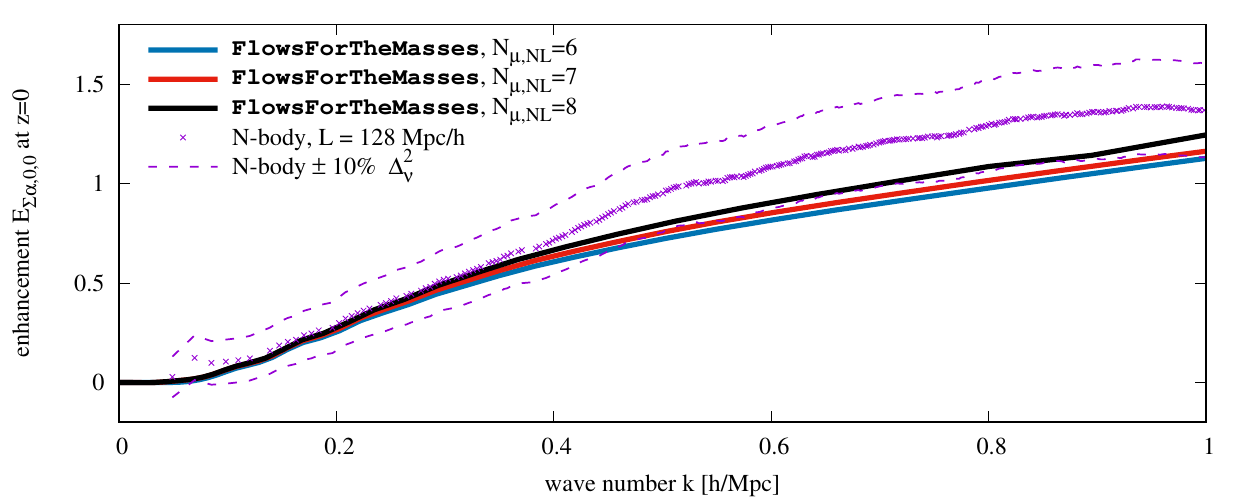}%
  \caption{
    Neutrino density monopole enhancement $E_{\Sigma\alpha,0,0}^k$ for the
    model of \EQ{e:nu05} at redshift $z=0$, using \FFTM{} with $N_\tau=50$
    flows and $N_\mu=20$ angular modes.  N-body points are smoothed using a
    $100$-point centered moving average.
    \label{f:Enu_all_vs_Nbody}
  }
\end{figure}

Shown in Fig.~\ref{f:Enu_all_vs_Nbody} is the corresponding neutrino enhancement $E_{\Sigma\alpha,0,0}^k$.  Up to simulation noise, the enhancements agree to $\approx 0.1$ at low $k$ and $\approx 0.2$ for $k \gtrsim 0.5~h/$Mpc, where $E_{\Sigma\alpha,0,0}^k \approx 1$.  This corresponds to an agreement of $< 10\%$ in the neutrino power spectra, which is impressive considering that perturbation theory was used for all neutrino flows, even the slowest.  We have therefore confirmed that \FFTM{} alone can predict the neutrino power spectrum to an accuracy of $10\%$ all the way up to $k=1~h/$Mpc for $\Ono h^2$ up to $0.005$, corresponding to $\sum m_\nu = 0.47$~eV. This includes nearly the entire $95\%$CL range of Ref.~\cite{Upadhye:2017hdl}, which allowed for substantial variations in the dark energy equation of state.

\subsection{Variation of neutrino fraction}
\label{subsec:enu_Omega_nu}

Now that we have confirmed the accuracy of a purely perturbative calculation of the non-linear neutrino power spectrum using \FFTM{}, we consider the impact of varying the neutrino fraction $\Ono h^2$.  We do so while keeping constant all other model parameters in \EQ{e:nu05}.  Since greater $\Ono h^2$ implies a greater non-linear enhancement, hence greater errors in perturbation theory, restricting $\Ono h^2 \leq 0.005$ means that the perturbative power spectrum should be accurate to $< 10\%$ across this entire range.

\begin{figure}[tbp]
  \includegraphics[width=155mm]{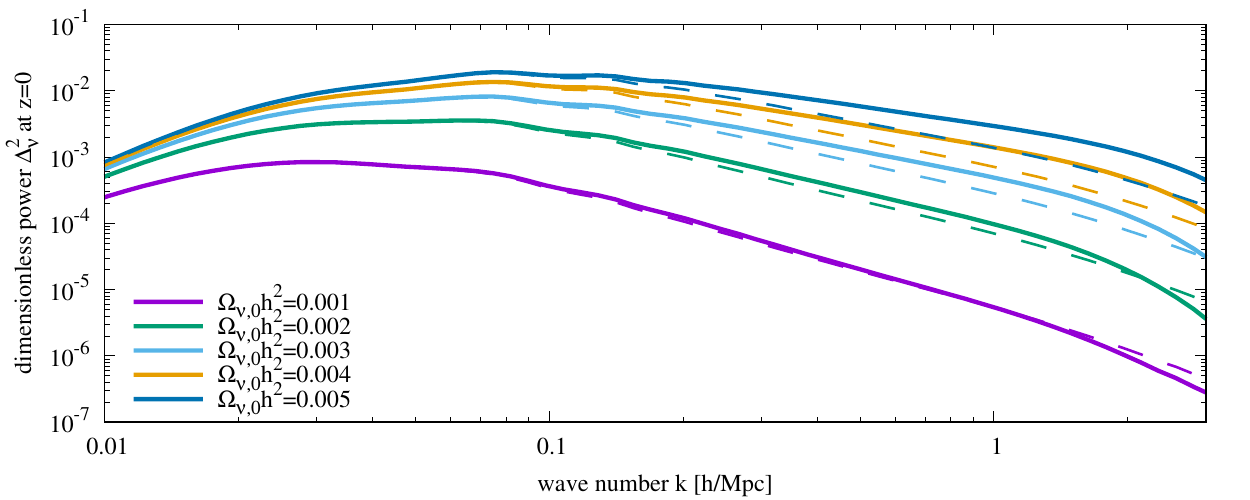}%
  \caption{
    Dimensionless neutrino density monopole power $\Delta^2_\nu(k)$  at redshift
    $z=0$, using \FFTM{} with $N_\tau=50$ flows, $N_\mu=20$ angular modes, and
    $\Nmunl=6$.  All cosmological parameters in \EQ{e:nu05} are held constant
    except for $\Ono h^2$.  Thick solid lines show a \FFTM{} perturbative
    calculation, while thin dashed lines show multi-fluid linear response.
    \label{f:D2nu_all_varyOnu}
  }
\end{figure}

Our results are displayed in Fig.~\ref{f:D2nu_all_varyOnu}.  At linear scales, $k \sim 0.01~h/$Mpc, all power spectra are similar due to having the same normalization $A_{\rm s}$.  Around the free-streaming scales, $k \sim 0.1~h/$Mpc for these models, significant differences in the power spectra are apparent, and non-linear enhancement has risen to $\sim 10\%$.  On non-linear scales, $k \gtrsim 0.5~h/$Mpc, non-linear corrections are significant, doubling the power spectrum for the larger $\Ono h^2$.  Non-linear enhancement is large enough at $k \sim 1~h/$Mpc that neglecting it would cause us to confuse $\Ono h^2$ of $0.004$ for $0.005$, a significant error in the neutrino mass fraction.

Meanwhile, for $\Ono h^2 = 0.001$, the non-linear correction peaks at $13\%$ around $k=0.2~h/$Mpc.  Thus a power spectrum computation with an accuracy goal of $\lesssim 10\%$ may neglect a particle treatment of neutrinos up to $\Ono h^2 = 0.005$, and may neglect non-linear corrections altogether up to $\Ono h^2 \approx 0.001$.  For $\Ono h^2 = 0.002$, the non-linear correction plateaus at $38\%$ around $k=1~h/$Mpc.  Our result is somewhat more pessimistic than that of Ref.~\cite{Chen:2020bdf}, which estimated that fewer than $1\%$ of neutrinos cluster non-linearly for $\Ono h^2 = 0.002$.

Well below the free-streaming length scale, Refs.~\cite{Ringwald:2004np,Wong:2008ws} show that the linear response neutrino power scales as $(\kfs/k)^4$ times the total matter power.  Assuming all parameters except for $\Ono h^2$ in \EQ{e:nu05} are held constant, this in turn scales as $(\Ono h^2)^4$.  Figure~\ref{f:D2nu_all_varyOnu} shows that the non-linear neutrino power spectra scale similarly.  For example, at $k=1~h/$Mpc, the ratio of power spectra with $\Ono h^2$ of $0.004$ and $0.005$ is $0.47 = 0.83^4$, compared with $0.8^4$ predicted by the scaling relation; at $k=2~h/$Mpc, the ratio is $0.404 = 0.797^4$.  This means that the $<10\%$ accuracy of our non-linear perturbative power spectrum is sufficient for a $< 2.5\%$ accuracy in $\Ono h^2$, and hence the sum of neutrino masses.

\section{Conclusions}
\label{sec:conclusion}

We have derived, implemented, and validated the accuracy of a new non-linear cosmological perturbation theory for massive neutrinos and other hot dark matter species, which we call \FFTM{}.  Deriving the equations of motion, \EQS{e:eom_P}{e:eom_chi}, based upon an extension of the Time-RG perturbation theory, and \EQS{e:eom_I}{e:def_Xita} for the bispectrum integrals, we identified the extended mode-coupling integrals $\Aa{acd,bef,j}{k}$ as the most computationally expensive components of the perturbation theory.  Sections~\ref{sec:mode-coupling_P22} and \ref{sec:mode-coupling_P13} constructed the computational machinery of \FFTM{} by using Fast Fourier Transforms to accelerate the mode-coupling computations by more than two orders of magnitude.

In Sec.~\ref{sec:nu_NL_enhancement} we confirmed the convergence and accuracy of \FFTM{} for neutrino power spectrum computations.  Even for neutrino fractions as high as $\Ono h^2 = 0.005$, around the maximum allowed by the current data when the dark energy is allowed substantial freedom to vary, we find in Figs.~\ref{f:D2nu_all_vs_Nbody} and \ref{f:Enu_all_vs_Nbody} that a purely perturbative computation of the neutrino power spectrum agrees with N-body simulations to better than $10\%$ up to $k=1~h/$Mpc, an impressive feat.  Errors will be even smaller for the lower-mass neutrinos preferred by the data under restrictive assumptions about the dark energy.

Moreover, the bulk of this error is due to the $\sim 10\%$ of the neutrinos which begin with the lowest velocities.  It is precisely these which are most amenable to an N-body treatment.  \FFTM{} holds the most promise in combination with particle simulations, since perturbation theory is most accurate for precisely those fast-moving, weakly-clustering neutrinos that present the most challenges to simulations.  Non-linear perturbation theory can accurately evolve $80\%-90\%$ of the neutrino population for $\Ono h^2$ as high as $0.005$.  It can also allow the remaining neutrinos to be initialized as particles later, when their thermal velocities are smaller, minimizing the resulting shot noise in their velocities.

\subsubsection*{Acknowledgments}
The authors are grateful to J.~Kwan, I.~G.~McCarthy, M.~Mosbech, and V.~Yankelevich for insightful discussions.  JZC acknowledges support from an Australian Government Research Training Program Scholarship.  AU is supported by the European Research Council (ERC) under the European Union’s Horizon 2020 research and innovation programme (grant agreement No.~769130).   Y$^3$W is supported by the Australian Research Council (ARC) Future Fellowship (project FT180100031).  This research is enabled by the ARC Discovery Project (project DP170102382) funding scheme, and includes computations using the computational cluster Katana supported by Research Technology Services at UNSW Sydney.

\appendix

\section{Expansions of $\Lc_{\ell m n}$ and $\Mc_{\ell m n}^{(j)}$}
\label{app:expansions_of_Lc_Mc}

Our goal is to expand the functions $\Lc_{\ell m n}$ and $\Mc_{\ell m n}^{(j)}$ of \EQS{e:def_Lc}{e:def_Mc}.  We begin with $\Mc_{\ell m n}^{(1)}(\mua)$.  The quantities $\sigmamnj{m}{n}{1}$ of \EQ{e:sigmamn1} may be evaluated exactly for several useful values. For $I := (n - m + 1)/2$ an integer,
\begin{eqnarray}
  \sigmamnj{m}{n}{1}
  &=&
  \sum_{i=0}^{2I+m} \frac{\bNi{2I+m-1}{i}}{2I-2i+1}
  =
  \sum_{i=0}^{2I+m-1} \frac{\binom{2I+m-1}{i}}{2I-2i+1}
  -\sum_{i=0}^{2I+m-1} \frac{\binom{2I+m-1}{i}}{2I-2i-1}
  \nonumber\\
  &=&
  \int d\xi  \sum_{i=0}^{2I+m-1} \binom{2I+m-1}{i} \xi^{2I-2i}
  - \int d\xi \sum_{i=0}^{2I+m-1} \binom{2I+m-1}{i} \xi^{2I-2i-2}
  \nonumber\\
  &=&
  \int d\xi \, \xi^{2I} (1+\xi^{-2})^{2I+m-1}
  - \int d\xi \, \xi^{2I-2} (1+\xi^{-2})^{2I+m-1}
  \nonumber\\
  &=&\!
  -\frac{{}_2F_1(\!1\!-2I\!-\!m,\!\tfrac{3}{2}\!-\!I\!-\!m,\!\tfrac{5}{2}\!-\!I\!-\!m,\!-1\!)}{2I+2m-3}
  +\frac{{}_2F_1(\!1\!-\!2I\!-\!m,\!\tfrac{1}{2}\!-\!I\!-\!m,\!\tfrac{3}{2}\!-\!I\!-\!m,\!-1\!)}{2I+2m-1}\qquad
\end{eqnarray}
where the integrals are evaluated at $\xi=1$.  For $m$ ranging from $1$ to $6$, the first hypergeometric function in the final expression is, respectively: $-2^{2I} \tfrac{2I-1}{2I+1}$; $0$; $2^{2I+2}$; $2^{2I+4}$; $2^{2I+5} (\tfrac{3}{2} - \tfrac{1}{2I+5})$; and $2^{2I+7} \tfrac{2I+5}{2I+7}$.  The second hypergeometric function may be obtained from the first by substituting $I'=I-1$ and $m'=m+2$ for $I$ and $m$, respectively.

Now $\Mc^{(1)}_{\ell m n}(\mua)$ can be written down exactly for all cases of interest here.  If the sum of $\ell$, $m$, and $n$ is even, then $\Mc^{(j)}_{\ell m n}(\mua)$ vanishes.  Restricting ourselves to even $\ell$ and defining the integers $L = \ell/2$ and $i = L + (m + n - 1)/2$, we find that
\begin{eqnarray}
  \Mc^{(1)}_{2L,1,n}(\mu)
  &=&
  \frac{\Gamma(L+\tfrac{1}{2}) \Gamma(i-L+\tfrac{1}{2})}{2\sqrt{\pi} \Gamma(i+\tfrac{3}{2})}
  s^{(L,i)}_\mathrm{e}(\mu^2)
  \label{e:Mc1_2L_1_n}
  \\
  \Mc^{(1)}_{2L,2,n}(\mu)
  &=&
  \frac{\Gamma(L+\tfrac{1}{2}) \Gamma(i-L+\tfrac{1}{2})}{\sqrt{\pi} \Gamma(i+\tfrac{3}{2})}
  s^{(L,i)}_\mathrm{e}(\mu^2)
  \label{e:Mc1_2L_2_n}
  \\
  \Mc^{(1)}_{2L,3,n}(\mu)
  &=&
  \frac{2\Gamma(L\!+\!\tfrac{1}{2}) \Gamma(i\!-\!L\!+\!\tfrac{1}{2})
  }{\sqrt{\pi} \Gamma(i+\tfrac{3}{2})}
  s^{(L,i)}_\mathrm{e}(\mu^2)
  -
  \frac{\Gamma(L\!+\!\tfrac{1}{2}) \Gamma(i\!-\!L\!-\!\tfrac{1}{2})
  }{\sqrt{\pi} \Gamma(i+\tfrac{1}{2})}
  s^{(L,i-1)}_\mathrm{e}(\mu^2)\qquad
  \label{e:Mc1_2L_3_n}
  \\
  \mathrm{where}\quad
  s^{(L,i)}_\mathrm{e}(\mu^2)
  &:=&
  \sum_{n=0}^{i-L} \binom{i-L}{n} \binom{L}{n}
  \frac{n! \sqrt{\pi}}{\Gamma(n+\tfrac{1}{2})} \mu^{2n}.
  \label{e:def_s_even}
\end{eqnarray}

Simplification of $\Lc_{\ell m n}(\mu,x)$ and $\Mc^{(0)}_{\ell m n}(\mu)$ is more difficult. Each consists of an outer sum over $r$ and an inner sum over $s$.  Begin with the outer sum of the quantity
\begin{equation}
  X_{\ell m n r}
  :=
  \left\{
  \begin{array}{cl}
    \sum_{s=0}^r (-1)^{r+s} \binom{r}{s}
    \sum_{i=0}^{\ell+n+1-2s}
    \frac{\bNi{\ell+n-2s}{i}\lambda_{\ell+m+n+1-2s-2i}(x)}{2^{\ell+n-2s}}
    & \textrm{for $\Lc_{\ell m n}$}
    \\
    \sum_{s=0}^r (-1)^{r+s} \binom{r}{s}
    \frac{\sigmamnj{|m|}{\ell+n-2s}{0}}{2^{\ell+n-2s}}
    & \textrm{for $\Mc^{(0)}_{\ell m n}$}
  \end{array}
  \right. .
\end{equation}
The outer summation is then $\sum_{r=0}^{\ell/2} \binom{\ell}{2r} \kappa_r \mua^{\ell-2r} \nua^{2r} X_{\ell m n r}$.  Defining the integer $L = (\ell - \Odd_\ell)/2$ and expanding in binomial series, we find $\mua^{\ell-2r}\nua^{2r} = \sum_{t=0}^r (-1)^{t-r} \binom{r}{t} \mua^{\Odd_\ell + 2L - 2t}$.  Writing factorials in terms of the Gamma function, we have $\binom{\ell}{2r} \kappa_r = \frac{1}{r!} \binom{L}{r} \frac{\Gamma(L+1/2+\Odd_\ell)}{\Gamma(L+1/2+\Odd_\ell-r)}$.  Thus each outer summation may be rearranged into
\begin{equation}
  \sum_{r=0}^{\ell/2}\binom{\ell}{2r} \kappa_r \mua^{\ell-2r} \nua^{2r} X_{\ell m n r}
  =
  \sum_{t=0}^L (-1)^t \binom{L}{t} \mua^{\Odd_\ell + 2t}
  \sum_{r=0}^t \frac{(-1)^r}{(L-r)!} \binom{t}{r}
  \frac{\Gamma(L+\Odd_\ell+\tfrac{1}{2})}{\Gamma(r+\Odd_\ell+\tfrac{1}{2})}
  X_{\ell,m,n,L-r}
  \qquad\quad
\end{equation}
making its expansion in powers of $\mua$ explicit.

Our strategy for the inner summation of $\Lc_{\ell m n}$ is to pull the $\lambda_N(x)$ function outside of the $i$ and $s$ summations.  The remaining quantities are wave-number-independent numerical constants which may be computed in advance.  Thus we change summation indices from $i$ to $v := \ell + n + 1 - s - i$, with the result that
\begin{eqnarray}
  X_{\ell m n r}^{(\Lc)}
  &=&
  \sum_{s=0}^r (-1)^{r-s} \binom{r}{s}
  \sum_{v=s}^{\ell+n+1} \frac{\bNi{\ell+n-2s}{\ell+n+1-s-v}}{2^{\ell+n-2s}}
  \lambda_{m-\ell-n-1+2v}(x)
  \nonumber\\
  &=&
  \sum_{v=0}^{\ell+n+1}   \lambda_{m-\ell-n-1+2v}(x)
  \sum_{s=0}^v (-1)^{r-s+1} \binom{r}{s}
  \frac{\bNi{\ell+n-2s}{v-s}}{2^{\ell+n-2s}}
\end{eqnarray}
where we have used the identity $\bNi{n}{i} = -\bNi{n}{n+1-i}$.  We may also pull $\lambda$ outside of one of the outer sums, the one over $r$.  Again restricting $\ell$ to even numbers, $\ell = 2L$, we have
\begin{eqnarray}
  &\!\!\!&
  \sum_{r=0}^t (-1)^r \binom{t}{r}
  \frac{(r+\tfrac{1}{2})_{L-r}}{(L-r)!}
  X_{2L,m,n,L-r}
  \nonumber\\
  &~&\qquad\quad
  =
  \sum_{r=0}^t \binom{t}{r}
  \frac{(r+\tfrac{1}{2})_{L-r}}{(L-r)!}
  \sum_{v=0}^{2L+n+1} \lambda_{m-2L-n-1+2v}(x)
  \sum_{s=0}^v (-1)^{L+s+1}\binom{L-r}{s}
  \frac{\bNi{2L+n-2s}{v-s}}{2^{2L+n-2s}}
  \nonumber\\
  &~&\qquad\quad=
  \sum_{v=0}^{2L+n+1} \lambda_{m-2L-n-1+2v}(x)
  \sum_{r=0}^t \sum_{s=0}^v (-1)^{L+s+1} \binom{t}{r}\binom{L-r}{s}
  \frac{(r+\tfrac{1}{2})_{L-r}}{(L-r)!}
  \frac{\bNi{2L+n-2s}{v-s}}{2^{2L+n-2s}}\qquad
  \nonumber\\
  &\!\!\!&
  \Rightarrow
  \Lc_{2L,m,n}(\mua,x)
  =
  \sum_{t=0}^L (-1)^t \binom{L}{t} \mua^{2t}
  \sum_{v=0}^{2L+n+1} {\hat g}_{Lntv} \, \lambda_{m-2L-n-1+2v}(x)
  \label{e:Lc}
  \\
  &\!\!\!&
  {\hat g}_{Lntv}
  :=
  \sum_{r=0}^t \sum_{s=0}^v (-1)^{L+s+1} \binom{t}{r}\binom{L-r}{s}
  \frac{(r+\tfrac{1}{2})_{L-r}}{(L-r)!}
  \frac{\bNi{2L+n-2s}{v-s}}{2^{2L+n-2s}}
  \label{e:def_hat_g}
\end{eqnarray}
where $(y)_z := \Gamma(y+z)/\Gamma(y)$ is the Pochhammer symbol.

Computation of $\Mc^{(0)}_{\ell m n}(\mua)$ with $\ell = 2L$ is simpler.  Since $\Mc^{(0)}_{2L, m, n}(\mua)$ vanishes unless $m+n$ is even, let $I := (m+n)/2$ be an integer.  Then we may replace $n$ by $2I-m$.  Since $X_{\ell m n r}^{(\Mc)}$ is now independent of $\mua$ and $x$, we immediately express $\Mc^{(0)}_{2L, m, 2I-m}(\mua)$ in terms of wave-number-independent numerical coefficients.

Collecting our results for $\Mc_{\ell m n}^{(j)} =  \sum_t {\hat d}^{(j)}_{LmIt} \mua^{2t}$, we have:
\begin{equation}
  {\hat d}^{(0)}_{LmIt}
  =
  (-1)^{t+L} \binom{L}{t} \sum_{r=0}^t \frac{(r+\tfrac{1}{2})_{L-r}}{(L-r)!}
  \binom{t}{r} \sum_{v=0}^{L+I} \frac{1}{1+2v}
  \sum_{s=0}^{L+I-v} (-1)^s \binom{L-r}{s}
  \frac{\bNi{2L+2I-2s-m}{L+I-s-v}}{2^{2L+2I-2s-m}}\qquad
  \label{e:def_hat_d0}
\end{equation}
\begin{equation}
  {\hat d}^{(1)}_{L1It}
  =
  \frac{t!}{2} \binom{I}{t} \binom{L}{t} \frac{(t+1/2)_{L-t}}{(I+1/2)_{L+1}},
  \quad
  {\hat d}^{(1)}_{L2It}
  =
  2 {\hat d}^{(1)}_{L1It},
  \quad
  {\hat d}^{(1)}_{L3It}
  =
  4 {\hat d}^{(1)}_{L1It} - 2 {\hat d}^{(1)}_{L,1,I-1,t}.
  \label{e:def_hat_d1}
\end{equation}
Since $j$ may be lowered using the recursion relation of Eq.~(\ref{e:sigmamnj_recursion}), the above completely specify all $\Mc_{\ell m n}^{(j)}$ of interest to us here.

\section{Divergence cancellation}
\label{app:divergence_cancellation}

Omitted from the catalog of $\Pot$-type $\Aa{acd,bef,j}{k}$ integrals in Sec.~\ref{subsec:p13_A13_catalog} are several divergent terms which cancel against similar divergences in the $\Ptt$-type terms $\Sa{\pm 2,2,e1,fb}{\vec k}$ and $\Sa{-\pm 2,2,eb,f1}{\vec k}$ in the catalog of Sec.~\ref{subsec:p22_A22_catalog}.  This cancellation was considered thorougly in Ref.~\cite{McEwen:2016fjn}, which argued that it was a necessary consequence of the translation invariance of observable power spectra.  One-loop corrections $\Ptt$ and $\Pot$ are not observable individually, but only together, thus IR divergences in each must cancel when they are added.

Since we need not modify their argument, our task here is one of bookkeeping to ensure that divergences in our $\Aa{acd,bef,j}{(2,2)k}$ and $\Aa{acd,bef,j}{(1,3)k}$ do actually cancel when these are combined into $\Aa{acd,bef,j}{k}$.  For convenience we expand the power in a basis of even powers of $\mua$:
\begin{equation}
  \Pa{ab}{\vec k}
  =
  \sum_L \mua^{2L} \Pa{abL}{k}
  \quad\textrm{with}\quad
  \Pa{abL}{k}
  =
  \sum_{\ell=L}^\infty \QLeg_{L\ell} \Pa{ab\ell}{k}.
\end{equation}
Mode-coupling integrals of $\Pot$ type have the following divergent terms:
\begin{eqnarray}
  \Aa{001,bef}{\vec k}
  &~:~&
  -\frac{1}{4}
  \sum_{L=0}^\infty \sum_{M=0}^\infty \mua^{2L} \Mc_{2M,1,2}^{(1)}(\mua)
  \Pa{beL}{k} \intq x^{-2} \Pa{f1M}{}(kx)
  \label{e:div_A13_001}
  \\
  \Aa{111,bef}{\vec k}
  &~:~&
    -\frac{1}{4}
  \sum_{L=0}^\infty \sum_{M=0}^\infty \mua^{2L} \Mc_{2M,1,2}^{(1)}(\mua)
  \Pa{beL}{k} \intq x^{-2} \Pa{f1M}{}(kx)
  \ldots
  \nonumber\\
  &~~~&
  -\frac{1}{4}
  \sum_{L=0}^\infty \sum_{M=0}^\infty \mua^{2L} \Mc_{2M,1,2}^{(1)}(\mua)
  \Pa{bfL}{k}  \intq x^{-2} \Pa{e1M}{}(kx)
  \label{e:div_A13_111}
\end{eqnarray}
where $x$ represents either $q/k$ or $p/k$.

Since $\Sa{BN,ab,cd}{\vec k} = \Sa{-B,N,cd,ab}{\vec k}$, without loss of generality we exclusively consider small-$p$ divergences in $\Sa{-2,N,ab,cd}{\vec k}$.  Such divergences arise in the monopole terms, $\lambda=0$, of \EQ{e:def_J}, and specifically, $\Ja{A,-2,0, abL, cdM}{k} = \intq q^A p^{-2} \PLeg_0(\mupq) \Pa{abL}{q} \Pa{cdM}{p}$ for integer $A$.  Near $p=0$, $q\rightarrow k$, so this integral behaves as $k^{A-2} \Pa{abL}{k} \intq x^{-2} \Pa{cdM}{}(kx)$ with $x=p/k$.  Already this resembles the divergent integrals in \EQS{e:div_A13_001}{e:div_A13_111}; all that remains is to compute the coefficients and $\mua$-dependences of these $J$ monopole terms.

We start with \EQ{e:avg_muaq2L_muap2M}.  Making the replacements $\vtc=(p\mupq+q)/k$, $\vtct=(q\mupq+p)/k$, $\vts=p\sqrt{1-\mupq^2}/k$, and $\vtst=-q\sqrt{1-\mupq^2}/k$, we see that
\begin{eqnarray}
  \vtc^{2L-i} \vtct^{2M+i-2r} \vts^i \vtst^{2r-i}
  &=&
  \left(\frac{q}{k}\right)^{2L} \left(\frac{p}{k}\right)^{2M}
  \sum_{a=0}^r \sum_{b=0}^{2L-i} \sum_{d=0}^{2M-2r+i}   \!\!(-1)^{i+a} 
  \binom{\!r\!}{\!a\!} \binom{\!2L\!-\!i\!}{b} \binom{\!2M\!-\!2r\!+\!i\!}{d}
  \ldots\nonumber\\
  &~&\qquad\qquad\qquad\qquad\qquad\qquad\qquad\times
  \mupq^{2a+b+d} (q/p)^{2r-2i-b+d}.
  \nonumber
\end{eqnarray}
The exponent of $p$ is smallest when $d$ takes its maximum value and $i$ and $b$ their minimum values, resulting in the cancellation of all powers of $p$, and a summand proportional to $(q/k)^{2L+2M}$, which is unity in the low-$p$ limit.  The resulting exponent of $\mupq$ is $2a + 2M - 2r$.

Inserting this into \EQ{e:avg_muaq2L_muap2M} for $\left<\muaq^{2L} \muap^{2M}\right>_\varphi$, and this in turn into \EQ{e:def_S_BN_ab_cd} for $\Sa{-2,N,ab,cd}{\vec k}$, we see that the divergent terms may only arise from
\begin{equation*}
  \sum_{r=0}^{L+M} \mua^{2L+2M-2r}\nua^{2r} \kappa_r \binom{\!2M\!}{\!2r\!}
  \intq \left(\frac{q}{k}\right)^{2L+2M}\left(\frac{p}{q}\right)^{-2}
  \Pa{abL}{q} \Pa{cdM}{p}
  \sum_{a=0}^r (-1)^a \binom{\!r\!}{\!a\!} \mupq^{2a+2M+N-2r}.
\end{equation*}
Furthermore, only even powers of $\mupq$ contribute to a monopole term, so divergences are only present for even $N$.  Assuming $N$ to be even, the projection of $\mupq^{2a+2M+N-2r}$ on the monopole is $(2a+2M+N-2r+1)^{-1}$.  Changing summation variables from $a$ to $s:=r-a$, we see that the divergent term in $\Sa{-2,N,ab,cd}{\vec k}$ is the summation over $L$ and $M$ of:
\begin{eqnarray}
  &&\!\!\!\!\!\!\!\!\!\!\!\!\!\!\!\!\!\!
  \frac{\Ja{2L+2M+2,-2,0,abL,cdM}{k}}{k^{2L+2M}}
  \sum_{r=0}^{L+M} \mua^{2L+2M-2r}\nua^{2r} \kappa_r \binom{\!2M\!}{\!2r\!}
  \sum_{s=0}^r \frac{(-1)^{r-s}\binom{r}{s}}{2M\!+\!N\!-\!2s\!+\!1}
  \nonumber\\
  &=&
  \frac{\Ja{2L+2M+2,-2,0,abL,cdM}{k}}{k^{2L+2M}}
  \sum_{r=0}^{L+M} \mua^{2L+2M-2r}\nua^{2r} \kappa_r \binom{\!2M\!}{\!2r\!}
  \sum_{s=0}^r \frac{(-1)^{r-s}\binom{r}{s} \sigma_1^{(1,2M+N-2s)}}{2^{2M+N-2s}}
  \nonumber\\
  &=&
  \frac{\Ja{2L+2M+2,-2,0,abL,cdM}{k}}{k^{2L+2M}}
  \mua^{2L} \Mc_{2M,1,2}^{(1)}(\mua).
  \nonumber
\end{eqnarray}
Thus our result is
\begin{equation}
  \left. \Sa{-2,N,ab,cd}{\vec k} \right|_{\rm div}
  =
  \sum_{L=0}^{\infty} \sum_{M=0}^{\infty}
  \frac{\Ja{2L+2M+2,-2,0,abL,cdM}{k}}{k^{2L+2M}}
  \mua^{2L} \Mc_{2M,1,N}^{(1)}(\mua).
  \label{e:S_div}
\end{equation}

Evidently the $\mua$-dependent factors $\mua^{2L} \Mc_{2M,1,N}^{(1)}(\mua)$ in \EQ{e:S_div} are the same as those in \EQS{e:div_A13_001}{e:div_A13_111} provided that $N=2$, which is the case for all divergent terms in \EQS{e:A22_001_0ef}{e:A22_111_1ef}.  Further, the numerical factor of $1/4$ multiplying $\Sa{-2,2,eb,f1}{\vec k}$ in \EQS{e:A22_001_0ef}{e:A22_001_1ef} is precisely what we need to cancel the divergence in \EQ{e:div_A13_001}.  The divergent terms in \EQS{e:A22_111_0ef}{e:A22_111_1ef}, $\frac{1}{4} \left.\Sa{-2,2,eb,f1}{\vec k}\right|_{\rm div}$ and $\frac{1}{4}  \left.\Sa{2,2,e1,fb}{\vec k}\right|_{\rm div} =  \frac{1}{4}\left.\Sa{-2,2,fb,e1}{\vec k}\right|_{\rm div}$, cancel those in \EQ{e:div_A13_111}.

\bibliographystyle{unsrt}
\bibliography{flowsForTheMasses}

\begin{thebibliography}{10}

\bibitem{Planck:2018vyg}
N.~Aghanim et~al.
\newblock {Planck 2018 results. VI. Cosmological parameters}.
\newblock {\em Astron. Astrophys.}, 641:A6, 2020.
\newblock [Erratum: Astron.Astrophys. 652, C4 (2021)].

\bibitem{Palanque-Delabrouille:2019iyz}
Nathalie Palanque-Delabrouille, Christophe Y\`eche, Nils Sch\"oneberg, Julien
  Lesgourgues, Michael Walther, Sol\`ene Chabanier, and Eric Armengaud.
\newblock {Hints, neutrino bounds and WDM constraints from SDSS DR14
  Lyman-$\alpha$ and Planck full-survey data}.
\newblock {\em JCAP}, 04:038, 2020.

\bibitem{eBOSS:2020yzd}
Shadab Alam et~al.
\newblock {Completed SDSS-IV extended Baryon Oscillation Spectroscopic Survey:
  Cosmological implications from two decades of spectroscopic surveys at the
  Apache Point Observatory}.
\newblock {\em Phys. Rev. D}, 103(8):083533, 2021.

\bibitem{DES:2022ccp}
T.~M.~C. Abbott et~al.
\newblock {Dark Energy Survey Year 3 results: Constraints on extensions to
  \ensuremath{\Lambda}CDM with weak lensing and galaxy clustering}.
\newblock {\em Phys. Rev. D}, 107(8):083504, 2023.

\bibitem{deSalas:2017kay}
P.F. de~Salas, D.V. Forero, C.A. Ternes, M.~T{\'o}rtola, and J.W.F. Valle.
\newblock {Status of neutrino oscillations 2018: 3$\sigma$ hint for normal mass
  ordering and improved CP sensitivity}.
\newblock {\em Phys. Lett. B}, 782:633--640, 2018.

\bibitem{Esteban:2018azc}
Ivan Esteban, M.C. Gonzalez-Garcia, Alvaro Hernandez-Cabezudo, Michele Maltoni,
  and Thomas Schwetz.
\newblock {Global analysis of three-flavour neutrino oscillations: synergies
  and tensions in the determination of $\theta_{23}$, $\delta_{CP}$, and the
  mass ordering}.
\newblock {\em JHEP}, 01:106, 2019.

\bibitem{Upadhye:2017hdl}
Amol Upadhye.
\newblock {Neutrino mass and dark energy constraints from redshift-space
  distortions}.
\newblock {\em JCAP}, 05:041, 2019.

\bibitem{KiDS:2020ghu}
Tilman Tr\"oster et~al.
\newblock {KiDS-1000 Cosmology: Constraints beyond flat
  \ensuremath{\Lambda}CDM}.
\newblock {\em Astron. Astrophys.}, 649:A88, 2021.

\bibitem{Sgier:2021bzf}
Raphael Sgier, Christiane Lorenz, Alexandre Refregier, Janis Fluri, Dominik
  Z\"urcher, and Federica Tarsitano.
\newblock {Combined $13\times2$-point analysis of the Cosmic Microwave
  Background and Large-Scale Structure: implications for the $S_8$-tension and
  neutrino mass constraints}.
\newblock 10 2021.

\bibitem{Ghosh:2019tab}
Subhajit Ghosh, Rishi Khatri, and Tuhin~S. Roy.
\newblock {Can dark neutrino interactions phase out the Hubble tension?}
\newblock {\em Phys. Rev. D}, 102(12):123544, 2020.

\bibitem{Lyu:2020lps}
Kun-Feng Lyu, Emmanuel Stamou, and Lian-Tao Wang.
\newblock {Self-interacting neutrinos: Solution to Hubble tension versus
  experimental constraints}.
\newblock {\em Phys. Rev. D}, 103(1):015004, 2021.

\bibitem{Escudero:2021rfi}
Miguel Escudero and Samuel~J. Witte.
\newblock {The hubble tension as a hint of leptogenesis and neutrino mass
  generation}.
\newblock {\em Eur. Phys. J. C}, 81(6):515, 2021.

\bibitem{Gu:2021lni}
Yuchao Gu, Lei Wu, and Bin Zhu.
\newblock {Axion dark radiation: Hubble tension and the Hyper-Kamiokande
  neutrino experiment}.
\newblock {\em Phys. Rev. D}, 105(9):095008, 2022.

\bibitem{Cai:2022dkh}
Rong-Gen Cai, Zong-Kuan Guo, Shao-Jiang Wang, Wang-Wei Yu, and Yong Zhou.
\newblock {No-go guide for late-time solutions to the Hubble tension: matter
  perturbations}.
\newblock 2 2022.

\bibitem{Leauthaud:2016jdb}
Alexie Leauthaud et~al.
\newblock {Lensing is Low: Cosmology, Galaxy Formation, or New Physics?}
\newblock {\em Mon. Not. Roy. Astron. Soc.}, 467(3):3024--3047, 2017.

\bibitem{Lange:2020mnl}
Johannes~U. Lange, Alexie Leauthaud, Sukhdeep Singh, Hong Guo, Rongpu Zhou,
  Tristan~L. Smith, and Francis-Yan Cyr-Racine.
\newblock {On the halo-mass and radial scale dependence of the lensing is low
  effect}.
\newblock {\em Mon. Not. Roy. Astron. Soc.}, 502(2):2074--2086, 2021.

\bibitem{Amon:2022ycy}
A.~Amon et~al.
\newblock {Consistent lensing and clustering in a low-$S_8$ Universe with BOSS,
  DES Year 3, HSC Year 1 and KiDS-1000}.
\newblock {\em Mon. Not. Roy. Astron. Soc.}, 518(1):477--503, 2023.

\bibitem{Abazajian:2012ys}
K.N. Abazajian et~al.
\newblock {Light Sterile Neutrinos: A White Paper}.
\newblock 4 2012.

\bibitem{Arguelles:2021meu}
C.~A. Arg\"uelles, I.~Esteban, M.~Hostert, Kevin~J. Kelly, J.~Kopp, P.~A.~N.
  Machado, I.~Martinez-Soler, and Y.~F. Perez-Gonzalez.
\newblock {MicroBooNE and the \ensuremath{\nu}e Interpretation of the MiniBooNE
  Low-Energy Excess}.
\newblock {\em Phys. Rev. Lett.}, 128(24):241802, 2022.

\bibitem{MiniBooNE:2022emn}
A.~A. Aguilar-Arevalo et~al.
\newblock {MiniBooNE and MicroBooNE Combined Fit to a 3+1 Sterile Neutrino
  Scenario}.
\newblock {\em Phys. Rev. Lett.}, 129(20):201801, 2022.

\bibitem{Brinckmann:2020bcn}
Thejs Brinckmann, Jae~Hyeok Chang, and Marilena LoVerde.
\newblock {Self-interacting neutrinos, the Hubble parameter tension, and the
  cosmic microwave background}.
\newblock {\em Phys. Rev. D}, 104(6):063523, 2021.

\bibitem{RoyChoudhury:2020dmd}
Shouvik Roy~Choudhury, Steen Hannestad, and Thomas Tram.
\newblock {Updated constraints on massive neutrino self-interactions from
  cosmology in light of the $H_0$ tension}.
\newblock {\em JCAP}, 03:084, 2021.

\bibitem{Oldengott:2019lke}
Isabel~M. Oldengott, Gabriela Barenboim, Sarah Kahlen, Jordi Salvado, and
  Dominik~J. Schwarz.
\newblock {How to relax the cosmological neutrino mass bound}.
\newblock {\em JCAP}, 04:049, 2019.

\bibitem{Alvey:2021sji}
James Alvey, Miguel Escudero, and Nashwan Sabti.
\newblock {What can CMB observations tell us about the neutrino distribution
  function?}
\newblock {\em JCAP}, 02(02):037, 2022.

\bibitem{LoVerde:2014pxa}
Marilena LoVerde.
\newblock {Halo bias in mixed dark matter cosmologies}.
\newblock {\em Phys. Rev. D}, 90(8):083530, 2014.

\bibitem{Chiang:2017vuk}
Chi-Ting Chiang, Wayne Hu, Yin Li, and Marilena Loverde.
\newblock {Scale-dependent bias and bispectrum in neutrino separate universe
  simulations}.
\newblock {\em Phys. Rev. D}, 97(12):123526, 2018.

\bibitem{Chiang:2018laa}
Chi-Ting Chiang, Marilena LoVerde, and Francisco Villaescusa-Navarro.
\newblock {First detection of scale-dependent linear halo bias in $N$-body
  simulations with massive neutrinos}.
\newblock {\em Phys. Rev. Lett.}, 122(4):041302, 2019.

\bibitem{Banerjee:2019omr}
Arka Banerjee, Emanuele Castorina, Francisco Villaescusa-Navarro, Travis Court,
  and Matteo Viel.
\newblock {Weighing neutrinos with the halo environment}.
\newblock {\em JCAP}, 06:032, 2020.

\bibitem{Zhu:2013tma}
Hong-Ming Zhu, Ue-Li Pen, Xuelei Chen, Derek Inman, and Yu~Yu.
\newblock {Measurement of Neutrino Masses from Relative Velocities}.
\newblock {\em Phys. Rev. Lett.}, 113:131301, 2014.

\bibitem{Inman:2016prk}
Derek Inman, Hao-Ran Yu, Hong-Ming Zhu, J.~D. Emberson, Ue-Li Pen, Tong-Jie
  Zhang, Shuo Yuan, Xuelei Chen, and Zhi-Zhong Xing.
\newblock {Simulating the cold dark matter-neutrino dipole with TianNu}.
\newblock {\em Phys. Rev. D}, 95(8):083518, 2017.

\bibitem{Zhu:2019kzb}
Hong-Ming Zhu and Emanuele Castorina.
\newblock {Measuring dark matter-neutrino relative velocity on cosmological
  scales}.
\newblock {\em Phys. Rev. D}, 101(2):023525, 2020.

\bibitem{Zhou:2021sgl}
Shuren Zhou et~al.
\newblock {Sensitivity tests of cosmic velocity fields to massive neutrinos}.
\newblock {\em Mon. Not. Roy. Astron. Soc.}, 512(3):3319--3330, 2022.

\bibitem{Yu:2018llx}
Hao-Ran Yu, Ue-Li Pen, and Xin Wang.
\newblock {Parity-odd Neutrino Torque Detection}.
\newblock {\em Phys. Rev. D}, 99(12):123532, 2019.

\bibitem{Zhu:2014qma}
Hong-Ming Zhu, Ue-Li Pen, Xuelei Chen, and Derek Inman.
\newblock {Probing Neutrino Hierarchy and Chirality via Wakes}.
\newblock {\em Phys. Rev. Lett.}, 116(14):141301, 2016.

\bibitem{Inman:2015pfa}
Derek Inman, J.~D. Emberson, Ue-Li Pen, Alban Farchi, Hao-Ran Yu, and Joachim
  Harnois-D\'eraps.
\newblock {Precision reconstruction of the cold dark matter-neutrino relative
  velocity from $N$-body simulations}.
\newblock {\em Phys. Rev. D}, 92(2):023502, 2015.

\bibitem{Liu:2020mzl}
Yu~Liu, Yu~Yu, Hao-Ran Yu, and Pengjie Zhang.
\newblock {Neutrino effects on the morphology of cosmic large-scale structure}.
\newblock {\em Phys. Rev. D}, 101(6):063515, 2020.

\bibitem{Wong:2021ats}
Hiu~Wing Wong and Ming-chung Chu.
\newblock {Effects of neutrino masses and asymmetries on dark matter halo
  assembly}.
\newblock {\em JCAP}, 03(03):066, 2022.

\bibitem{Yu:2016yfe}
Hao-Ran Yu et~al.
\newblock {Differential Neutrino Condensation onto Cosmic Structure}.
\newblock {\em Nature Astronomy}, 1:0143, 9 2017.

\bibitem{Brandbyge:2008js}
Jacob Brandbyge and Steen Hannestad.
\newblock {Grid Based Linear Neutrino Perturbations in Cosmological N-body
  Simulations}.
\newblock {\em JCAP}, 05:002, 2009.

\bibitem{AliHaimoud:2012vj}
Yacine Ali-Haimoud and Simeon Bird.
\newblock {An efficient implementation of massive neutrinos in non-linear
  structure formation simulations}.
\newblock {\em Mon. Not. Roy. Astron. Soc.}, 428:3375--3389, 2012.

\bibitem{Archidiacono:2015ota}
Maria Archidiacono and Steen Hannestad.
\newblock {Efficient calculation of cosmological neutrino clustering in the
  non-linear regime}.
\newblock {\em JCAP}, 06:018, 2016.

\bibitem{Dakin:2017idt}
Jeppe Dakin, Jacob Brandbyge, Steen Hannestad, Troels Haugbølle, and Thomas
  Tram.
\newblock {$\nu$CO$N$CEPT: Cosmological neutrino simulations from the
  non-linear Boltzmann hierarchy}.
\newblock {\em JCAP}, 02:052, 2019.

\bibitem{Mccarthy:2017yqf}
Ian~G. Mccarthy, Simeon Bird, Joop Schaye, Joachim Harnois-Deraps, Andreea~S.
  Font, and Ludovic Van~Waerbeke.
\newblock {The BAHAMAS project: the CMB\textendash{}large-scale structure
  tension and the roles of massive neutrinos and galaxy formation}.
\newblock {\em Mon. Not. Roy. Astron. Soc.}, 476(3):2999--3030, 2018.

\bibitem{Bird:2018all}
Simeon Bird, Yacine Ali-Ha\"\i{}moud, Yu~Feng, and Jia Liu.
\newblock {An Efficient and Accurate Hybrid Method for Simulating Non-Linear
  Neutrino Structure}.
\newblock {\em Mon. Not. Roy. Astron. Soc.}, 481(2):1486--1500, 2018.

\bibitem{Chen:2020bdf}
Joe~Zhiyu Chen, Amol Upadhye, and Yvonne Y.~Y. Wong.
\newblock {The cosmic neutrino background as a collection of fluids in
  large-scale structure simulations}.
\newblock {\em JCAP}, 03:065, 2021.

\bibitem{Chen:2020kxi}
Joe~Zhiyu Chen, Amol Upadhye, and Yvonne Y.~Y. Wong.
\newblock {One line to run them all: SuperEasy massive neutrino linear response
  in $N$-body simulations}.
\newblock {\em JCAP}, 04:078, 2021.

\bibitem{Inman:2020oda}
Derek Inman and Hao-ran Yu.
\newblock {Simulating the Cosmic Neutrino Background using Collisionless
  Hydrodynamics}.
\newblock {\em Astrophys. J. Suppl.}, 250(1):21, 2020.

\bibitem{Brandbyge:2008rv}
Jacob Brandbyge, Steen Hannestad, Troels Haugb\o{}lle, and Bjarne Thomsen.
\newblock {The Effect of Thermal Neutrino Motion on the Non-linear Cosmological
  Matter Power Spectrum}.
\newblock {\em JCAP}, 08:020, 2008.

\bibitem{Viel:2010bn}
Matteo Viel, Martin~G. Haehnelt, and Volker Springel.
\newblock {The effect of neutrinos on the matter distribution as probed by the
  Intergalactic Medium}.
\newblock {\em JCAP}, 06:015, 2010.

\bibitem{Villaescusa-Navarro:2013pva}
Francisco Villaescusa-Navarro, Federico Marulli, Matteo Viel, Enzo Branchini,
  Emanuele Castorina, Emiliano Sefusatti, and Shun Saito.
\newblock {Cosmology with massive neutrinos I: towards a realistic modeling of
  the relation between matter, haloes and galaxies}.
\newblock {\em JCAP}, 03:011, 2014.

\bibitem{Castorina:2015bma}
Emanuele Castorina, Carmelita Carbone, Julien Bel, Emiliano Sefusatti, and
  Klaus Dolag.
\newblock {DEMNUni: The clustering of large-scale structures in the presence of
  massive neutrinos}.
\newblock {\em JCAP}, 07:043, 2015.

\bibitem{Banerjee:2016zaa}
Arka Banerjee and Neal Dalal.
\newblock {Simulating nonlinear cosmological structure formation with massive
  neutrinos}.
\newblock {\em JCAP}, 11:015, 2016.

\bibitem{Banerjee:2018bxy}
Arka Banerjee, Devon Powell, Tom Abel, and Francisco Villaescusa-Navarro.
\newblock {Reducing Noise in Cosmological N-body Simulations with Neutrinos}.
\newblock {\em JCAP}, 09:028, 2018.

\bibitem{Brandbyge:2018tvk}
Jacob Brandbyge, Steen Hannestad, and Thomas Tram.
\newblock {Momentum space sampling of neutrinos in $N$-body simulations}.
\newblock {\em JCAP}, 03:047, 2019.

\bibitem{Bayer:2020tko}
Adrian~E. Bayer, Arka Banerjee, and Yu~Feng.
\newblock {A fast particle-mesh simulation of non-linear cosmological structure
  formation with massive neutrinos}.
\newblock {\em JCAP}, 01:016, 2021.

\bibitem{Elbers:2020lbn}
Willem Elbers, Carlos~S. Frenk, Adrian Jenkins, Baojiu Li, and Silvia Pascoli.
\newblock {An optimal non-linear method for simulating relic neutrinos}.
\newblock {\em Mon. Not. Roy. Astron. Soc.}, 507(2):2614--2631, 2021.

\bibitem{Elbers:2022xid}
Willem Elbers.
\newblock {Geodesic motion and phase-space evolution of massive neutrinos}.
\newblock {\em JCAP}, 11:058, 2022.

\bibitem{Parimbelli:2022pmr}
G.~Parimbelli, C.~Carbone, J.~Bel, B.~Bose, M.~Calabrese, E.~Carella, and
  M.~Zennaro.
\newblock {DEMNUni: comparing nonlinear power spectra prescriptions in the
  presence of massive neutrinos and dynamical dark energy}.
\newblock {\em JCAP}, 11:041, 2022.

\bibitem{Brandbyge:2009ce}
Jacob Brandbyge and Steen Hannestad.
\newblock {Resolving Cosmic Neutrino Structure: A Hybrid Neutrino N-body
  Scheme}.
\newblock {\em JCAP}, 01:021, 2010.

\bibitem{Crocce:2005xy}
Martin Crocce and Roman Scoccimarro.
\newblock {Renormalized cosmological perturbation theory}.
\newblock {\em Phys. Rev. D}, 73:063519, 2006.

\bibitem{Crocce:2005xz}
Martin Crocce and Roman Scoccimarro.
\newblock {Memory of initial conditions in gravitational clustering}.
\newblock {\em Phys. Rev. D}, 73:063520, 2006.

\bibitem{McDonald:2006hf}
Patrick McDonald.
\newblock {Dark matter clustering: a simple renormalization group approach}.
\newblock {\em Phys. Rev. D}, 75:043514, 2007.

\bibitem{Valageas:2006bi}
Patrick Valageas.
\newblock {Large-N expansions applied to gravitational clustering}.
\newblock {\em Astron. Astrophys.}, 465:725, 2007.

\bibitem{Crocce:2007dt}
Martin Crocce and Roman Scoccimarro.
\newblock {Nonlinear Evolution of Baryon Acoustic Oscillations}.
\newblock {\em Phys. Rev. D}, 77:023533, 2008.

\bibitem{Matsubara:2007wj}
Takahiko Matsubara.
\newblock {Resumming Cosmological Perturbations via the Lagrangian Picture:
  One-loop Results in Real Space and in Redshift Space}.
\newblock {\em Phys. Rev. D}, 77:063530, 2008.

\bibitem{Taruya:2007xy}
Atsushi Taruya and Takashi Hiramatsu.
\newblock {A Closure Theory for Non-linear Evolution of Cosmological Power
  Spectra}.
\newblock {\em Astrophys. J.}, 674:617, 2008.

\bibitem{Matsubara:2008wx}
Takahiko Matsubara.
\newblock {Nonlinear perturbation theory with halo bias and redshift-space
  distortions via the Lagrangian picture}.
\newblock {\em Phys. Rev. D}, 78:083519, 2008.
\newblock [Erratum: Phys.Rev.D 78, 109901 (2008)].

\bibitem{Pietroni:2008jx}
Massimo Pietroni.
\newblock {Flowing with Time: a New Approach to Nonlinear Cosmological
  Perturbations}.
\newblock {\em JCAP}, 10:036, 2008.

\bibitem{Lesgourgues:2009am}
Julien Lesgourgues, Sabino Matarrese, Massimo Pietroni, and Antonio Riotto.
\newblock {Non-linear Power Spectrum including Massive Neutrinos: the Time-RG
  Flow Approach}.
\newblock {\em JCAP}, 06:017, 2009.

\bibitem{Garny:2020ilv}
Mathias Garny and Petter Taule.
\newblock {Loop corrections to the power spectrum for massive neutrino
  cosmologies with full time- and scale-dependence}.
\newblock {\em JCAP}, 01:020, 2021.

\bibitem{Fuhrer:2014zka}
Florian F\"uhrer and Yvonne Y.~Y. Wong.
\newblock {Higher-order massive neutrino perturbations in large-scale
  structure}.
\newblock {\em JCAP}, 03:046, 2015.

\bibitem{Dupuy:2013jaa}
H\'el\`ene Dupuy and Francis Bernardeau.
\newblock {Describing massive neutrinos in cosmology as a collection of
  independent flows}.
\newblock {\em JCAP}, 01:030, 2014.

\bibitem{Dupuy:2014vea}
H\'el\`ene Dupuy and Francis Bernardeau.
\newblock {Cosmological Perturbation Theory for streams of relativistic
  particles}.
\newblock {\em JCAP}, 03:030, 2015.

\bibitem{Dupuy:2015ega}
H\'el\`ene Dupuy and Francis Bernardeau.
\newblock {On the importance of nonlinear couplings in large-scale neutrino
  streams}.
\newblock {\em JCAP}, 08:053, 2015.

\bibitem{McEwen:2016fjn}
Joseph~E. McEwen, Xiao Fang, Christopher~M. Hirata, and Jonathan~A. Blazek.
\newblock {FAST-PT: a novel algorithm to calculate convolution integrals in
  cosmological perturbation theory}.
\newblock {\em JCAP}, 09:015, 2016.

\bibitem{Fang:2016wcf}
Xiao Fang, Jonathan~A. Blazek, Joseph~E. McEwen, and Christopher~M. Hirata.
\newblock {FAST-PT II: an algorithm to calculate convolution integrals of
  general tensor quantities in cosmological perturbation theory}.
\newblock {\em JCAP}, 02:030, 2017.

\bibitem{Schmittfull:2016jsw}
Marcel Schmittfull, Zvonimir Vlah, and Patrick McDonald.
\newblock {Fast large scale structure perturbation theory using one-dimensional
  fast Fourier transforms}.
\newblock {\em Phys. Rev. D}, 93(10):103528, 2016.

\bibitem{McDonald:2009dh}
Patrick McDonald and Arabindo Roy.
\newblock {Clustering of dark matter tracers: generalizing bias for the coming
  era of precision LSS}.
\newblock {\em JCAP}, 08:020, 2009.

\bibitem{Chen:2022dsv}
Joe~Zhiyu Chen, Markus~R. Mosbech, Amol Upadhye, and Yvonne Y.~Y. Wong.
\newblock {Hybrid multi-fluid-particle simulations of the cosmic neutrino
  background}.
\newblock {\em JCAP}, 03:012, 2023.

\bibitem{Ma:1995ey}
Chung-Pei Ma and Edmund Bertschinger.
\newblock {Cosmological perturbation theory in the synchronous and conformal
  Newtonian gauges}.
\newblock {\em Astrophys. J.}, 455:7--25, 1995.

\bibitem{Upadhye:2013ndm}
Amol Upadhye, Rahul Biswas, Adrian Pope, Katrin Heitmann, Salman Habib, Hal
  Finkel, and Nicholas Frontiere.
\newblock {Large-Scale Structure Formation with Massive Neutrinos and Dynamical
  Dark Energy}.
\newblock {\em Phys. Rev. D}, 89(10):103515, 2014.

\bibitem{Audren:2011ne}
Benjamin Audren and Julien Lesgourgues.
\newblock {Non-linear matter power spectrum from Time Renormalisation Group:
  efficient computation and comparison with one-loop}.
\newblock {\em JCAP}, 10:037, 2011.

\bibitem{Scoccimarro:1999ed}
Roman Scoccimarro, H.~M.~P. Couchman, and Joshua~A. Frieman.
\newblock {The Bispectrum as a Signature of Gravitational Instability in
  Redshift-Space}.
\newblock {\em Astrophys. J.}, 517:531--540, 1999.

\bibitem{Yankelevich:2018uaz}
Victoria Yankelevich and Cristiano Porciani.
\newblock {Cosmological information in the redshift-space bispectrum}.
\newblock {\em Mon. Not. Roy. Astron. Soc.}, 483(2):2078--2099, 2019.

\bibitem{Vollmer:2014pma}
Adrian Vollmer, Luca Amendola, and Riccardo Catena.
\newblock {Efficient implementation of the Time Renormalization Group}.
\newblock {\em Phys. Rev. D}, 93(4):043526, 2016.

\bibitem{Ringwald:2004np}
Andreas Ringwald and Yvonne~Y.Y. Wong.
\newblock {Gravitational clustering of relic neutrinos and implications for
  their detection}.
\newblock {\em JCAP}, 12:005, 2004.

\bibitem{Wong:2008ws}
Yvonne~Y.Y. Wong.
\newblock {Higher order corrections to the large scale matter power spectrum in
  the presence of massive neutrinos}.
\newblock {\em JCAP}, 10:035, 2008.

\bibitem{Galassi_2009}
M.~Galassi et~al.
\newblock {\em GNU Scientific Library Reference Manual - Third Edition}.
\newblock 2009.

\bibitem{Moran:2022iwe}
Kelly~R. Moran, Katrin Heitmann, Earl Lawrence, Salman Habib, Derek Bingham,
  Amol Upadhye, Juliana Kwan, David Higdon, and Richard Payne.
\newblock {The Mira\textendash{}Titan Universe \textendash{} IV. High-precision
  power spectrum emulation}.
\newblock {\em Mon. Not. Roy. Astron. Soc.}, 520(3):3443--3458, 2023.

\bibitem{KatanaCluster}
Katana, 2010.
\newblock Published online at {\tt https://doi.org/10.26190/669x-a286}~.

\bibitem{Springel:2020plp}
Volker Springel, R\"udiger Pakmor, Oliver Zier, and Martin Reinecke.
\newblock {Simulating cosmic structure formation with the gadget-4 code}.
\newblock {\em Mon. Not. Roy. Astron. Soc.}, 506(2):2871--2949, 2021.

\bibitem{Elbers:2022tvb}
Willem Elbers, Carlos~S. Frenk, Adrian Jenkins, Baojiu Li, and Silvia Pascoli.
\newblock {Higher order initial conditions with massive neutrinos}.
\newblock {\em Mon. Not. Roy. Astron. Soc.}, 516(3):3821--3836, 2022.

\end{thebibliography}
\end{document}